# The Physics of Fast Radio Bursts

Di Xiao[1,2*], Fayin Wang[1,2*], and Zigao Dai[1,2*]

[1] School of Astronomy and Space Science, Nanjing University, Nanjing 210023, China;
[2] Key Laboratory of Modern Astronomy and Astrophysics (Nanjing University), Ministry of Education, China

In 2007, a very bright radio pulse was identified in the archival data of the Parkes Telescope in Australia, marking the beginning of a new research branch in astrophysics. In 2013, this kind of millisecond bursts with extremely high brightness temperature takes a unified name, fast radio burst (FRB). Over the first few years, FRBs seemed very mysterious because the sample of known events was limited. With the improvement of instruments over the last five years, hundreds of new FRBs have been discovered. The field is now undergoing a revolution and understanding of FRB has rapidly increased as new observational data increasingly accumulates. In this review, we will summarize the basic physics of FRBs and discuss the current research progress in this area. We have tried to cover a wide range of FRB topics, including the observational property, propagation effect, population study, radiation mechanism, source model, and application in cosmology. A framework based on the latest observational facts is now under construction. In the near future, this exciting field is expected to make significant breakthroughs.

## 1 Introduction

Fast Radio Bursts (FRBs) are transient radio pulses of millisecond durations that flash randomly in the sky. They were first discovered by the Parkes Telescope in 2007 [1]. Despite being confused with rotating radio transients [2] and "perytons" [3] in the beginning, FRBs were finally deemed to be new astrophysical signals, thanks to a new sample of FRBs identified in 2013 [4]. Since then, the scientific community has been increasingly focusing on them and theorists have been investigating the underlying physics. Knowledge regarding FRBs has taken a great leap forward since the landmark discovery of the first repeating FRB 121102 [5, 6] and further the identification of its host galaxy [7-9]. In the following years, an increasing number of events have been collected [10], benefiting from the improved observational techniques and new facilities. At the time of writing about twenty repeaters have been reported mainly by the Canadian Hydrogen Intensity Mapping Experiment (CHIME) [11-13] which has a large field of view and is extremely powerful in finding new FRBs. The interferometric arrays, represented by Australian Square Kilometre Array Pathfinder (ASKAP), can accurately localize individual bursts [14, 15]. The recent two breakthroughs in 2020 are even more exciting: the periodical activity has first been found for the well-known repeater FRB 180916 [16] and later for FRB 121102 [17, 18]. Moreover, the association of FRB 200428 with short X-ray bursts (XRBs) from a Galactic magnetar SGR 1935+2154 has been confirmed [19-25], shedding new light on the origin of FRBs.

*Corresponding authors (Di Xiao, email: dxiao@nju.edu.cn; Fayin Wang, email: fayinwang@nju.edu.cn; Zigao Dai, email: dzg@nju.edu.cn)

As we deduce from the above milestones, the field of FRB is currently flourishing. The readers may refer to some previous good reviews [26-29] for the progress till 2019. As this field is evolving fast and our understanding is being constantly updated, we feel it is a proper time for the latest review to appear. At the time of writing, we noticed a newly-published, well-written review article on the FRB radiation mechanism [30], and a brief introduction of recent observational progress [31]. With better data resolution and a larger sample of FRBs, we are now reaching a comprehensive understanding of this phenomenon. First, we get to know some individual bursts better, including their pulse structure, spectrum, polarization properties, and radiation mechanism. Second, the population research is also possible and we could study the classification, event rate and a few statistical properties of FRBs. Last, there is a growing trend that FRBs are treated as powerful probes of fundamental physics and cosmology. In this review, we will discuss all these aspects on the basis of the latest discoveries.

This review is organized as follows. In Section 2 we introduce the basic observational facts of FRBs. Section 3 & 4 present the recent progress on statistical properties and population study of FRBs. The possible radiation mechanisms and source models of FRBs are discussed in Section 5 & 6 respectively. The application of FRBs in cosmology is explored in Section 7. Finally, we provide a summary and some thoughts regarding future prospects in Section 8.

# 2 General Observational Properties

There is no strict definition for FRBs right now. Generally, an FRB is defined as a radio pulse with duration $T\sim$ ms and extremely high brightness temperature $T_{\rm B}$, which can be obtained by equaling the observed intensity $I_\nu$ with blackbody luminosity at the Rayleigh-Jeans end,

$$I_\nu = 2kT_{\rm B}/\lambda^2, \tag{1}$$

where $k$ is Boltzmann constant and $\lambda = c/\nu$ is the radio wavelength. $I_\nu$ can be related to flux density $F_\nu$ as $I_\nu \sim F_\nu/\Omega$, where the observed solid angle is determined by the source size and angular distance, i.e., $\Omega\sim\pi(cT)^2/d_{\rm A}^2$ [30]. Thus, the brightness temperature of an FRB with typical scaling is

$$\begin{aligned} T_{\rm B} &\sim F_\nu d_{\rm A}^2/2\pi k(\nu T)^2 \\ &\simeq 1.1\times10^{35}\,{\rm K}\left(\frac{F_\nu}{\rm Jy}\right)\left(\frac{\nu}{\rm GHz}\right)^{-2}\left(\frac{T}{\rm ms}\right)^{-2}\left(\frac{d_{\rm A}}{\rm Gpc}\right)^2, \end{aligned} \tag{2}$$

which is several orders of magnitude higher than that of the normal pulsar radio emission.

Radio waves can be easily affected by the intervening medium between the source and Earth observer. First, the radio pulse is dispersed in the plasma, i.e., the high frequency photons arrive earlier than the low energy ones. Second, the pulse can be broadened due to scattering with an extended screen. Third, diffraction and refraction by turbulent gas make the radio signal scintillate occasionally. Fourth, these refractive plasma could lead to lensing effect. Fifth, sometimes the absorption of radio emission is important if the plasma is dense enough. Last but not the least, the polarization property can be changed as radio wave propagates in the magnetized plasma. We will introduce these propagation effects of FRBs respectively in the following subsections.

## 2.1 Dispersion

The dispersion relation in a cold plasma is

$$\omega^2 = \omega_{\rm p}^2 + k^2c^2,\ \ \text{where}\ \ \omega_{\rm p} = \sqrt{4\pi n_e e^2/m_e}, \tag{3}$$

so the group velocity of the electromagnetic (EM) wave is

$$v_{\rm g} = \frac{{\rm d}\omega}{{\rm d}k} = c\sqrt{1-(\omega_{\rm p}/\omega)^2}. \tag{4}$$

The delay time of a signal traveling through the plasma is

$$\begin{aligned} t - t_0 &= \int_0^d {\rm d}l\left(\frac{{\rm d}k}{{\rm d}\omega} - \frac{1}{c}\right) \\ &= \int_0^d \frac{{\rm d}l}{c}\left(\frac{1}{2}\left(\frac{\omega_{\rm p}}{\omega}\right)^2 + \frac{3}{8}\left(\frac{\omega_{\rm p}}{\omega}\right)^4 + \frac{5}{16}\left(\frac{\omega_{\rm p}}{\omega}\right)^6 + ...\right). \end{aligned} \tag{5}$$

In radio astronomy the dispersion measure (DM) is defined as the electron number density integrated along the traveled path ${\rm DM} = \int n_e {\rm d}l/(1+z)$ where $z$ is the redshift, so that the delayed arrival time of two photons with frequencies $\nu_1$, $\nu_2$ ($\nu_1 < \nu_2$) can be expressed as [28, 32],

$$\begin{aligned} \Delta t &= \frac{e^2}{2\pi m_e c}\left(\frac{1}{\nu_1^2} - \frac{1}{\nu_2^2}\right)\int\frac{n_e}{1+z}{\rm d}l \\ &\simeq 4.15\,{\rm s}\left[\left(\frac{\nu_1}{1\,{\rm GHz}}\right)^{-2} - \left(\frac{\nu_2}{1\,{\rm GHz}}\right)^{-2}\right]\frac{\rm DM}{10^3\,{\rm pc\,cm^{-3}}}, \end{aligned} \tag{6}$$

Figure 1 shows the dynamic spectrum ("waterfall" plot) of the first-discovered event, FRB 010724 (also know as "Lorimer burst") [1]. The dispersive "sweep" clearly indicates that high energy photons arrive earlier than low energy ones. This burst has a total DM of about 375 $\rm pc\,cm^{-3}$, which exceeds the DM budget that could be contributed by electrons in the Milky Way, suggesting an extragalactic origin.

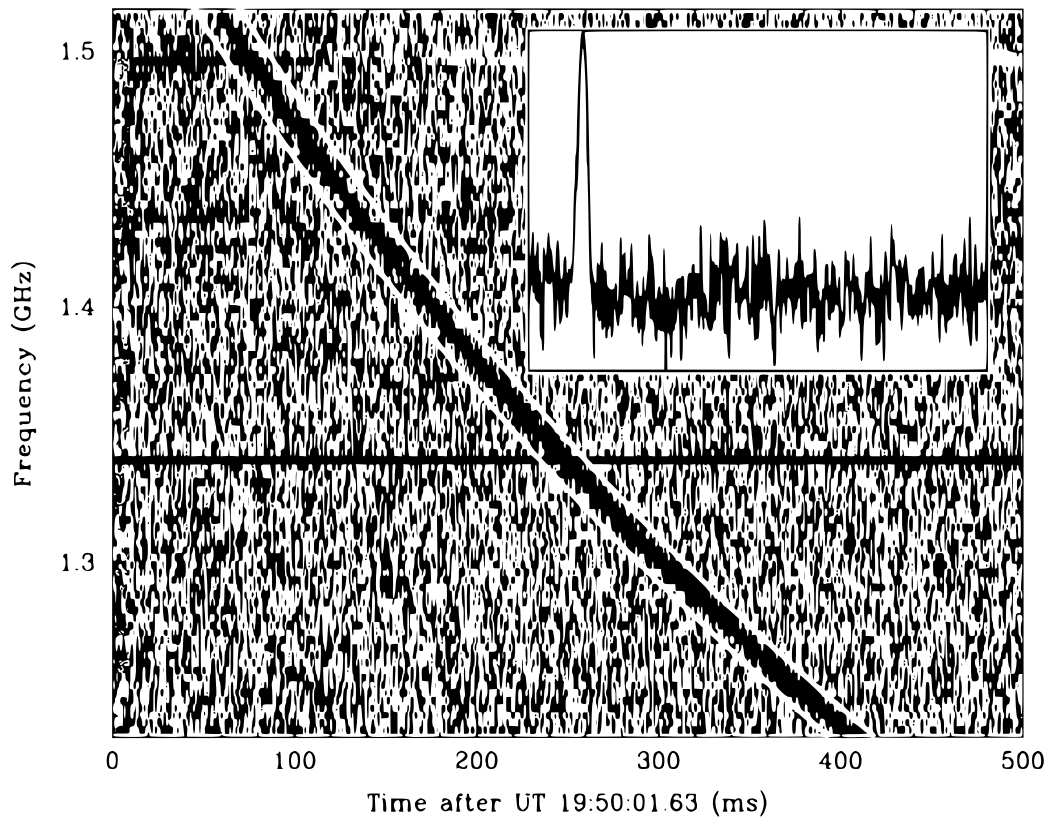

**Figure 1** Dynamic spectrum of the Lorimer burst. (Adapted from [1]. Reprinted with permission from AAAS.)

Generally, the total DM of an FRB at redshift $z$ can be separated as

$$\mathrm{DM} = \mathrm{DM_{MW}} + \mathrm{DM_{IGM}} + \frac{\mathrm{DM_{host}} + \mathrm{DM_{source}}}{1+z}, \tag{7}$$

where the subscripts MW, IGM, host, source refer to the contributions by the plasma of Milky Way, intergalactic medium, host galaxy and source local environment respectively. $\mathrm{DM_{MW}}$ can be obtained from Galactic electron density models such as NE2001 [33] and YMW16 [34]. It is also useful to define an excess of DM with respect to the Milky Way value, i.e.,

$$\mathrm{DM_E} = \mathrm{DM} - \mathrm{DM_{MW}} = \mathrm{DM_{IGM}} + \frac{\mathrm{DM_{host}} + \mathrm{DM_{source}}}{1+z}. \tag{8}$$

The average $\langle \mathrm{DM_{IGM}} \rangle$ as a function of redshift is given by [32]:

$$\langle \mathrm{DM_{IGM}} \rangle = \frac{3c\Omega_b H_0}{8\pi G m_p} \int_0^z \frac{F(z')}{E(z')} \mathrm{d}z', \tag{9}$$

where $H_0$ is the Hubble constant, $\Omega_b$ is the cosmic baryon mass density fraction, $E(z) = H(z)/H_0$ and $F(z) = (1+z) f_{\rm IGM}(z) f_e(z)$ with $f_{\rm IGM}$ being the fraction of the baryon mass in the IGM. Further, $f_e = Y_{\rm H} X_{\rm e,H}(z) + \frac{1}{2} Y_{\rm He} X_{\rm e,He}(z)$, where $Y_{\rm H}$, $Y_{\rm He}$ are the mass fractions of hydrogen and helium, and $X_{\rm e,H}$, $X_{\rm e,He}$ are the ionization fractions of intergalactic hydrogen and helium, respectively. It is worth mentioning that the $\mathrm{DM_{IGM}}$ term has been applied widely in cosmology, which will be discussed in detail later in Section 7.

## 2.2 Pulse Width and Scattering Effect

The pulse width is commonly defined as the full width at half maximum. The observed FRB width is a combined result of the intrinsic width $W_i$, instrumental broadening and, propagation effect [28],

$$W = \sqrt{W_i^2 + t_{\rm samp}^2 + \Delta t_{\rm DM}^2 + \Delta t_{\rm DMerr}^2 + \tau^2}, \tag{10}$$

where $t_{\rm samp}$ is the data sampling interval and $\tau$ is the scattering timescale. The dispersion smearing due to finite channel bandwidth $\Delta\nu$ of the receiver is [35]

$$\Delta t_{\rm DM} = 8.3\,\mu{\rm s}\,\mathrm{DM}\Delta\nu_{\rm MHz}\nu_{\rm GHz}^{-3}. \tag{11}$$

Also, an error of trial DM (i.e., $\mathrm{DM_{err}}$) with respect to the true value in dedispersion could cause additional smearing, which is expressed as $\Delta t_{\rm DMerr} = \Delta t_{\rm DM}(\mathrm{DM_{err}}/\mathrm{DM})$ [35].

A radio photon can be easily scattered by the particles on its path. As a result, the moving direction of this photon could be changed to some extent. For an FRB event, a portion of its photons could be scattered and travel a longer way then arrive later than the unscattered ones, giving rise to a tail feature of the pulse profile. Usually, this tail appears as an exponential decay, which can be well modeled by the scattering with a thin extended screen. The scatter-broadening time scales with frequency as $\tau \propto \nu^{-\alpha}$ and $\alpha \simeq 4.0$ for the above screen [36]. However, if a Kolmogorov spectrum of scattering medium is assumed, the index could be $\alpha \simeq 4.4$ if the minimum turbulence scale is smaller than the diffractive length scale [35, 37].

Practically, the scattering timescale of an FRB is determined by fitting its pulse profile. First we need to make a few assumptions on its intrinsic pulse shape (usually assumed a Gaussian shape), the receiver noise and instrumental smearing. Distinctions in assumptions lead to a diversity in FRB signal models. Then these models are applied to each individual FRB through a Markov-Chain Monte-Carlo (MCMC) fitting of the pulse flux. Last, the goodness of fitting and the preferred model is determined by the Bayes Information Criterion. Ravi (2019) collected a sample of 17 Parkes FRBs and studied their scattering properties [38]. He found the typical scattering timescale $\tau$ ranges from submilliseconds to tens of milliseconds. Further, FRBs are underscattered in the plane of $\tau - \mathrm{DM_E}$, compared with the $\tau - \mathrm{DM}$ relation of Galactic pulsars [39-41]. These have been confirmed by a latest analysis on a sample of 33 ASKAP FRBs [42].

## 2.3 Scintillation

The scintillation of FRBs is akin to the twinkling of stars. However, the latter is caused by turbulent atmosphere of the Earth, while the former is due to turbulent plasma medium. Once the turbulent intervening plasma moves perpendicularly to our line of sight (LOS) with a velocity $V$, the interference fringes and diffraction fringes would vary accordingly and observer on Earth could be in the position of either maximum or minimum intensity. This leads to a variation of observed flux density, depending strongly on the frequency.

Also, the variability of scintillation pattern depends on turbulence spectrum and the relative motion between the medium and observer.

The physics of pulsar scintillation has been well reviewed before [43, 44]. Random phase fluctuations $\phi(x, y)$ are generated when the plane waves pass through a turbulent scattering screen. The diffractive length scale $r_{\rm diff}$ is defined as the transverse radius at which the root mean square phase difference is 1 rad [44]. If $r_{\rm diff}$ is larger than the Fresnel scale $r_{\rm F}$, which is expressed as $r_{\rm F} = \sqrt{\lambda D/2\pi}$ with $D$ being the distance between the screen and observer, the perturbations to wavefronts are weak and this is called the weak scattering regime. The observed scintillation timescale is the Fresnel timescale $t_{\rm scint} \simeq t_{\rm F} = r_{\rm F}/V$. In the opposite strong scattering regime ($r_{\rm diff} < r_{\rm F}$), diffractive interstellar scintillation (DISS) plays an important role and the scintillation timescale is $t_{\rm scint} \simeq r_{\rm diff}/V$. Detailed discussion of the dependence of DISS timescale on velocities of the source, scattering medium and observer was given by Cordes & Rickett (1998) [45]. They found that the scintillation bandwidth $\Delta\nu_{\rm scint}$ is related to scattering $\tau$ by $2\pi\Delta\nu_{\rm scint}\tau = C_1$ where $C_1$ is a factor of order unity. Combined with the previous discussions on $\tau$, we get $\Delta\nu_{\rm scint} \propto \nu^4$. Scintillation phenomena that have been observed in a few cases like FRB 150807 and FRB 121102 match the predictions well [46, 47]. Note that the spectral structure of the recent Galactic FRB 200428 could be potentially explained by scintillation [48]. Besides, the refractive interstellar scintillation (RISS) may occur on a larger scale $r_{\rm ref} = r_{\rm F}^2/r_{\rm diff}$, of which the scintillation timescale $t_{\rm scint} \simeq r_{\rm ref}/V$ could be much longer than the FRB burst duration [6].

## 2.4 Plasma Lensing

Ionized plasma is refractive that radio waves are deflected while passing through. The plasma refractive index is given by $n_{\rm p} = \sqrt{1 - \omega_{\rm p}^2/\omega^2}$. In many ways plasma lensing is analogous to gravitational lensing, however, the differences between them are also obvious. In general, gravitational lenses are converging and achromatic [49], while plasma lenses are diverging and highly chromatic [50]. If the diverging plasma lens is on the axis between the source and the observer, the observed flux will be in its minimum. However, this lens can also cause large magnification (caustic spots) if it lies offset. Therefore the variability of light curves is enriched due to the source-lens-observer geometry.

The mathematical prescription of plasma lensing is very similar to that of gravitational lensing. The major difference lies in the formulae of effective deflection potential and a detailed comparison can be found in Wagner & Er (2020) [51]. The electron density distribution on the lens plane will influence the deflection angle to a great extent and different lens models have been developed. The most commonly-adopted model is one-dimension Gaussian lens [50, 52] and more complicated light curves can be realized if other lens models are assumed [53, 54].

Plasma lensing effect could be relevant for FRBs in various aspects. It can introduce an additional frequency-dependent delay of the arrival time so that the inferred DM should be examined carefully [55]. The impact of plasma lensing on the observed FRB luminosity and spectrum has been discussed [52]. Moreover, the complex time-frequency pulse structure of FRB 121102 could be possibly explained by plasma lensing [56]. The downward frequency drift in a sequence of sub-bursts ("sad trombone" effect) can be attributed to plasma lensing. However, upward drifting is also expected in this scenario but has not been observed in the majority of FRBs. Even though the sign of upward drifting appears for FRB 190611 [57] and FRB 200428 [20], it happens much less frequently than downward drifting. Note that several investigations have been done to explain the sad trombone effect in distinct scenarios. Within the framework of coherent curvature emission in the magnetosphere, Wang et al. (2019) proposed a generic geometrical model that EM waves of different frequencies are emitted by bunched particles streaming along field lines of different curvature radii [58]. Alternatively, Metzger et al. (2019) suggested that the sad trombone effect could be attributed to the deceleration of the blast wave, assuming the FRB mechanism is synchrotron maser emission [59]. A detailed discussion of this model can be found in Section 6.1.

## 2.5 Polarization and Faraday Rotation Effect

Polarization refers to the behaviour of the electric field vector $\boldsymbol{E}$ with time. Owing to the variety of the trajectory that $\boldsymbol{E}$ draws on the plane perpendicular to wave propagation, we have linearly, circularly, elliptically polarized or totally unpolarized light. Suppose an EM wave propagates in the same $z$-axis direction with the magnetic field of magnetized plasma ($\boldsymbol{B} = B_0\boldsymbol{e}_z$), then $\boldsymbol{E}$ oscillates in the $x - y$ plane and the equation of motion for an electron is

$$m_e\ddot{\boldsymbol{r}}_{\rm p} = -e\boldsymbol{E} - e\dot{\boldsymbol{r}}_{\rm p} \times (B_0\boldsymbol{e}_z). \tag{12}$$

For the plane wave $\boldsymbol{E} = \boldsymbol{E}_0 e^{i(\boldsymbol{k}\cdot\boldsymbol{r}-\omega t)}$, and the electric polarization intensity $\boldsymbol{P} = -n_e e\boldsymbol{r}_{\rm p} = (\epsilon-\epsilon_0)\boldsymbol{E}$, we have $\boldsymbol{r}_{\rm p} = -\frac{(\epsilon-\epsilon_0)}{n_e e}\boldsymbol{E}$. Substituting into Eq. (12), we get

$$\begin{aligned} &\left[\frac{m_e\omega^2}{n_e e}(\epsilon - \epsilon_0) + e\right]E_x + {\rm i}\omega B_0\frac{\epsilon - \epsilon_0}{n_e}E_y = 0 \\ &\left[\frac{m_e\omega^2}{n_e e}(\epsilon - \epsilon_0) + e\right]E_y - {\rm i}\omega B_0\frac{\epsilon - \epsilon_0}{n_e}E_x = 0. \end{aligned} \tag{13}$$

The determinant of coefficient should be zero, therefore

$$\left[\frac{m_e\omega^2}{n_e e}(\epsilon-\epsilon_0)+e\right]^2=\left(\omega B_0\frac{\epsilon-\epsilon_0}{n_e}\right)^2 \tag{14}$$

Since the refractive index $n^2=\epsilon/\epsilon_0$, we obtain

$$n_\pm=\left(1-\frac{\omega_{\rm p}^2}{\omega^2\pm\omega\Omega_e}\right)^{1/2}, \tag{15}$$

where $+$, $-$ represents left and right handed wave respectively and $\Omega_e=eB_0/m_ec$ is the cyclotron frequency. A linearly polarized wave can be decomposed into left- and right-hand circularly polarized components

$$\boldsymbol{E}=E_+(\boldsymbol{e}_x-{\rm i}\boldsymbol{e}_y)e^{{\rm i}(k_+z-\omega t)}+E_-(\boldsymbol{e}_x+{\rm i}\boldsymbol{e}_y)e^{{\rm i}(k_-z-\omega t)}, \tag{16}$$

The Faraday rotation effect is the rotation of polarization direction of linearly polarized light under the influence of a magnetic field. Suppose that the magnetized plasma extends from $z=0$ to $L$ and without loss of generality we assume the EM wave is initially linearly polarized in the $x$ direction, i.e., $\boldsymbol{E}(z=0)=E_0e^{-{\rm i}\omega t}\boldsymbol{e}_x$, so $E_+=E_-=(1/2)E_0$. The electric field vector at the exit of the medium is

$$\begin{aligned}\boldsymbol{E}(z=L) &= \frac{1}{2}E_0\left[e^{{\rm i}k_+L}(\boldsymbol{e}_x-{\rm i}\boldsymbol{e}_y)+e^{{\rm i}k_-L}(\boldsymbol{e}_x+{\rm i}\boldsymbol{e}_y)\right]e^{-{\rm i}\omega t}\\ &= \frac{1}{2}E_0e^{-{\rm i}\omega t}\left[(e^{{\rm i}k_+L}+e^{{\rm i}k_-L})\boldsymbol{e}_x+{\rm i}(e^{{\rm i}k_-L}-e^{{\rm i}k_+L})\boldsymbol{e}_y\right]\\ &= E_0\left(\cos\frac{k_+-k_-}{2}L\,\boldsymbol{e}_x+\sin\frac{k_+-k_-}{2}L\,\boldsymbol{e}_y\right)e^{{\rm i}(\frac{k_++k_-}{2}L-\omega t)}\end{aligned} \tag{17}$$

Now this wave is linearly polarized at angle

$$\Psi=\frac{k_+-k_-}{2}L=\frac{\omega}{c}(n_+-n_-)\frac{L}{2} \tag{18}$$

In the limit $\omega\gg\Omega_e$, Eq. (15) can be approximated as

$$n_\pm\approx 1-\frac{1}{2}\frac{\omega_{\rm p}^2}{\omega^2\pm\omega\Omega_e}, \tag{19}$$

thus in general case, the polarization direction after traveling an infinitesimal length d$l$ will rotate:

$${\rm d}\Psi=\frac{\omega{\rm d}l}{2c}(n_+-n_-)\approx\frac{\omega_{\rm p}^2\Omega_e{\rm d}l}{2c\omega^2}=\frac{2\pi e^3}{\omega^2m_e^2c^2}n_eB_0{\rm d}l. \tag{20}$$

Therefore, the observed polarization angle (PA)

$$\Psi_{\rm obs}(\lambda)=\Psi_0+\Delta\Psi=\Psi_0+\frac{e^3\lambda^2}{2\pi m_e^2c^4}\int_0^L n_e(l)B_\parallel(l){\rm d}l, \tag{21}$$

where $\Psi_0$ is the initial PA and $B_\parallel$ is the component of the magnetic field along the LOS. In terms of the rotation measure (RM),

$${\rm RM}=\frac{e^3}{2\pi m_e^2c^4}\int_0^L n_e(l)B_\parallel(l){\rm d}l, \tag{22}$$

$\Psi_{\rm obs}$ can be written as

$$\Psi_{\rm obs}(\lambda)=\Psi_0+\lambda^2{\rm RM}. \tag{23}$$

For cosmological sources, RM value is usually scaled as

$${\rm RM}\left[\frac{\rm rad}{\rm m^2}\right]=0.812\int_0^L\frac{n_e[{\rm cm^{-3}}]B_\parallel[\mu{\rm G}]}{(1+z)^2}{\rm d}l[{\rm pc}]. \tag{24}$$

A positive RM means the magnetic field direction is towards the observer and inversely it can be negative.

Orthogonal feeds of radio telescopes can measure the full Stokes parameters ($I$, $Q$, $U$, $V$) and PA can be determined by $\Psi_{\rm obs}=0.5\tan^{-1}(U/Q)$. Then one can fit the relation with wavelength in Eq. (23) and estimate RM. More precise RM value can be obtained by means of RMFIT, RM Synthesis or QU-fitting [60-62]. Further, if DM and RM originate from the same plasma medium, we can estimate its magnetic field strength as [63]

$$\langle B_\parallel\rangle=\frac{\rm RM}{0.812{\rm DM}}\simeq 1.232\left(\frac{\rm RM}{\rm rad\,m^{-2}}\right)\left(\frac{\rm DM}{\rm pc\,cm^{-3}}\right)^{-1}\mu{\rm G}. \tag{25}$$

The polarization properties of FRBs vary from case to case. The repeaters like FRB 121102 and FRB 180916 are nearly 100% linearly polarized and have a flat PA curve (i.e., PA does not vary obviously with time) [12, 47, 64-66]. Note that the RM of FRB 121102 is the highest among all FRBs ($\sim 10^5$ rad/m$^2$) and is decreasing with time [64, 67]. The non-repeaters like FRB 181112 and FRB 190102 are usually partial linearly or circularly polarized and their PAs vary significantly with time (named "PA swing") [57, 68]. The polarization fractions are found to evolve with time for non-repeaters [57]. However, it remains questionable whether the difference in polarization favors the classification of FRBs by repeating or not. In particular, a recent observation challenges this viewpoint, i.e., the repeating FRB 180301 shows diverse PA swings [69]. It is possible that with the enlarging of the FRB sample in the future, repeaters and non-repeaters are not so different in polarization.

### 2.6 Absorption Processes

The most relevant absorption process in radio band is free-free absorption. Free electrons jump to higher energy states after absorbing photons and the optical depth of this process depends on the cross section and electron density. Generally the IGM is sparse and the free-free absorption in IGM is unimportant. Still, significant absorption could possibly happen in the vicinity of the FRB source, where the plasma is dense enough. For instance, the FRB progenitor may locate in pulsar wind nebulae where there are numerous electrons [59, 70-72]. In this case, synchrotron self-absorption (SSA) may also play an important role [73, 74]. FRBs can escape only if the optical depths $\tau_{ff}$ and $\tau_{SSA}$ are less than unity.

Early in 2020 the lowest frequency of FRB detections is still near 330 MHz for FRB 180916 [75, 76] and FRB 200125A [77]. Very recently, searches conducted at even lower frequencies (i.e., 150 MHz by LOFAR) found a few new bursts for FRB 180916 [78, 79]. This means that the low-energy spectrum cutoff lies below $\sim$ 100 MHz. It is unclear whether the low-energy cutoff is caused by absorption or intrinsic FRB radiation spectrum. However, a general constraint on the intrinsic spectrum can be given for particular bursts [75, 80].

### 2.7 Host galaxies

The host galaxies of FRBs are crucial for understanding the environment and origin of FRBs. The first localized event is repeating FRB 121102 [7]. It was localized to a low-metallicity, dwarf galaxy with stellar mass $(1.4 \pm 0.7) \times 10^8$ $M_\odot$. A radio continuum spectrum has also been found in this galaxy. The second localized repeating FRB is FRB 180916 [81], the host galaxy of which is dramatically different from that of FRB 121102. The stellar mass of this galaxy is about $2 \times 10^9$ $M_\odot$ and the star formation rate (SFR) of the birth region is about $0.06\,M_\odot yr^{-1}$. The vast difference between these two galaxies suggests that the host galaxies of repeating FRBs span a large range. FRB 190711 has also been localized and its host galaxy is also distinct [82]. Recently, it has been identified as a repeating FRB [83].

The localization of non-repeating FRBs is a big challenge. FRB 180924 was localized to a luminous galaxy at redshift 0.3212 [14]. Using the Deep Synoptic Array ten-antenna prototype (DSA-10), Ravi et al. (2019) reported the localization of FRB 190523 [84]. These two host galaxies are similar to each other. The stellar mass is about $10^{10} - 10^{11} M_\odot$ and the SFR is about $< 2\,M_\odot yr^{-1}$. Several non-repeating FRBs has also been localized [15, 82, 85-88]. We list the properties of host galaxies in Table 1 and more information can be found in FRB host database [1].

The properties of host galaxy can give clues to the progenitors of FRBs. For example, the host galaxy of FRB 121102 is similar to that of superluminous supernovae and long gamma-ray bursts (GRBs), which has led to the hypothesis that FRBs are produced by young active magnetars [89-95]. The magnetars are mainly formed through the core-collapses of massive stars that can be simultaneously observed by CHIME and third observing run of the Advanced LIGO/Virgo gravitational-wave detectors [96]. Meanwhile, FRBs may be produced by pre-merger NS-NS interactions [97] or activities of newborn magnetars from binary neutron star (BNS) mergers [98, 99], motivated by the large offsets of FRB 180924 and FRB 190523. The offset distribution of BNS mergers derived from new BSE code is consistent with the offsets of FRB 180924, FRB 190523 and short GRBs [98]. Besides rapidly spinning magnetars, the connection of FRB 200428 with SGR 1935+2154 demonstrates that regular magnetars can produce FRBs. Therefore, the FRB host galaxy could be Milky Way-type with moderate SFR. Recent works showed that both intermediate hosts and old hosts are consistent with the majority of the FRB hosts [100-102]. There is no clear sign of a bimodal distribution in SFR.

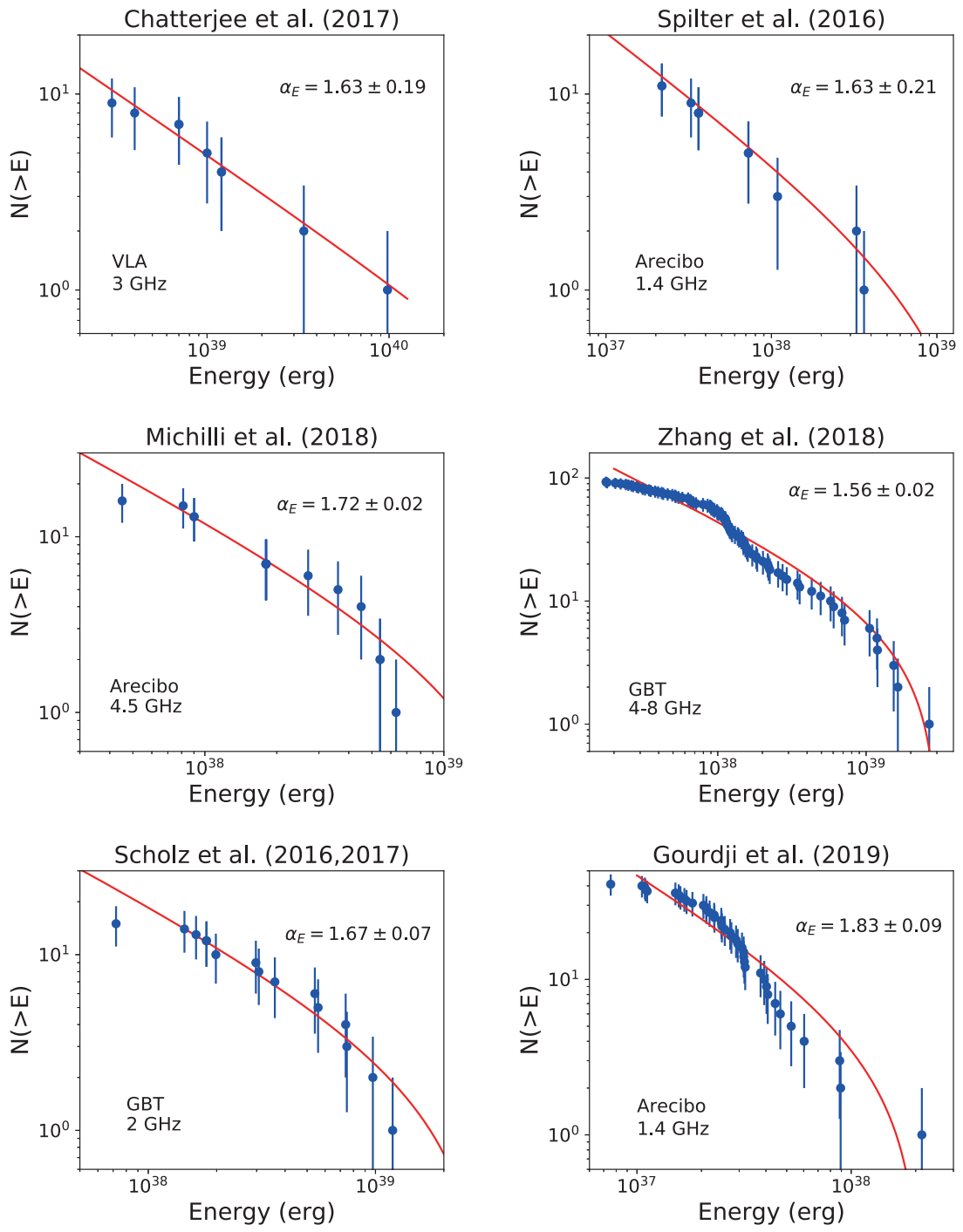

**Figure 2** The cumulative energy distributions of FRB 121102 for different observations. The value of $\alpha_E$ is in a narrow range $1.6 < \alpha_E < 1.8$. (Adapted from [103].)

## 3 Statistical properties of FRBs

As the number of FRBs increases, the statistical study of observational properties becomes important, especially for the classification and comparison with model predictions. Many works have been carried out to study FRBs statistically, which can give important constraints on the progenitors

1) http://frbhosts.org

| | FRB name | redshift | DM (pc cm$^{-3}$) | Stellar Mass ($\times 10^8$ M$_\odot$) | SFR (M$_\odot$ yr$^{-1}$) | offset (kpc) | References |
|---|---|---|---|---|---|---|---|
| Repeating FRB | FRB 121102 | 0.1927 | 557 | $1.4 \pm 0.7$ | $0.15 \pm 0.04$ | $0.6 \pm 0.3$ | [7] |
| | FRB 180916 | 0.0337 | 348.76 | $21.5 \pm 3.3$ | $0.06 \pm 0.02$ | $5.5 \pm 0.1$ | [81] |
| | FRB 190711 | 0.5220 | 593.1 | $8.1 \pm 2.9$ | $0.42 \pm 0.12$ | $3.2 \pm 2.1$ | [82] |
| Non-Repeating FRBs | FRB 180924 | 0.3212 | 362.4 | $132 \pm 51$ | $0.88 \pm 0.26$ | $3.4 \pm 0.5$ | [14] |
| | FRB 181112 | 0.4755 | 589.0 | $39.8 \pm 20.2$ | $0.37 \pm 0.11$ | $1.7 \pm 19.2$ | [85] |
| | FRB 190102 | 0.2912 | 364.5 | $33.9 \pm 10.2$ | $0.86 \pm 0.26$ | $2.0 \pm 2.2$ | [82] |
| | FRB 190523 | 0.6600 | 760.8 | $612 \pm 401$ | $< 0.09$ | $27 \pm 23$ | [84] |
| | FRB 190608 | 0.1178 | 338.7 | $116 \pm 28$ | $0.69 \pm 0.01$ | $6.6 \pm 0.6$ | [82] |
| | FRB 190611 | 0.3778 | 321.4 | $\sim 8$ | $0.27 \pm 0.08$ | $11 \pm 4$ | [82] |
| | FRB 191001 | 0.2340 | 507.9 | $464 \pm 188$ | $8.06 \pm 2.42$ | $11 \pm 1$ | [15] |
| | FRB 190614 | $\sim 0.6$ | 959.2 | $\sim 10$ | | | [88] |
| | FRB 190714 | 0.2365 | 504.1 | $149 \pm 71$ | $0.65 \pm 0.20$ | $1.9 \pm 1.1$ | [87] |
| | FRB 200430 | 0.1600 | 380.0 | $13 \pm 6$ | $\sim 0.2$ | $3.0 \pm 2.4$ | [87] |

**Table 1** Properties of FRB host galaxies.

[92, 104-111]. It has been found that the statistics of FRBs are similar to those of magnetar bursts [106, 111] and solar radio bursts [112]. This similarity also supports the association of magnetar bursts and FRBs.

### 3.1 Energy distribution

From observation, burst energy can be calculated through

$$E = \frac{4\pi d_{\rm L}^2 F \nu_{\rm c}}{1+z}, \tag{26}$$

where $d_{\rm L}$ is luminosity distance, $F$ is the burst fluence. Zhang (2018) suggested that it should be the central frequency $\nu_{\rm c}$ here instead of the bandwidth of the receiver $\Delta\nu$ [113], which is more reasonable since the FRB spectrum is unknown and the emission could extend outside the range of $\Delta\nu$.

Wang & Yu (2017) first found that the energies of 17 bursts of FRB 121102 show a power-law distribution with an index of $-1.80 \pm 0.15$ [106]. Similar energy distribution of $dN/dE \propto E^{-1.7}$ from the Very Large Array (VLA), Arecibo, and Green Bank Telescope (GBT) bursts are found by Law et al. (2017) [114]. However, a much steeper index of -2.8 is obtained by Gourdji et al. (2019) using 41 bursts observed by Arecibo [115]. Further, a universal energy distribution is found by Wang & Zhang (2019) [103]. They collected the bursts of FRB 121102 observed by different telescopes, including VLA at 3 GHz [7], Arecibo at 1.4 GHz [5], Arecibo at 4.5 GHz [64], GBT at 4-8 GHz [116], GBT at 2 GHz [6, 117] and Arecibo at 1.4 GHz [115]. The threshold power-law distribution with a high-energy cutoff is adopted to fit the cumulative distribution, which is

$$N(>E) = A(E^{1-\alpha_E} - E_{\rm max}^{1-\alpha_E}), \tag{27}$$

where $E_{\rm max}$ is the maximumal energy and $\alpha_E$ is the index of differential distribution $dN/dE \propto E^{-\alpha_E}$. The fitting results for different samples are shown in Figure 2. It is interesting that the value of $\alpha_E$ is in a narrow range of $1.6-1.8$ for bursts at different frequencies, which supports a universal energy distribution for FRB 121102. Moreover, similar $\alpha_E$ for non-repeating FRBs observed by Parkes and ASKAP is also found [110, 118]. This energy distribution is also similar to those of magnetar bursts [106, 111, 119, 120], and solar type III radio bursts [121]. A few works have discussed the energy distribution across different sources [108, 122-125], and the detailed method is introduced in Section 4.1.

### 3.2 Width distribution

Several studies have been performed on the distribution of observed pulse widths. Wang & Yu (2017) found that pulse widths of 17 bursts of FRB 121102 show a power-law distribution with an index of $-1.95 \pm 0.32$ [106]. Using more bursts of FRB 121102 observed by GBT, a shallower index of $-1.57 \pm 0.13$ is found [111], which is shown in Figure 3. The width distribution of FRB 121102 is consistent with that of magnetar bursts [106, 111, 119].

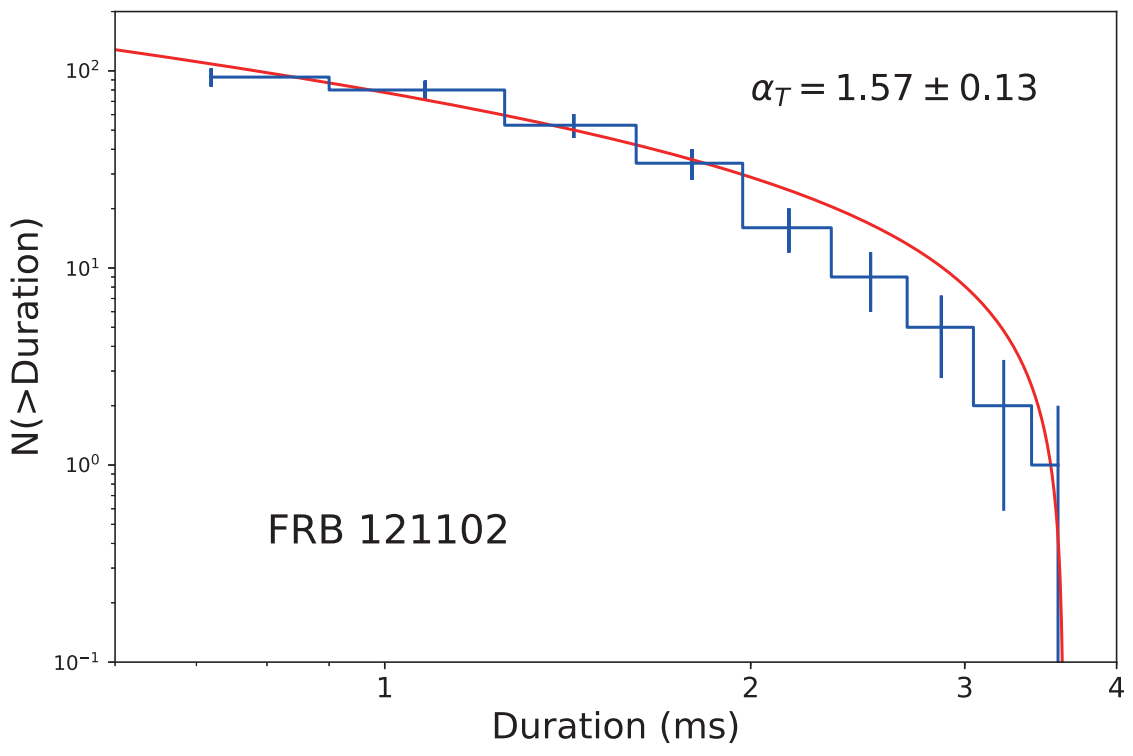

**Figure 3** Cumulative distribution of pulse width for 93 bursts of FRB 121102. The fit is shown as red line. (Adapted from [111].)

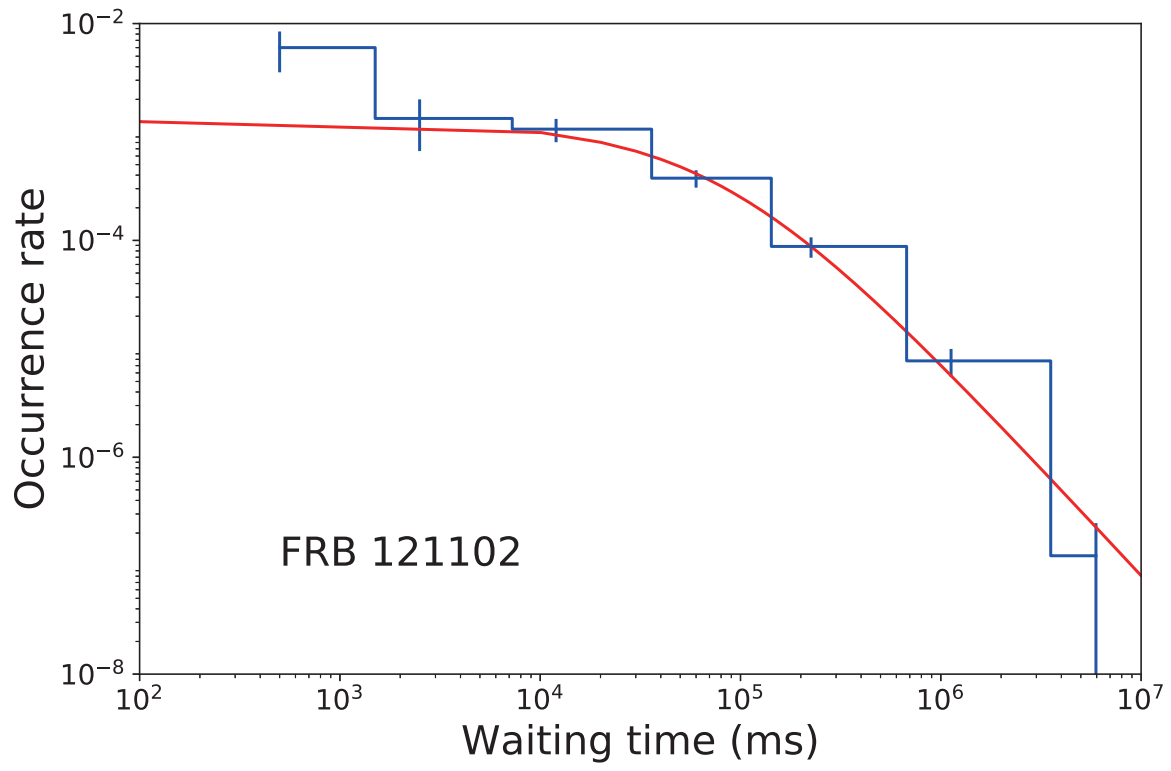

**Figure 4** Differential distribution of waiting times for 93 bursts of FRB 121102. The fit is shown as red line. (Adapted from [111].)

### 3.3 Waiting time distribution

The waiting time $\Delta t_w$ is defined as the difference between occurrence times for two adjacent bursts. It can give the information of the burst rate and period. If the burst rate is constant, the occurrence rate of waiting time should follow the Poisson distribution [126]

$$P(\Delta t_w) = \lambda e^{-\lambda \Delta t_w}, \tag{28}$$

where $\lambda$ is the burst rate. If the burst rate is time varying, the waiting time distribution can be treated as a combination of piecewise constant Poisson processes. Generally, for most forms of $\lambda(t)$, the waiting time distribution shows a power-law form. Wang & Yu (2017) first found that the waiting time distribution of FRB 121102 shows a power-law form, not a Poissonian distribution [106]. Subsequent work by Oppermann et al. (2018) confirmed this result [127]. Figure 4 gives the cumulative distribution of waiting times of FRB 121102 [111], which is similar to that of magnetar bursts [106, 111, 119] and solar type III radio bursts [121]. It shows power-law-like distributions with a slope close to −2 for large waiting times. If the variability of burst rate shows spikes like $\delta-$functions, the occurrence rate $\lambda$ is exponentially growing and fulfills the normalization requirement $\int_0^\infty \lambda f(\lambda) d\lambda = \lambda_0$,

$$f(\lambda) = \lambda^{-1} \exp\left(-\frac{\lambda}{\lambda_0}\right). \tag{29}$$

The waiting time distribution in a given time interval can be written as

$$P(\Delta t_w) = \frac{\lambda_0}{(1 + \lambda_0 \Delta t_w)^2}. \tag{30}$$

The fit of red line in Figure 4 gives $\lambda_0 = 1.23^{+0.80}_{-0.38} \times 10^{-5}\ \mathrm{ms}^{-1}$ [111].

## 4 Population Study of FRBs

### 4.1 Global Statistical Properties

The all-sky rate is very high for FRBs, typically within the range of $10^3 \sim 10^4\ \mathrm{sky}^{-1}\ \mathrm{day}^{-1}$. Different surveys give different rates since their sensitivities and central frequencies vary [28]. Basically, the event rate density is more important since it is directly related to the source system. It depends strongly on FRB luminosity. A recent study suggests that the event rate density above $10^{42}\ \mathrm{erg\ s}^{-1}$ is $\sim 3.5 \times 10^4\ \mathrm{Gpc}^{-3}\ \mathrm{yr}^{-1}$ [122]. The precise cosmological volumetric rate is waiting to be revealed since the FRB redshift distribution $\Psi(z)$ is still largely unknown. Except for 13 FRBs that have been localized with identified host galaxies [9, 14, 81, 82, 84-87], the majority of FRBs are lacking in direct distance information. Several pioneering works have put constraints on $\Psi(z)$ with current statistical properties of FRBs [107, 128-131].

The event rate theory is briefly introduced as follows. Other than $\Psi(z)$, the frequency-dependent FRB detection rate is largely determined by the intrinsic luminosity function. We can define a function $\Theta(L_\nu, z)$ to incorporate both factors [107]. The form of $\Theta(L_\nu, z)$ is unclear from the first principle, however we can model it with some simplified assumptions. If we assume that the luminosity function does not evolve with redshift, $\Theta(L_\nu, z)$ is therefore separable $\Theta(L_\nu, z) \equiv \Phi(L_\nu)\Psi(z)$. Further, different models could be applied to $\Phi(L_\nu)$ and $\Psi(z)$ respectively. For instance, simple luminosity function models like standard candle, power-law function and log-normal distribution have been adopted in previous studies [129, 131]. More specially, the Schechter function has been discussed [108, 132]. For the redshift evolution, the most commonly adopted model is that $\Psi(z)$ traces the cosmic SFR [129, 131]. Note that there might be a time delay of $\Psi(z)$ from SFR if an FRB is produced by compact binary mergers [107, 118, 133]. This delay effect has been taken into account for short GRB before [134, 135]. Also, other models such as a constant comoving density or $\Psi(z)$ in proportion to cosmic stellar mass density have been discussed [131].

The generic mathematical formalism that relates the expression of $\Theta(L_\nu, z)$ with observational properties has been completed in Macquart & Ekers (2018b) [107]. Usually, the cumulative distribution of FRB flux density/fluences (known as $\log N - \log S$ distribution) is devoted to constrain $\Theta(L_\nu, z)$. If most of FRBs are in our local universe, the slope of $\log N - \log S$ plot should be $-1.5$ [136]. However, different surveys give different power-law indexes in a wide range [129, 130, 137-140]. Meanwhile, the cosmological $\langle V/V_{\rm max}\rangle$ tests of different samples deviate from 0.5, which disfavor an Euclidean geometric distribution [133, 141]. Furthermore, the DM distribution of FRBs can constrain $\Theta(L_\nu, z)$ due to the relation between DM and $z$ in Eq. (7). Assuming the models of $\Psi(z)$, several works have constrained the FRB luminosity function, or equivalently, the enengy distribution function [108, 110, 118, 123, 128, 131, 142, 143]. Recently a Bayesian method has been applied to give both the best-fitting values and error-bars of the parameters in Schechter luminosity function [110, 122]. Besides, the intrinsic spectra of FRBs can influence the detection rate, since the emission frequency in source frame differs from that in the observer frame due to cosmological expansion. This effect is embedded in a $k$-correction factor $(1+z)^{-\alpha}$ if the power-law spectrum $L_\nu \propto \nu^{-\alpha}$ is assumed [107, 113, 131]. There are a few works that have discussed the impact of index $\alpha$ [128, 131, 133, 143]. However, currently it is poorly constrained from observations except for some small samples [144].

### 4.2 Repeating and Non-repeating: A Single Population?

There are hundreds of FRBs from different surveys now recorded by the Transient Name Server [2)], among which most are one-off events and only twenty are identified as repeaters. Generally, an FRB is recognized as a repeater if a second burst has been detected in the same position of the sky, which relies much on the precision of localization. However, there are several factors that make a repeater look non-repeating. First one is the selection bias of the receiver. If the pulse width of the second burst is close to the instrumental sampling or smearing timescale, then this signal is probably being missed [145, 146]. Also, the beam shape could play a role [147]. Second, the energy distribution function of individual FRBs matters a lot. If the second burst is much dimmer than the original one, it may fall below the detection threshold. This has been implied by the repetition of FRB 171019, for which two fainter bursts by a factor of $\sim 590$ than the original burst that have been detected [148]. It may be just a lucky case and we could probably have missed many faint bursts of other FRBs. Third, the repetition statistics could differ a lot for each repeater. The waiting times for the majority of FRBs may be much longer than those of the frequent repeaters like FRB 121102. This problem will be alleviated as the observing time accumulates.

These above lead to a debate: do all FRBs repeat [145, 149, 150]? There is no firm conclusion but some emerging clues seem disfavoring a single population. The burst morphology looks different for repeaters and non-repeaters. Repeaters like FRB 121102 have wider pulse width and show complex sub-burst structures [11-13, 56, 151], while pulses of non-repeaters are usually narrower [13, 143]. If the temporal and frequency structure is caused by self-modulation during propagation [152], the circumburst environment of repeaters and non-repeaters should be very different. Further, the polarization properties seem different, though not prominent in consideration of the presence of FRB 180301 [69]. Moreover, there are some differences in the mass and SFR of their host galaxies, however, a clear sign of dichotomy has not been found [15, 100]. Note that these clues suffer from a small number of repeaters. With more repeaters coming soon, bimodal distributions are expected in these statistical properties if repeaters and non-repeaters are physically different.

Still, it is fair to argue that all FRBs repeat on the supposition that their repeating modes are diverse. Several works have tried to reproduce the observed statistical properties with different models of repeating [153, 154]. For instance, the difference in pulse width distribution can be ascribed to FRB beaming [155]. Two factors are decisive in repetition modeling: the burst energy function and the distribution of waiting time [123, 147, 150, 156]. Further, assuming the cosmic evolution of comoving FRB number density, we can calculate how many repeaters will be identified for a given survey. The energy function is commonly assumed as power-law or Schechter forms [123, 147, 150, 156]. Since the distribution of waiting time of FRB 121102 is non-Poissonian [18, 111, 115, 127, 157], a Weibull distribution is usually adopted [123, 127, 147, 150, 156]. A lack of repeaters in several surveys suggest that most FRBs should not repeat as frequently as FRB 121102 [145, 158, 159]. Generally, with the accumulation of the observing time, the observed repeating fraction (defined as the number ratio of repeaters among all FRBs) is expected to increase to nearly 100% [123, 150, 156]. However, if there exist physically non-repeating FRBs, this

2) https://wis-tns.weizmann.ac.il/

fraction will peak at a certain observing time and the peak fraction is less than 100% [153, 156]. The current total observing time of CHIME may be still insufficient to bring out the peak. Also, the repeating fraction will decrease with the increasing redshift since the repetitions of FRBs with closer distances are easier to be detected [123,153]. This means that the DM distribution of repeaters should concentrate on a lower value than non-repeaters. Observationally, this trend has not been confirmed yet since the number of repeaters is limited. A bottom line is that the volumetric rate of apparently non-repeating FRBs exceeds those of candidate cataclysmic progenitor events [160], therefore, there are many underlying repeaters waiting to be discovered in the near future. However, the existence of a physically one-off population can be confirmed once the direct association of an FRB with a cataclysmic event is observed.

### 4.3 Periodicity

The periodicity of FRBs is a new character discovered in early 2020. The repeat bursts of FRB 180916 arrive regularly with a period of 16.35±0.15 days [16]. More intriguingly, the activity window is found to be narrower and earlier at higher frequencies recently [78]. Further, a period of $\sim$ 157 days for FRB 121102 has been claimed by two different groups using different burst data [17, 18]. Currently, period finding has only been performed to three relatively "bursty" repeaters (the above two together with FRB 180814), however, no significant periodicity has been found for FRB 180814 yet [161]. Usually at least tens of repeat bursts are needed to achieve high enough confidence level after period folding process. This implies that there may be underlying periodicity for other repeaters. The situation is very much analogous to that of repetition two years before. Future debate might arise like this: do all repeaters have a periodicity? One thing is obvious that the unique period is distinct for each individual case, which may vary from seconds to years. Learning from repetition, we can do period modeling by assuming a distribution of period, which is left for future work.

Theorists have proposed several possible physical scenarios for the periodical activity [162]. Generally the existing models can be classified into two categories. First kind of models attributes the periodicity to the orbital motion of a binary system [97, 163-170]. This binary usually involves a neutron star (NS) as FRB emitter together with a companion. The companion can be an O/B type star [166, 168] or a compact object [97, 164, 165]. There are multiple ways to achieve periodicity and one possibility was proposed that the wind of the companion can pave a way for the FRBs to escape [165, 166]. This opened window would point to us periodically due to the orbital motion. However, this scenario has recently been disfavored since it predicts a wider activity window for higher frequencies [78]. Alternatively, there may be an asteroid belt surrounding the companion and the NS collide with them in a certain orbital phase [163, 171]. The second kind of models suggests that the periodicity is due to NS precession [172-177]. The NS could either undergo a free precession [172, 174, 175] or forced precession [173], while the latter may be caused by the surrounding disk [176, 177]. However, one important issue for precession models should be addressed properly: why are FRBs produced at a fixed region on the NS surface? Instead of an NS, the precession of a jet produced by a massive black hole has been discussed [178]. Note that a third option that associates the periodicity with the rotation of a magnetar has been proposed [179], however, there is no evidence that such a slowly-rotating magnetar exists and is able to produce FRBs. The periodical behavior is closely related to the radiation mechanism and source model of FRBs, which will be discussed in next section.

## 5 Radiation Mechanism for FRBs

The radiation mechanism is always the core issue of a new astrophysical signal. It may take a while before we can figure out the mechanism of FRBs. However, we can learn a lesson from similar phenomena like pulsar radio emission and GRB. Very recently, Zhang (2020) gives a very nice review on this issue [30]. The high brightness temperature of FRBs from Eq. (2) implies that the radiation must be coherent. Basically there are three main approaches to generate coherence (for a review, see [180]). The first one is coherent curvature emission by bunched particles (also known as antenna mechanism) [95, 181-183]. The second mechanism is relativistic plasma emission by reactive instabilities (also called plasma maser in some literatures). The particles' kinetic energy is transferred to plasma waves through a beam instability and is finally converted to escaping radio emission [184, 185]. The third mechanism is maser (microwave amplification by stimulated emission of radiation). The radiation can be amplified if "negative absorption" is reached. The key point is that the number of high-energy electrons should exceed that of low-energy ones under certain physical conditions, which is called "population inversion". Till now various versions of maser has been proposed for FRBs, including plasma synchrotron maser, vacuum maser and synchrotron maser from magnetized shocks [59, 90, 186-188].

### 5.1 Coherent Curvature Emission

*5.1.1 Synchro-curvature emission from a single charge*

In vacuum, the trajectory of a relativistic electron in the presence of a magnetic field can be decomposed into two components: a motion along the field line and a gyration in the perpendicular plane. The latter leads to synchrotron emission and the former could produce curvature emission if the field line is curved (e.g., a dipole field geometry). Therefore, the realistic emission of relativistic electrons in a curved field can be generalized as synchro-curvature radiation, which is illustrated in Fig 5.

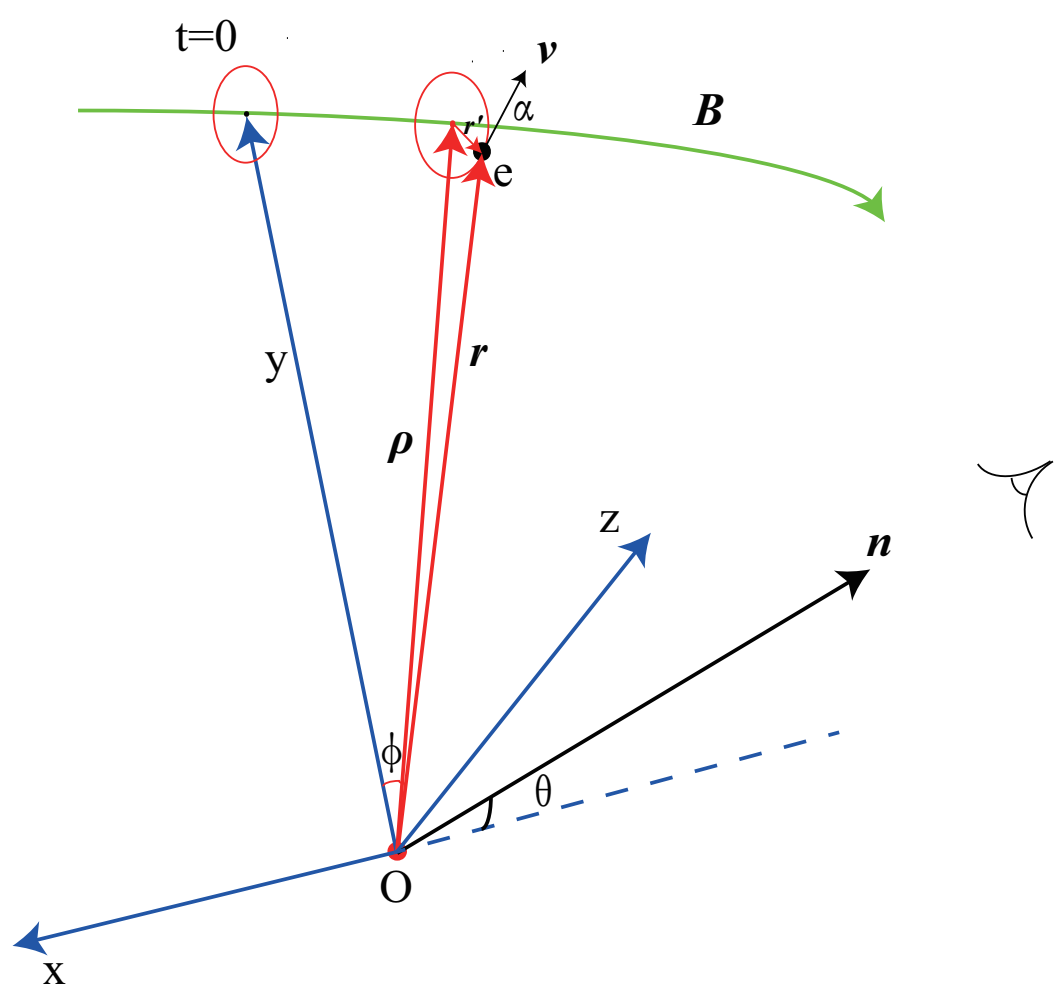

**Figure 5** Schematic picture of synchro-curvature radiation.

Suppose the magnetic field is along $\phi$-direction $\boldsymbol{B} = B_0\boldsymbol{e}_\phi$ and its curvature radius is $\rho$. A relativistic electron moves with a Lorentz factor $\gamma$ and the angle between $\boldsymbol{v}$ and $\boldsymbol{B}$ is $\alpha$. The guiding center drifts along the field line with an angular velocity $\Omega_0$. The gyro-radius is defined as

$$r_B = \frac{\gamma v_z m_e c}{eB_0} = \frac{\gamma\beta\sin\alpha m_e c^2}{eB_0} = c\sin\alpha/\omega_B, \quad (31)$$

where $\omega_B = eB_0/(\gamma m_e c)$ is Larmor frequency.

The energy radiated per unit solid angle per unit frequency interval of a moving charge is [189]

$$\frac{\mathrm{d}E}{\mathrm{d}\omega\mathrm{d}\Omega} = \frac{e^2\omega^2}{4\pi^2 c}\left|\int_{-\infty}^{+\infty}\boldsymbol{n}\times(\boldsymbol{n}\times\boldsymbol{\beta})e^{i\omega(t-\boldsymbol{n}\cdot\boldsymbol{r}(t)/c)}\mathrm{d}t\right|^2. \quad (32)$$

We can construct a Cartesian coordinate system, making sure that $x$−$y$ plane overlaps with $\rho$−$\phi$ plane and initially the guiding center is on the $y$-axis (Fig 5). Without loss of generality, the unit vector pointing to the observer $\boldsymbol{n}$ is in $x-z$ plane and makes an angle $\theta$ with the $x$-axis. Substituting the expressions of $\boldsymbol{n}$, $\boldsymbol{\beta}$, $\boldsymbol{r}$ in the $xyz$ coordinate and integrating the solid angle, we can obtain the characteristic frequency and radiation spectrum of synchro-curvature radiation [190,191],

$$\omega_c = \frac{3}{2}\gamma^3 c\frac{1}{\rho}\left[\frac{(r_B^3+\rho r_B^2-3\rho^2 r_B)}{\rho r_B^2}\cos^4\alpha + \frac{3\rho}{r_B}\cos^2\alpha + \frac{\rho^2}{r_B^2}\sin^4\alpha\right]^{1/2}.$$

$$\frac{\mathrm{d}P}{\mathrm{d}\omega} = \frac{\sqrt{3}e^2\gamma}{4\pi r_c^*}\frac{\omega}{\omega_c}\left\{\left[\int_{\omega/\omega_c}^{\infty}K_{5/3}(y)\mathrm{d}y - K_{2/3}\left(\frac{\omega}{\omega_c}\right)\right] + \frac{1}{r_c^{*2}Q_2^2}\left[K_{2/3}\left(\frac{\omega}{\omega_c}\right) + \int_{\omega/\omega_c}^{\infty}K_{5/3}(y)\mathrm{d}y\right]\right\}, \quad (33)$$

where $r_c^* \approx c^2/[r_B\omega_B^2 + (\rho + r_B)\Omega_0^2]$, and

$$Q_2^2 \equiv \left(\frac{r_B^2+\rho r_B-3\rho^2}{\rho^3}\cos^3\alpha\cos\theta + \frac{3}{\rho}\cos\alpha\cos\theta + \frac{1}{r_B}\sin^3\alpha\sin\theta\right)\frac{1}{r_B}. \quad (34)$$

The above formulae can be easily degraded to synchrotron and curvature radiation case. For pure synchrotron radiation, $\rho = \infty$, $\alpha \neq 0$, $\Omega_0 = 0$, $Q_2 = \sin^2\alpha/r_B$, and $r_c^{*2}Q_2^2 = 1$. So,

$$\omega_c = \frac{3}{2}\gamma^3\omega_B\sin\alpha,$$
$$\frac{\mathrm{d}P}{\mathrm{d}\omega} = \frac{\sqrt{3}e^2\gamma\omega_B\sin\alpha}{2\pi c}\frac{\omega}{\omega_c}\int_{\omega/\omega_c}^{\infty}K_{5/3}(y)\mathrm{d}y. \quad (35)$$

Then for pure curvature radiation, $\alpha = 0$, $r_B = 0$, $\Omega_0 = c/\rho$, $Q_2 = 1/\rho$, and $r_c^{*2}Q_2^2 = 1$. So we have

$$\omega_c = \frac{3}{2}\gamma^3\frac{c}{\rho},$$
$$\frac{\mathrm{d}P}{\mathrm{d}\omega} = \frac{\sqrt{3}e^2\gamma}{2\pi\rho}\frac{\omega}{\omega_c}\int_{\omega/\omega_c}^{\infty}K_{5/3}(y)\mathrm{d}y. \quad (36)$$

If FRBs are produced inside the magnetosphere of NS where $B > 10^9$ G, the gyration damps very quick since the synchrotron cooling timescale is proportional to $B^{-2}$. The electron moves along the field line and only curvature emission is important [192]. However, outside in the low $B$ region, both curvature and synchrotron emission can be effective and Eq. (33) should be applied.

*5.1.2 Coherent curvature emission by bunches*

Now we can discuss the radiation spectrum from a bunch of $N$ particles in the magnetosphere. Coherence can be achieved if the phases of EM waves emitted by each individual electron are near the same. According to Eq. (32), the total radiation intensity of $N$ electrons is

$$\frac{\mathrm{d}E_{\mathrm{tot}}}{\mathrm{d}\omega\mathrm{d}\Omega} = \frac{e^2\omega^2}{4\pi^2 c}\left|\int_{-\infty}^{+\infty}\sum_j^N\boldsymbol{n}\times(\boldsymbol{n}\times\boldsymbol{\beta}_j)e^{i\omega(t-\boldsymbol{n}\cdot\boldsymbol{r}_j(t)/c)}\mathrm{d}t\right|^2, \quad (37)$$

where the subscript $j$ represents the $j$-th electron. This integration can be simplified under certain assumptions and has been discussed in detail by Yang & Zhang (2018) [95]. Here we list some of their important results below.

Considering a one-dimensional (1-D) bunch with finite length $L$ moving along a magnetic field line of curvature radius $\rho$, the position vector from the $j$-th electron to the first one is $\Delta\boldsymbol{r}_j(t) = \boldsymbol{r}_j(t) - \boldsymbol{r}(t)$. Since the emission is highly beamed, this vector could be considered as time-independent. Eq. (37) can be rewritten as

$$\frac{\mathrm{d}E_{\rm tot}}{\mathrm{d}\omega\mathrm{d}\Omega} \simeq \frac{e^2\omega^2}{4\pi^2 c}\left|\int_{-\infty}^{+\infty}\boldsymbol{n}\times(\boldsymbol{n}\times\boldsymbol{\beta})e^{i\omega(t-\boldsymbol{n}\cdot\boldsymbol{r}(t)/c)}\mathrm{d}t\right|^2 \times\left|\sum_j^N e^{-i\omega(\boldsymbol{n}\cdot\Delta\boldsymbol{r}_j/c)}\right|^2. \quad (38)$$

The phase stacking term $F_\omega \equiv \left|\sum_j^N e^{-i\omega(\boldsymbol{n}\cdot\Delta\boldsymbol{r}_j/c)}\right|^2$ can be evaluated analytically. Defining a typical frequency $\omega_L \equiv 2c/(L\cos\theta)$ where $\theta$ is the observing angle, we have [95]

$$F_\omega \simeq \begin{cases} N^2, & \omega \ll \omega_L, \\ N^2\left(\frac{\omega}{\omega_L}\right)^{-2}, & \omega_L \ll \omega \ll \omega_{\rm coh}. \end{cases} \quad (39)$$

where $\omega_{\rm coh} \sim (\rho/L)^2\omega_L$ is a frequency boundary beyond which the emission becomes incoherent. Further, if we assume the electrons have a power-law energy distribution, i.e., $N_e(\gamma) \propto \gamma^{-p}$ where $\gamma_{\rm m} < \gamma < \gamma_{\rm max}$, the resulting radiation spectrum has a break at $\omega_{\rm m} \equiv \omega_c(\gamma_{\rm m})$ given by Eq. (36). The power-law index of the spectrum turns from $2/3$ (for $\omega \ll \omega_{\rm m}$) to $-(2p-4)/3$ (for $\omega \gg \omega_{\rm m}$). For more complicated scenario, a 3-D bunch with a half opening angle $\varphi$ has been considered. If $\varphi$ is larger than the angle of beamed cone, only part of the radiation by this bunch can be observed. This also leads to a spectral break from $2/3$ to 0, at the frequency $\omega_\varphi \equiv 3c/(\rho\varphi^3)$.

Taking all these above into account, the total radiation spectrum of a 3-D bunch filled with power-law distributed electron has a four-segments-broken-power-law shape, characterized by three frequencies $\omega_L$, $\omega_{\rm m}$, $\omega_\varphi$. The indices of four segments change according to the relative values of $\omega_L$, $\omega_{\rm m}$, $\omega_\varphi$ and the complete analytical expressions can be found in Yang & Zhang (2018) [95].

#### *5.1.3 Bunch formation*

Coherent curvature emission by bunches has long been proposed as a possible mechanism for pulsar radio emission [193-195]. However, it remains controversial how the bunches could form effectively, and consequently some critiques has been put on this mechanism [196-198]. It seems unlikely to solve this leftover problem completely in FRB field but some possible ways could be explored.

For instance, the possibility of bunch formation by two stream instability under the physical condition of FRBs has been discussed [181, 188]. Instability caused by counter-streaming electrons and positrons leads to density fluctuations. Assuming that electrons and positrons have the same density and the four velocity distribution $f(u)$ is symmetric, the dispersion relation can be written as [188]

$$\int_{-\infty}^{+\infty}\frac{\omega_{\rm p}^2}{\gamma^3}\frac{f(u)}{(\omega-\beta k)^2}\mathrm{d}u = \int_{-1}^{+1}\omega_{\rm p}^2\frac{f(\beta)}{(\omega-\beta k)^2}\mathrm{d}\beta = 1, \quad (40)$$

where $k$ is wave-number and $\beta = u/\gamma$ is velocity. Further, assuming a step-function distribution $f(u)$ within velocity range $\beta_{\rm min} < \beta < \beta_{\rm max}$, the instability growth rate ${\rm Im}(\omega)$ can be determined, which is given by the imaginary part of complex frequency $\omega$. Defining an effective plasma frequency $\omega_{\rm p,eff} = \omega_{\rm p}\langle\gamma^{-3}\rangle^{1/2}$ where $\langle\gamma^{-3}\rangle = \int_{-\infty}^{+\infty}\gamma^{-3}f(u)\mathrm{d}u$, the effective skin depth is then $\ell_{\rm skin} = c/\omega_{\rm p,eff}$. Large density fluctuations can be produced on length-scales larger than $\ell_{\rm skin}$, i.e., bunches with $L > \ell_{\rm skin}$ could possibly form via two stream instability. Note that above treatment is greatly simplified since $f(u)$ is highly uncertain in reality. More future work is needed to address the issue of bunch formation.

### 5.2 Relativistic Plasma Emission (Plasma Maser)

The theory of plasma emission was first proposed to explain solar radio bursts [199]. Later its relativistic version was applied to pulsar radio emission [200]. Generally plasma emission process includes two stages: Langmuir-like waves are generated by a beam instability and then part of the wave energy is converted to escaping radiation [184].

The beam instability could occur when a beam of relativistic particles runs into a background plasma. The growth rate of beam instability depends on the dispersive properties of the plasma. To ensure effective plasma emission, the growth rate should be relatively fast. Assuming the four-velocity distributions of the beam $f_{\rm b}(u_{\rm b})$ and target plasma $f(u)$, the dielectric tensor of the plasma can be obtained and the dispersion relation of Langmuir-like waves can be determined [184]. The growth rate is enhanced at the resonant frequencies. The Cherenkov resonance condition reads

$$1 - n\beta\cos\theta = 0, \quad (41)$$

where $n$ is the refraction index. Also, the cyclotron-Cherenkov (or anomalous Doppler) resonance condition is

$$\omega(1 - n\beta\cos\theta) + \omega_B = 0. \quad (42)$$

Plasma maser is effective if the growth rate at resonant conditions is fast enough. However, there are two main difficulties for this mechanism. First, it remains questionable whether Langmuir-like waves through beam instabilities can be generated in the NS magnetosphere [185]. Second, usually the growth rates for beam instabilities seem too small to allow effective wave growth [180]. In the case of FRB 121102, cyclotron-Cherenkov resonance condition requires unrealistically high beam Lorentz factors and the growth rate of Cherenkov instaiblity is too small [188]. Thus, plasma maser can hardly explain the emission of FRB 121102. For completeness, however, we list this mechanism here since it represents the second way of generating coherence. The living space is limited even it might be applicable to some particular low-energy bursts.

### 5.3 Plasma Synchrotron Maser

Synchrotron self-absorption can happen when the emitted synchrotron radio waves propagate in a weakly magnetized relativistic plasma. The self-absorption coefficient can be expressed as [201, 202]

$$\begin{aligned}\mu &= -\frac{c^2}{8\pi\nu^2}\int_0^\infty E^2\frac{\mathrm{d}}{\mathrm{d}E}\left[\frac{N(E)}{E^2}\right]\frac{\mathrm{d}P}{\mathrm{d}\nu}\mathrm{d}E\\ &= \frac{8\pi^3c^2}{\omega^2}\int_0^\infty\frac{N(E)}{E^2}\frac{\mathrm{d}[E^2\mathrm{d}P/\mathrm{d}\omega]}{\mathrm{d}E}\mathrm{d}E,\end{aligned}\tag{43}$$

where $E=\gamma m_ec^2$ is electron energy. Normally $\mu$ is positive but it can turns negative if the plasma is dense and relativistic. A direct requirement is that $\frac{\mathrm{d}}{\mathrm{d}E}\left[\frac{N(E)}{E^2}\right]>0$, i.e., the energy distribution of electrons grows faster than $E^2$. This means that population inversion is necessary. Further we need to explore at which frequencies this negative absorption is significant.

The synchrotron radiation power in the presence of a background plasma is [203].

$$\frac{\mathrm{d}P}{\mathrm{d}\omega}=\frac{\sqrt{3}e^2\gamma\omega_B\sin\alpha}{2\pi c\sqrt{1+\gamma^2\frac{\omega_\mathrm{p}^2}{\omega^2}}}\frac{\omega}{\omega_c'}\int_{\omega/\omega_c'}^\infty K_{5/3}(x)\mathrm{d}x,\tag{44}$$

where the critical frequency $\omega_c'$ is then

$$\omega_c'=\frac{3}{2}\gamma^3\omega_B\sin\alpha(1+\gamma^2\omega_\mathrm{p}^2/\omega^2)^{-3/2}.\tag{45}$$

If $\gamma^2\omega_\mathrm{p}^2/\omega^2\ll 1$, Eq. (44)(45) turn back to Eq. (35). On the contrary, the influence of relativistic dense plasma ($\gamma$, $\omega_\mathrm{p}$ are large) should be taken into account. Substituting Eq. (44)(45) into Eq. (43), we get [204]

$$\mu\propto E^{-2}\frac{\mathrm{d}}{\mathrm{d}E}[E^2\mathrm{d}P/\mathrm{d}\omega]\propto z^{-2}\Phi(z),\tag{46}$$

where

$$z=\frac{\omega}{\omega_c'}\simeq\frac{2}{3}\frac{\omega_\mathrm{p}^3}{\omega^2\omega_B\sin\alpha}$$

$$\Phi(z)\equiv 2z\int_z^\infty K_{5/3}(x)\mathrm{d}x-z^2K_{5/3}(z).\tag{47}$$

The function $\Phi(z)$ has a negative minimum value of $\sim -0.24$ at the frequency $\nu_{\mathrm{R}^*}\equiv\nu_\mathrm{p}(\nu_\mathrm{p}/\nu_B)^{1/2}$ [205]. For $\nu<\nu_{\mathrm{R}^*}$, $\mu$ is always negative.

The intensity $I_\omega$ after radiation transferring in a source of size $L$ is

$$I_\omega=\frac{j_\omega}{\mu}(1-e^{-\tau}),\tag{48}$$

where $j_\omega=\int(\mathrm{d}P/\mathrm{d}\omega)N(E)\mathrm{d}E$ is the total emissivity and $\tau=\mu L$ is the optical depth. As long as $\mu$ is negative and $|\mu|L\gg 1$, the radiation intensity is amplified exponentially [204]

$$I_\omega\simeq(j_\omega/|\mu|)\exp(|\mu|L).\tag{49}$$

This indicates that a significant fraction of electron energy is converted to the strong synchrotron maser emission. Note that $|\mu|L\gg 1$ is achieved only for a narrow range of frequencies, so the maser emission has a narrow spectra [204, 205]. Also, its typical frequency is lower than synchrotron self-absorption frequency [204]. In some circumstances, this plasma synchrotron maser could arise around GHz, thus being regarded as a possible mechanism for FRBs [186].

### 5.4 Vacuum Maser

For synchrotron emission in vacuum, if Eq. (35) is substituted into Eq. (43), we always have $\frac{\mathrm{d}[E^2\mathrm{d}P/\mathrm{d}\omega]}{\mathrm{d}E}>0$ and thus $\mu>0$. However, we should notice that Eq. (35) has been integrated over solid angle. Actually, the synchrotron radiation power is angular-dependent and $\frac{\mathrm{d}P}{\mathrm{d}\omega\mathrm{d}\Omega}$ is relevant. At some certain observing angle $\theta$, the self-absorption coefficient can be negative [187, 188, 206].

The emissivity for a single electron is [207]

$$\begin{aligned}j_\nu(\theta)\equiv\frac{\mathrm{d}P}{\mathrm{d}\nu\mathrm{d}\Omega}&=\frac{9\sigma_\mathrm{T}cB^2}{64\pi^4\nu_\mathrm{B}}\left(\frac{\nu}{\nu_c}\right)^2(1+\gamma^2\theta^2)\\ &\times\left[(1+\gamma^2\theta^2)K_{2/3}^2(y)+\gamma^2\theta^2K_{1/3}^2(y)\right],\end{aligned}\tag{50}$$

where $\nu_B=\omega_B/2\pi$, $\nu_c=\omega_c/2\pi$ and $y\equiv\frac{\nu}{2\nu_c}(1+\gamma^2\theta^2)^{3/2}$. The angular-dependent absorption coefficient can be rewritten as

$$\mu(\theta)=\frac{1}{2m_e\nu^2}\int_1^\infty\frac{N(\gamma)}{\gamma p}\frac{\partial}{\partial\gamma}[\gamma p j_\nu(\theta)]\mathrm{d}\gamma,\tag{51}$$

where $p=(\gamma^2-1)^{1/2}$ is the electron momentum in units of $m_ec$. A differential absorption cross section has been defined as [206]

$$\frac{\mathrm{d}\sigma_\mathrm{s}(\theta)}{\mathrm{d}\Omega}=\frac{1}{2m_e\nu^2}\frac{1}{\gamma p}\frac{\partial}{\partial\gamma}[\gamma p j_\nu(\theta)].\tag{52}$$

Substituting the expression of $j_\nu(\theta)$ and letting $t\equiv\gamma^2\theta^2$,

$$\frac{\mathrm{d}\sigma_\mathrm{s}(\theta)}{\mathrm{d}\Omega}=\frac{2}{9}\frac{e}{B}\frac{1}{\gamma^4\sin^2\alpha}\{(13t-11)(1+t)K_{2/3}^2(y)$$

$$+t(11t-1)K_{1/3}^2(y)-6y(t-2)$$
$$\times[(1+t)K_{2/3}(y)K_{5/3}(y)+tK_{1/3}(y)K_{4/3}(y)]\}. \quad (53)$$

This cross section is negative only if both $\theta > \sqrt{2}/\gamma$ and $\nu/\nu_c \gg (\gamma\theta)^{-3}$ are satisfied [187, 206]. The maser emission can be observed outside the $1/\gamma$ emission cone of the relativistic electrons. For lower frequencies $\nu/\nu_c \ll (\gamma\theta)^{-3}$ or in the $1/\gamma$ cone the cross section turns positive with a larger absolute value [187]. This implies that the distribution of pitch angles should be narrower than $1/\gamma$, otherwise any photon will be preferentially absorbed by a particular electron whose emission cone wraps the photon's moving direction. However, it is unclear whether such narrow pitch angle distribution can be achieved.

Other than vacuum synchrotron maser, the vacuum curvature maser inside the magnetosphere has been considered. Since the expressions of emissivity for synchrotron and curvature radiation are very similar, the above procedure of calculation has been repeated [188]. The curvature self-absorption cross section could be negative for $\theta \lesssim (2\gamma^2)^{-1}$. However, since electrons move along curved field lines in the magnetosphere, the curvature photons emitted by an electron will be inevitably absorbed by electrons on other field lines since their intersection angles are large. Therefore, it is unlikely for vacuum curvature maser to be responsible for FRBs.

### 5.5 Synchrotron Maser from Magnetized Shocks

Relativistic perpendicular magnetized shocks has been studied extensively in literatures. Traditionally, a set of conservation equations of physical quantities between upstream and downstream plasma can be written. However, these macroscopic conservation laws fall short of describing the microscopic behavior of the plasma, such as the distributions of particle momentum, energy partition between different species and so on. These kinetic issues shall be addressed by collisionless plasma physics. The development of particle-in-cell (PIC) simulations since early 1990s greatly helps understand the plasma properties. In principle, particles reflected at magnetized shock front will achieve bunching in gyration phase and synchrotron maser (or relativistic cyclotron) instability will develop [208]. Strong coherent EM signal can be produced and the reabsorption of these cyclotron waves serves as the main heating mechanism of downstream plasma [209]. This EM signal occurs before the radiation from heated downstream particles so that it is called a precursor. The FRB emission is probably just this kind of EM precursor [90]. Here we will revisit the results of 1-D PIC simulations and briefly introduce the underlying physics.

First we present the analytical expressions of shock jump condition for an ideal MHD plasma. The conservation of particle number, energy, momentum and magnetic flux are respectively expressed as [210]

$$\frac{N_2}{N_1} = \frac{1+\beta_{\rm shock}}{\beta_{\rm shock}}, \quad (54)$$

$$(1+\beta_{\rm shock})N_1 m\gamma_1 c^2(1+\sigma) = \beta_{\rm shock}\left(e_2+\frac{B_2^2}{8\pi}\right), \quad (55)$$

$$(1+\beta_{\rm shock})N_1 m\gamma_1 c^2(1+\sigma) = p_2+\frac{B_2^2}{8\pi}, \quad (56)$$

$$\frac{B_2}{B_1} = \frac{1+\beta_{\rm shock}}{\beta_{\rm shock}}, \quad (57)$$

where subscripts 1, 2 represents upstream and downstream respectively. The upstream is assumed to be cold and relativistic, $p_1 \simeq 0$, $\beta_1 \simeq 1$. For the hot relativistic downstream plasma, $e_2 \simeq p_2/(\hat{\Gamma}-1)$ where $\hat{\Gamma}$ is adiabatic index. The magnetization parameter is defined as $\sigma = \frac{B_1^2}{4\pi N_1 mc^2\gamma_1}$. From above equations we can obtain the equation for shock velocity $\beta_{\rm shock}$:

$$\left(1+\frac{1}{\sigma}\right)\beta_{\rm shock}^2-\left[\frac{\hat{\Gamma}}{2}+\frac{1}{\sigma}(\hat{\Gamma}-1)\right]\beta_{\rm shock}-\left(1-\frac{\hat{\Gamma}}{2}\right)=0, \quad (58)$$

which has a solution,

$$\beta_{\rm shock} = \frac{1}{2(1+1/\sigma)}\left\{\left(\frac{\hat{\Gamma}}{2}+\frac{\hat{\Gamma}-1}{\sigma}\right)+\left[\left(\frac{\hat{\Gamma}}{2}+\frac{\hat{\Gamma}-1}{\sigma}\right)^2 +4\left(1-\frac{\hat{\Gamma}}{2}\right)\left(1+\frac{1}{\sigma}\right)\right]^{1/2}\right\}. \quad (59)$$

The ratio of the shock jump is

$$\frac{N_2}{N_1} = \frac{B_2}{B_1} = 1+\frac{1}{\beta_{\rm shock}}, \quad (60)$$

$$\frac{kT_2}{m\gamma_1 c^2} = \beta_{\rm shock}\left[1+\sigma\left(1-\frac{1+\beta_{\rm shock}}{2\beta_{\rm shock}^2}\right)\right], \quad (61)$$

where $T_2$ is downstream temperature. Substituting Eq. (59) we can express this ratio as a function of $\sigma$ and $\hat{\Gamma}$.

However, the results from PIC simulations shows that both $N_2/N_1$ and $kT_2/m\gamma_1c^2$ deviate from Eq. (60)(61). The reason is that the ideal MHD formalism does not account for the additional wave fluctuations that can dissipate the flow energy. The upstream field fluctuation is due to the EM precursor, which can be expressed with a parameter $\xi$,

$$\frac{1}{8\pi}\langle\delta E_1^2+\delta B_1^2\rangle \approx -\frac{1}{4\pi}\langle\delta E_1\delta B_1\rangle \equiv \frac{1}{4\pi}\xi B_1^2, \quad (62)$$

Besides, there are two sub-dominant fluctuations in the downstream. The parameters $\zeta$, $\eta$ are introduced to describe its EM and electrostatic fluctuations respectively,

$$\frac{1}{8\pi}\langle\delta E_2^2+\delta B_2^2\rangle \equiv \frac{1}{4\pi}\zeta B_1^2, \quad \frac{1}{8\pi}\langle\delta E_{es2}^2\rangle \equiv \frac{1}{4\pi}\eta B_1^2. \quad (63)$$

Taking these modifications into account, Eq. (55) and (56) can be rewritten as

$$N_1 m\gamma_1 c^2(1+\beta_{\rm shock})+\frac{1}{4\pi}B_1^2[(1+\beta_{\rm shock})-\xi(1-\beta_{\rm shock})]$$

$$= \left[e_2 + \frac{1}{8\pi}B_2^2 + \frac{1}{4\pi}(\zeta+\eta)B_1^2\right]\beta_{\rm shock}, \quad (64)$$

$$\left(N_1 m\gamma_1 c^2 + \frac{1}{4\pi}B_1^2\right)(1+\beta_{\rm shock}) + \frac{1}{4\pi}\xi B_1^2(1-\beta_{\rm shock}) = p_2 + \frac{1}{8\pi}B_2^2 + \frac{1}{4\pi}(\zeta-\eta)B_1^2. \quad (65)$$

while Eq. (54) and (57) remain unchanged. Then the equation for $\beta_{\rm shock}$ becomes

$$\left(1+\frac{1}{\sigma}-\xi\right)\beta_{\rm shock}^3 + \left[(2-\hat{\Gamma})\left(\frac{1}{2}+\frac{1}{\sigma}+\xi-\zeta\right)+\hat{\Gamma}\eta\right]\beta_{\rm shock}^2 - \left[1+(\hat{\Gamma}-1)\left(\frac{1}{\sigma}-\xi\right)\right]\beta_{\rm shock}^3 - \frac{2-\hat{\Gamma}}{2} = 0.(66)$$

The form of density jump Eq. (60) remains while the downstream temperature is now

$$\frac{kT_2}{m\gamma_1 c^2} = \beta_{\rm shock}\left[1+\sigma\left(1-\frac{1+\beta_{\rm shock}}{2\beta_{\rm shock}^2}\right)\right] + \frac{\sigma\beta_{\rm shock}}{1+\beta_{\rm shock}}[\xi(1-\beta_{\rm shock})-\zeta+\eta]. \quad (67)$$

Three efficiencies are defined as the fractions of flow energy that carried away by these fluctuations,

$$f_\xi \equiv \frac{\xi}{1+1/\sigma}\frac{1-\beta_{\rm shock}}{1+\beta_{\rm shock}} = \frac{\xi}{1+1/\sigma}\left(1-\frac{2N_1}{N_2}\right), \quad (68)$$

$$f_{\zeta,\eta} \equiv \frac{\zeta,\eta}{1+1/\sigma}\frac{\beta_{\rm shock}}{1+\beta_{\rm shock}} = \frac{\zeta,\eta}{1+1/\sigma}\frac{N_1}{N_2}. \quad (69)$$

Simulation results show that $f_{\zeta,\eta} \ll f_\xi$ [210], and we are only interested in the EM precursor. $f_\xi$ has a peak value around 10% at $\sigma \sim 0.1$, and increasing $\sigma$ leads to decreasing $f_\xi$. For very high magnetization $\sigma \gg 1$, $f_\xi$ has an asymptotic form of $7\times10^{-4}/\sigma^2$ [211]. The spectrum of EM precursor peaks at $\omega_{\rm peak} \simeq 3\omega_{\rm p}\max[1, \sqrt{\sigma}]$ (in the post-shock frame) and is relatively narrow-banded with $\Delta\omega/\omega_{\rm peak} \sim$ a few [211].

The synchrotron maser instability can be understood as follows. Bunching in gyration phase of incoming particles has been verified by PIC simulations [208, 210-212]. For pure electron-positron plasma, this means $e^\pm$ gyrates in the magnetic field with the same Lorentz factor. Therefore, the distribution of pairs in momentum space can be described as a cold ring, which reads

$$f(u_\perp, u_\parallel) = \frac{1}{2\pi u_0}\delta(u_\perp - u_0)\delta(u_\parallel), \quad (70)$$

where the symbols $\parallel$ and $\perp$ are with respect to the magnetic field direction. The population inversion has been reached since low-energy pairs are absent. If the magnetic field is assumed on $z$-axis and cyclotron waves from pairs propagate on $x$-axis, the dispersion relation can be written as

$$\frac{c^2k^2}{\omega^2} = \varepsilon_{yy} - \frac{\varepsilon_{xy}\varepsilon_{yx}}{\varepsilon_{xx}}, \quad (71)$$

where $\boldsymbol{\varepsilon}$ is the dielectric tensor. Sustituting Eq. (70), the four components of $\boldsymbol{\varepsilon}$ can be expressed in analytical forms [209]. The dispersion relation shows two unstable branches corresponding to EM waves and magnetosonic waves [209, 212]. It has been demonstrated that the growth of EM waves can be very effective at low harmonics thus a synchrotron maser (EM precursor) can be produced [209, 212]. Simultaneously, the energy spectrum of downstream pairs is heated to a relativistic Maxwellian distribution [209, 210]. Note that the above discussion can be generalized to an ion-pair plasma though the expression for $\boldsymbol{\varepsilon}$ becomes more complicated. In the presence of ions, a high-energy non-thermal tail of energy spectrum of pairs appears, which is due to the absorption of the ion cyclotron waves [212, 213].

To conclude this section, we should be aware of the fact that above mechanisms do not include all, since more than a dozen of pulsar radiation models have been proposed and developed. It is unclear which mechanism would work for FRBs. According to the site where these mechanisms take place, they are classified into two types: close-in (i.e., inside the NS magnetosphere) and far-away (i.e., outside the NS magnetosphere) (see the latest review of Zhang (2020) [30]). A few works have been devoted to discuss the applicability of different mechanisms to individual bursts. As an example, a critical analysis on radiation mechanism of FRB 121102 has been nicely performed by Lu & Kumar (2018) [188]. They concluded that given the properties of FRB 121102, plasma maser and vacuum curvature maser have been excluded. Vacuum synchrotron maser may survive, but needs fine-tuning. Plasma synchrotron maser and synchrotron maser from magnetized shocks suffer from the problem of low radiation efficiency. Coherent curvature emission satisfies the observations best, however, the relevant magnetic reconnection physics is highly uncertain. For other FRBs with lower energies, the constraints on radiation mechanism may be totally different. For recent FRB 200428, models based on both coherent curvature emission and synchrotron maser emission have been applied successfully [124, 214-216]. However, diverse PA swings of FRB 180301 clearly favor the close-in mechanism [69]. One thing that we know from the GRB field is that the radiation mechanism is the kernel of FRB physics and it is not easy to be made clear completely. Similar to GRBs, the controversy on radiation mechanism may last a while before reaching a final conclusion. A good mechanism needs to explain FRB energetics, duration, spectrum, pulse structure, polarization, repetition and periodicity, hence plenty of work should be done in the future.

## 6 Magnetar Source Models

The origin of FRBs has been an outstanding issue ever since its discovery. Various source models have been proposed, most of which involve compact objects (see [217] for a theory catalogue). Different models predict different EM counterparts and sometimes multi-messenger signals [218]. Note some progenitor models are catastrophic and can only be applied to one-off FRBs. The recent discovery of FRB 200428 has confirmed that magnetars are capable of producing FRBs. Moreover, localized FRBs seem to be consistent with magnetar progenitors in host galaxy properties [101, 154]. Although it is still early to say that all FRBs are from magnetars [102], at this time magnetar models attract the most attention of the community. In this section we will introduce four well-developed magnetar models and their predictions [59, 171, 219, 220].

### 6.1 A far-away model of Metzger et al. (2017, 2019)

The model proposed by Metzger et al. is based on synchrotron maser emission from magnetized shocks, which is far away from the central magnetar. The picture is shown in Figure 6. A magnetar is embedded in a supernova (SN) environment and the wind blows a bubble (nebula) behind the SN ejecta. Matter ejected during magnetar flare is composed of a highly relativistic shell and a subrelativistic ion tail. The relativistic ejecta shell collides with the leftover ion tail from the previous flare and forward-reverse shock system is formed. The ion tail is magnetized with assumed $\sigma \sim 0.1 - 1$, which is beneficial for producing synchrotron maser by the forward shock. Assuming the ion tail has a radial density profile $n_{\rm ext} \propto r^{-k}$, the dynamical evolution is then quite similar to that of GRB afterglow [221].

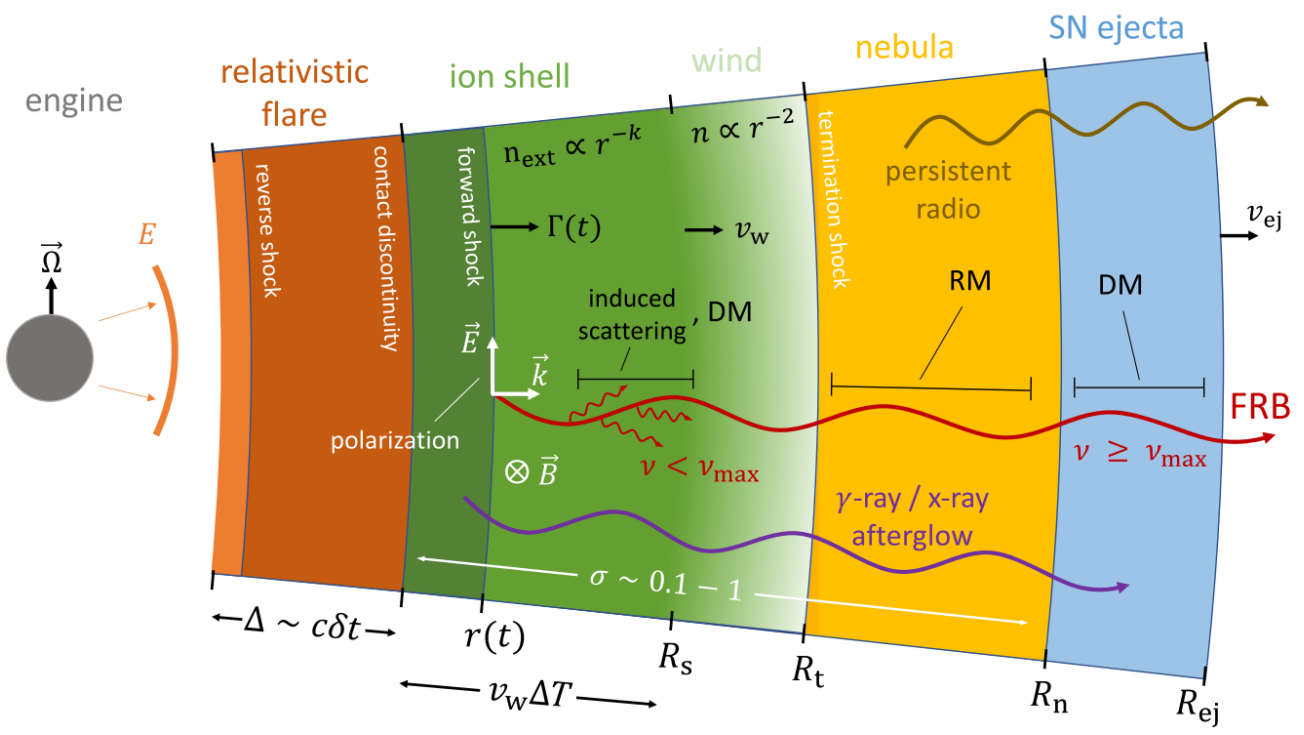

**Figure 6** The schematic picture of Metzger et al. model. (Adapted from [59]).

The density of the unshocked ejecta shell is

$$n_{\rm ej} \simeq \frac{\mathcal{E}}{\delta t} \frac{1}{4\pi r^2 m_p c^3 \Gamma_{\rm ej}^2}, \tag{72}$$

where $\Gamma_{\rm ej}$ is the initial Lorentz factor of the shell and $\mathcal{E}$, $\delta t$ is the energy, duration of the flare. The reverse shock crosses the shell at a time $t_{\rm dec} \sim \delta t$ [221]. For the typical flare parameters, the reverse shock is always relativistic. Before the crossing, the Lorentz factor of the shocked region is

$$\Gamma = \left( \frac{f \Gamma_{\rm ej}^2}{4} \right)^{1/4} \propto r^{\frac{(k-2)}{4}}, \tag{73}$$

where $f \equiv n_{\rm ej}/n_{\rm ext} \propto r^{k-2}$. After crossing, the Lorentz factor approaches the Blandford-McKee self-similar form $\Gamma \propto r^{(k-3)/2}$ [222]. Substituting $t \simeq r/(2c\Gamma^2)$, the time evolution of $\Gamma$ is

$$\Gamma \propto \begin{cases} t^{\frac{(k-2)}{2(4-k)}}, & t \ll \delta t \\ t^{\frac{(k-3)}{2(4-k)}}, & t \gg \delta t \end{cases}, \tag{74}$$

As discussed in Section 5.5, the synchrotron maser peaks at $\omega_{\rm peak} \simeq 3\omega_{\rm p} \max[1, \sqrt{\sigma}]$ in the comoving frame. For the assumed $\sigma < 1$, the peak frequency in lab frame is

$$\nu_{\rm pk} \approx \frac{1}{2\pi}(3\Gamma\omega_{\rm p}), \tag{75}$$

However, sometimes the observed FRB frequency may be not this intrinsic peak frequency. The opacity of induced Compton Scattering (ICS) can be very large for the low energy part of the maser spectrum. The observed peak frequency is shifted to a higher value set by $\nu_{\rm max} \equiv \nu(\tau_c = 3)$. Substituting Eq. (74), time evolution of the observed FRB frequency is then

$$\nu_{\rm max} \propto \nu_{\rm pk}^{5/4} t^{1/4} \propto \begin{cases} t^{-\frac{2+7k}{4(8-2k)}}, & t \lesssim \delta t, \\ t^{-\frac{2k+7}{4(8-2k)}}, & t \gtrsim \delta t, \end{cases} \tag{76}$$

This naturally predicts the down-drifting of the FRB emission. Since there is strong ICS absorption, the effective radiation efficiency is even lower than maser efficiency of $\sim 10^{-3}$, typically $\sim 10^{-6}$–$10^{-5}$. Another prediction is that the accelerated electrons in the downstream can produce a high-energy (keV-MeV) counterpart. Its duration is of order $\sim$ milliseconds, and the luminosity and spectrum highly depends on the flare parameters.

This model has been successfully applied to FRB 200428 in different details [215, 216, 223, 224]. The observed low efficiency $E_{\rm radio}/E_X \sim 10^{-5}$ of FRB 200428 matches the model prediction well. The duration and luminosity of the associated XRBs can be explained [215]. Especially, by introducing a density-jump structure of upstream medium, the double-peaked character, luminosity ratio and emission frequency of two radio pulses can be well explained [216].

Moreover, three X-ray peaks are expected [216], which is also consistent with observation [22, 23]. However, observationally the X-ray peak energy seems to increase with time [23], which needs further consideration. Non-thermal acceleration of electrons can be realized if sufficient ions exist in the downstream [212]. The superposition of two components could change the peak energy. Besides, the XRBs may simply have a different origin from FRB [124, 223, 224]. This model also requires a large amount of baryonic matter (about 0.005 solar mass), which may be larger than the typical mass of a magnetar outer crust [224]. Note that this model needs upgrade in order to explain the complex PA behaviour of FRB 180301, since the magnetic field configuration is basically fixed and a flat PA curve is expected.

### 6.2 A far-away model of Beloborodov (2017, 2020)

The radiation mechanism in Beloborodov model is also synchrotron maser emission, however a main difference from Metzger et al. (2019) is the upstream medium. In this model, the emission region is also far away from a central magnetar. The ejecta shell should decelerate in the long-lasting wind before it can hit the ion tail. A magnetized shock is likely to form in the wind instead of the ion tail, because the magnetization should be very low for the tail. Synchrotron maser can be produced as the blast wave sweeps the cold wind, which is illustrated in Figure 7.

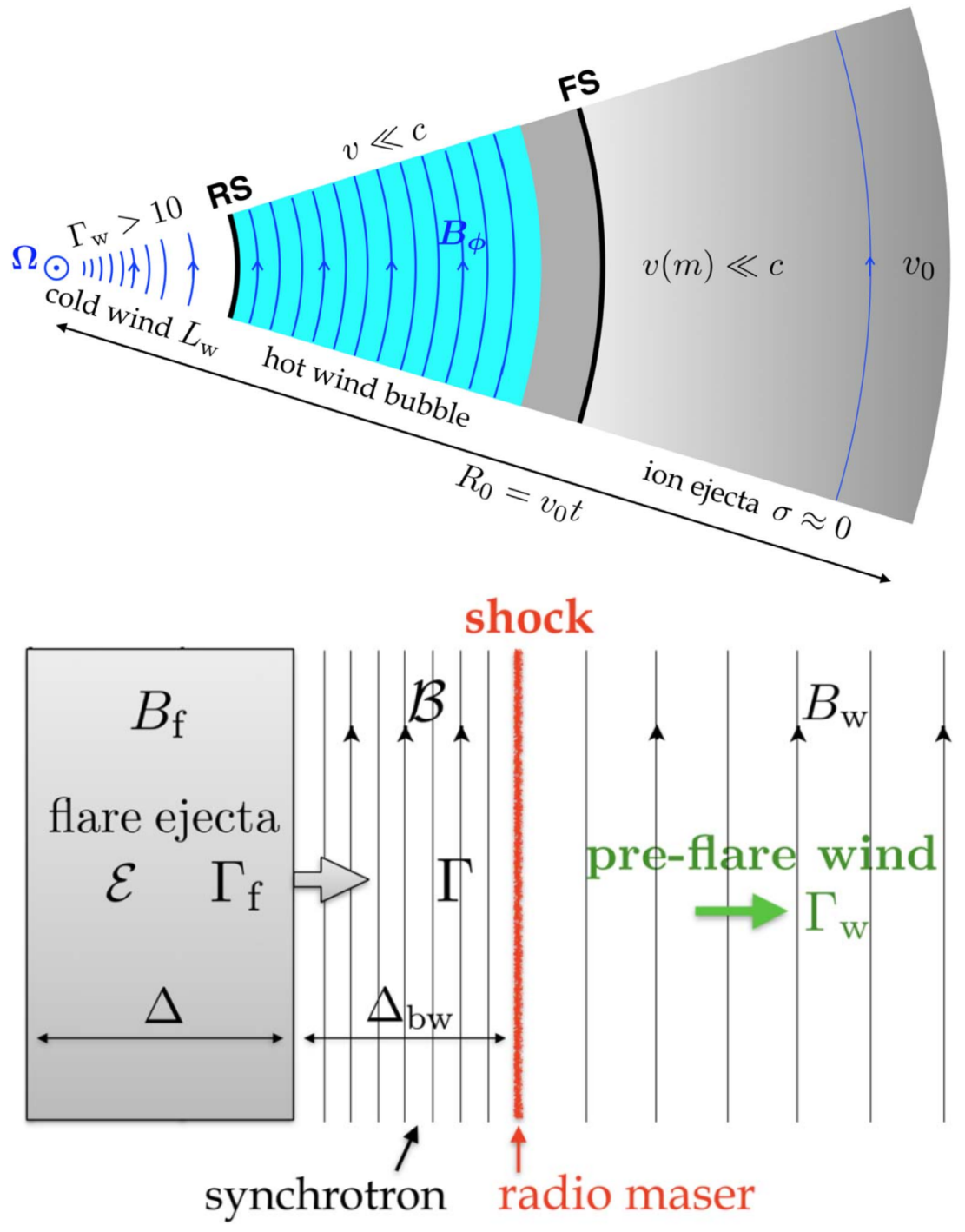

**Figure 7** The schematic picture of Beloborodov model. (Adapted from [219]. Reproduced by permission of the AAS.)

The evolution of $\Gamma$ also has two stages separated by the crossing radius $R_{\rm dec} \sim 2\Gamma^2 c\delta t$,

$$\Gamma \simeq \begin{cases} \Gamma_{\rm w}\left(\frac{L_{\rm f}}{L_{\rm w}}\right)^{1/4}, & r < R_{\rm dec}, \\ \Gamma_{\rm w}^2\left(\frac{2c\mathcal{E}}{rL_{\rm w}}\right)^{1/2}, & r > R_{\rm dec}, \end{cases} \tag{77}$$

where $L_{\rm w}$, $\Gamma_{\rm w}$ is the luminosity, Lorentz factor of the unshocked wind and $L_{\rm f} = \mathcal{E}/\delta t$ is the flare luminosity. Since the magnetization of the wind $\sigma_{\rm w} > 1$, the peak frequency of maser emission is

$$\nu_{\rm pk} = 3\frac{\Gamma\tilde{\omega}_c}{\pi} = \frac{3}{\pi}\frac{e}{m_e c}\left(\frac{L_{\rm w}}{cr^2}\right)^{1/2}\frac{\Gamma}{\Gamma_{\rm w}}. \tag{78}$$

Substituting $r$ with $t$ and using Eq. (77), the time evolution of $\nu_{\rm pk}$ is then

$$\nu_{\rm pk} = \nu_\diamond \times \begin{cases} t_\diamond/t_{\rm obs}, & t_{\rm obs} < t_\diamond, \\ (t_\diamond/t_{\rm obs})^{3/4}, & t_{\rm obs} > t_\diamond, \end{cases} \tag{79}$$

where

$$\nu_\diamond \sim \frac{e\,L_{\rm w}^{3/4}}{2m_e c^{5/2}\,\mathcal{E}^{1/4}\,\delta t^{3/4}\,\Gamma_{\rm w}^2} \approx 1.4\,\frac{L_{\rm w,39}^{3/4}}{\mathcal{E}_{44}^{1/4}\,\delta t_{-3}^{3/4}}\left(\frac{\Gamma_{\rm w}}{20}\right)^{-2}\,{\rm GHz},$$
$$t_\diamond \sim \frac{\tau}{2\sigma_{\rm w}} = \frac{1\ {\rm ms}}{2\sigma_{\rm w}}\,\delta t_{-3}. \tag{80}$$

Different from Metzger et al. (2019) scenario, He found that ICS is not important so that this intrinsic $\nu_{\rm pk}$ is responsible for the observed peak frequency. Also, Eq. (79) predicts a different downward drifting rate compared to Eq. (76). These distinctions originate from different physical conditions of upstream medium. Note that due to high magnetization of the wind, a very low maser efficiency is also expected in this scenario, $\mathcal{E}_{\rm FRB}/\mathcal{E} \sim 10^{-3}\,\sigma_{\rm w}^{-2} \sim 10^{-5} - 10^{-6}$. Moreover, FRB emission is almost 100% linearly polarized with polarization direction being aligned to the magnetar rotation axis since the helical wind is nearly toroidal. This is also disfavored by the observation of FRB 180301. A bubble structure can be formed behind the slow ion tail (Figure 7), and a strong optical flash of duration $\sim 1$ s is predicted as the shock sweeps the bubble. However, no optical emission has been detected for FRB 200428 and all other events yet.

### 6.3 A close-in model of Kumar & Bošnjak (2020) and Lu et al. (2020)

Different from the previous two scenarios, the radiation mechanism in this model is coherent curvature radiation from bunches in the magnetosphere. A sudden disturbance (e.g., crustal quake) of the magnetar can launch Alfvén waves which moves along the field lines. In the polar cap region (i.e., close to the stellar surface), these Alfvén waves propagate outward where the density is decreasing. At a certain radius $R_{\rm c}$, the plasma density is too low to support the electric current required by the wave. This means charge starvation occurs at $R > R_c$. Strong parallel electric field forms and accelerates electron-positron pairs. The counter-streaming of pairs leads to two-stream instability and then bunch formation. Afterwards, FRBs are produced by the coherent curvature radiation from bunches. This model is illustrated in Figure 8.

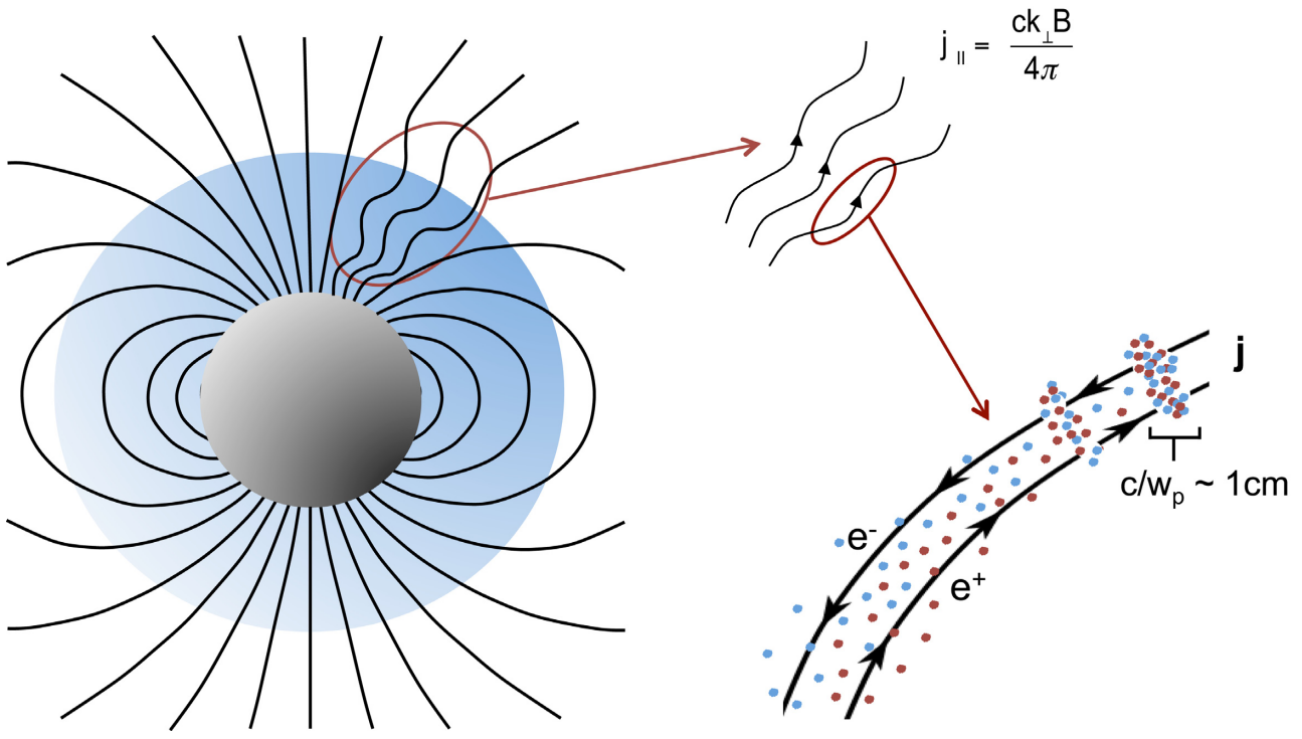

**Figure 8** The schematic picture of Kumar & Bošnjak 2020 model. (Adapted from [220]).

Assuming the Alfvén waves have an amplitude $\delta B$ and wavelength $\lambda$, the critical density as a function of $R$ is given by Ampere's equation,

$$n_c(R) = \frac{|\nabla \times \delta \boldsymbol{B}|}{8\pi e} = \frac{k_\perp \delta B}{8\pi e} = (10^{16}\,\mathrm{cm}^{-3})\frac{\delta B_{11}}{\lambda_{\perp,4}}\left(\frac{R_*}{R}\right)^3, \quad (81)$$

where $\perp$ represents the component perpendicular to the magnetic field and $R_*$ is the radius of the magnetar. A parallel electric field develops in the charge starvation region and electrons and positrons are separated. According to the discussion in Section 5.1.3, bunches of lengthscale $l_\parallel \sim c/\omega_p$ can form via two-stream instability. The transverse size of the coherent region is of order $l_\perp \sim R/\gamma$, therefore the number of particles in the coherent region is

$$N_{\rm coh} \sim n_c l_\parallel (R/\gamma)^2, \quad (82)$$

The total number of observed particles is larger by a factor $(R/\gamma^2)/l_\parallel$, so the power of coherent curvature radiation is

$$P_c = N_{\rm coh}^2 (R/\gamma^2 l_\parallel) p_c = \frac{2e^2 c R^5 l_\parallel n_c^2}{3\rho^2\gamma^2}, \quad (83)$$

where the single particle power $p_c = 2\gamma^4 e^2 c/3\rho^2$ is used. The isotropic luminosity of FRB is then

$$L_{\rm iso} \simeq 8\gamma^4 P_c \simeq \frac{16 e^2 c R^5 \gamma^2 n_c^2 l_\parallel}{3\rho^2}. \quad (84)$$

Due to the uncertainties in these parameters, FRB luminosity could span several orders of magnitude. The magnetosphere-origin models have been applied to FRB 200428 and the association with XRBs is also expected [124, 214, 225, 226]. The derived spectra from CHIME and STARE2 bursts can be explained [214]. Using the luminosity $L \sim 3 \times 10^{38}\,\mathrm{erg\,s^{-1}}$, the emission radius can be constrained as $R_c/R_* \sim 20(\delta B_{10}/\lambda_{\perp,4})^{6/11}$ [124]. Note that these models predict that X-rays should appear earlier than an FRB, since the FRB is not yet produced until Alfvén waves reach $R_c$. However, observationally there is a time lag of the X-ray peak respect to the radio peak, which is about several milliseconds [22, 23]. Further work needs to be done to explain this time lag as well as the non-thermal component of X-ray emission [22]. Also, the radio pulses should be able to break out from XRB fireballs [227].

### 6.4 A close-in model of Dai et al. (2016) and Dai (2020)

Based on coherent curvature emission, Dai et al. (2016) proposed a different model, in which a repeating FRB source could arise from a strongly magnetized NS encountering an extragalactic asteroid belt (EAB) around a stellar-mass object [171]. The central source could either be a normal pulsar or a magnetar and the FRB emission region is also close to the stellar surface. In this model, an FRB could be produced by a collision of the NS with an asteroid of mass $\sim 10^{17}$ g to $\sim 10^{21}$ g in the EAB. Following Colgate & Petscheck [228], Dai et al. (2016) analyzed the collision physics [171]. It is assumed that an asteroid as a solid body falls freely in the gravitational field of the NS. This asteroid is originally approximated as a sphere. It will first be distorted tidally by the NS at some break-up radius and subsequently elongated in the radial direction and compressed in the transverse direction. The timescale of such a bar-shaped asteroid accreted on the NS surface is estimated by

$$\Delta t \simeq 1.6 m_{18}^{4/9}\ \mathrm{ms}, \quad (85)$$

where and hereafter $m_{18} = m/10^{18}$ g, the NS mass $M_{\rm NS} = 1.4 M_\odot$, and the asteroidal tensile strength and original mass density are taken for iron-nickel matter (see Equation 2 of [171]). The average rate of gravitational energy release near the stellar surface during $\Delta t$ is approximated by

$$\dot{E}_G \simeq G m M_{\rm NS}/(R_{\rm NS}\Delta t) = 1.2 \times 10^{41} m_{18}^{5/9}\ \mathrm{erg\,s^{-1}}, \quad (86)$$

where and hereafter the NS radius $R_{\rm NS} = 10^6$ cm is adopted. These simple estimates of $\Delta t$ and $\dot{E}_G$ are well consistent with the observations of FRBs. This is why asteroid-NS collisions have been proposed as an origin model of FRBs [171, 229].

Further, they also explored the radiation physics during an asteroid-neutron star collision in detail and found that an electric field induced outside of the asteroid has such a strong component parallel to the stellar magnetic field that electrons are torn off the asteroidal surface and accelerated to ultra-relativistic energies instantaneously [171]. Subsequent movement of these electrons along magnetic field lines will cause coherent curvature radiation. From Equation (15) of [171], the isotropic-equivalent emission luminosity is estimated by

$$L_{\rm iso} \sim 2.6 \times 10^{40} (f\rho_{\rm c,6})^{-1} m_{18}^{8/9} \mu_{30}^{3/2}\,{\rm erg\,s^{-1}}, \tag{87}$$

where the beaming factor $f$ is introduced based on the emission beaming geometry [230], $\rho_{\rm c,6}$ is the curvature radius of a magnetic field line near the stellar surface in units of $10^6$ cm, $\mu_{30}$ is the NS magnetic dipole moment in units of $10^{30}$ G cm$^3$, and the other parameters are taken for an iron-nickel asteroid.

This model was shown to be able to interpret many aspects of current observations under reasonable conditions. First, Dai et al. (2016) suggested that this model can not only explain the observed emission frequency, duration and luminosity but also account for the repeating rate of FRB 121102 for the EAB's mass $\sim$ 10 times the earth mass [171]. The downward drifting rate and linear polarization of FRB 121102 is well consistent with model prediction [231].

Second, the periodical activity of the repeating FRB 180916 can be understood. Repeating bursts from this source show a $\sim$ 16 day period with an active phase of $\sim$ 4.0 days and a broken power-law form in differential energy distribution. Dai & Zhong (2020) proposed that FRB 180916-like periodical FRBs could provide a unique probe of EABs [232], following the model in which repeating FRBs arise from an old-aged, slowly spinning, strongly magnetized NS traveling through an EAB around another stellar-mass object [171]. These two objects form a binary and thus the observed period is in fact the orbital period. Furthermore, the EAB's physical properties were constrained and it was found that (1) the outer radius of the EAB is at least an order of magnitude smaller than that of its analog in the solar system, (2) the differential size distribution of the EAB's asteroids at small diameters (large diameters) is shallower (steeper) than that of solar system small objects, and (3) the two belts have a comparable mass.

Third, Dai (2020) proposed a specific model for FRB 200428 and its associated XRB, in which a magnetar encounters an asteroid of mass $\sim 10^{20}$ g [230]. This asteroid in the magnetar's gravitational field is first possibly disrupted tidally into a great number of fragments at radius $\sim$ a few times $10^{10}$ cm, and then slowed down around the magnetic interaction radius by an ultra-strong magnetic field and in the meantime two major fragments of mass $\sim 10^{17}$ g that cross magnetic field lines give rise to two pulses of FRB 200428, as shown in [171]. Dai (2020) also argued that the whole asteroid is eventually accreted onto the poles along magnetic field lines, colliding with the stellar surface, generating a photon-e$^{\pm}$ pair fireball trapped initially in the stellar magnetosphere, and further leading to an XRB [230]. This model was shown to explain all of the observed features of FRB 200428 and its associated XRB self-consistently. In addition, this model can interpret the time delay $\sim 3-6$ ms of the XRB's two peaks with respect to FRB 200428's two pulses, which was indicated by the observations of the *Insight*-HXMT and INTEGRAL satellites. This time delay disfavors a few models, e.g., a starquake-induced close-in model [124], a synchrotron-maser far-away model [215], and also the other magnetar-asteroid model of Geng et al. (2020) [233]. Finally, this model is well consistent with the constraints on bursting sites inferred from both the *Insight*-HXMT's detection of the time interval between the XRB's two peaks and the time interval between FRB 200428's two pulses.

## 7 FRBs as Cosmological Probes

In the near future, an FRB sample enlarged by orders of magnitude will be obtained, with a considerable fraction being localized and covering a wide range of redshift. It can be used to probe cosmological parameters such as the energy density of matter, dark energy, cosmic curvature, the baryon fraction of the IGM, the dark energy equation of state and the reionization histories for helium and hydrogen [234]. This can be realized by means of the DM(z) relation given by Eq.(9).

### 7.1 Cosmic Baryons

For the current universe, the significant discrepancy between the detected baryonic matter and the theoretically predicted baryonic matter is called the missing baryon problem [235, 236] . It is difficult for us to detect baryonic matter in the diffuse gas of the universe. Big Bang nucleosynthesis and Cosmic Microwave Background (CMB) can be used to constrain the density of cosmic baryons [237, 238]. There are many works that use simulated FRB data samples to constrain $f_{\rm IGM}$ [32,239-241] and try to alleviate the missing baryon problem [242-244]. Due to the large DMs of the cosmological FRBs, the localized events can be used to measure the baryonic matter of the universe. Once the DM value and the redshift of the localized FRBs are obtained, the $\rm DM_{IGM}$ value contributed by the intergalactic medium can be calculated. Only $\rm DM_{IGM}$ is related to baryon density, the value of which can be determined by combining redshift. Macquart et al. (2020) used the five ASKAP localized FRBs to derive a cosmic baryon density of $\Omega_{\rm b} = 0.05^{+0.021}_{-0.025}$, and their result shows a great consistency with the Big Bang nucleosynthesis and CMB measurements, indicating that it is feasible to measure the density of cosmic baryons using localized FRBs (see

Figure 3 of [82]).

### 7.2 Dark Energy

In cosmology, dark energy is devoted to explain the accelerating expansion of the universe. The extragalactic FRBs can be used as a cosmic probe to study the evolution of the universe and constrain cosmological parameters. A Combination of FRB with other cosmological probes will be a powerful tool in constraining the dark energy equation of state [245-250]. The Friedmann equation can be written as [251]:

$$\frac{H^2(z)}{H_0^2} = \Omega_{\rm M}(1+z)^3 + \Omega_{\rm DE}\exp\int 3(1+\omega(z))\,{\rm d}\ln(1+z) + \Omega_{\rm k}(1+z)^2, \quad (88)$$

where $\Omega_{\rm M}$ and $\Omega_{\rm DE}$ are the matter and dark energy density of the universe, $\Omega_{\rm k}$ is the spatial curvature density. $\omega(z)$ is the dark energy equation of state. The constraints on dark energy can be obtained using the DM($z$) relation.

Zhou et al. (2014) constrained the equation of state of dark energy using type Ia supernovae (SNe Ia), simulated FRBs and BAO data in the optimistic case that in each narrow-redshift-bin tens of FRBs have been detected [245]. FRB can also be associated with other cosmological probes to constrain the dark energy equation of state, such as GRB/FRB system [247], lensed FRBs/CMB/SNe Ia system [246], GW/FRB system [248], CMB+GW+FRB [250] and FRB/standard distance probes [249]. However, the effects of DMs contributed by host galaxies and fluctuations of IGM should be treated properly.

### 7.3 Proper Distance

The proper distance $d_{\rm p}$ is the length from the source to the observer along the light of sight at a specific moment of cosmological time. In the frame of the Friedmann-Lemaître-Robertson-Walker (FLRW) metric, it is expressed as

$$d_{\rm p}(r) = a\int_0^r \frac{{\rm d}r'}{\sqrt{1-Kr'^2}} \quad (89)$$

where $a$ is the scale factor, $r$ is the co-moving coordinate of the source and $K$ is the curvature parameter.

The proper distance is difficult to measure because it depends on the cosmic curvature. FRB is a good probe that is independent of the cosmic curvature and record information about the expansion of the universe. The proper distance-redshift relation can be derived from the DMs of FRBs with measured redshifts. Yu & Wang (2017) proposed that $d_{\rm p} - z$ relation can be derived from 500 mocked FRBs with ($\mathrm{DM_{IGM}}, z$) data through Monte Carlo simulation [252]. Figure 9 shows the ability of FRBs in measuring proper distance from simulations. 500 FRBs with DM and redshift measurements can tightly constrain cosmic proper distance. This $d_{\rm p}-z$ relation can constrain the cosmic curvature as well. Of course, this depends on the number of FRB detections in the near future. With the advancement of observational technique, this will become an effective method to measure the inherent distance.

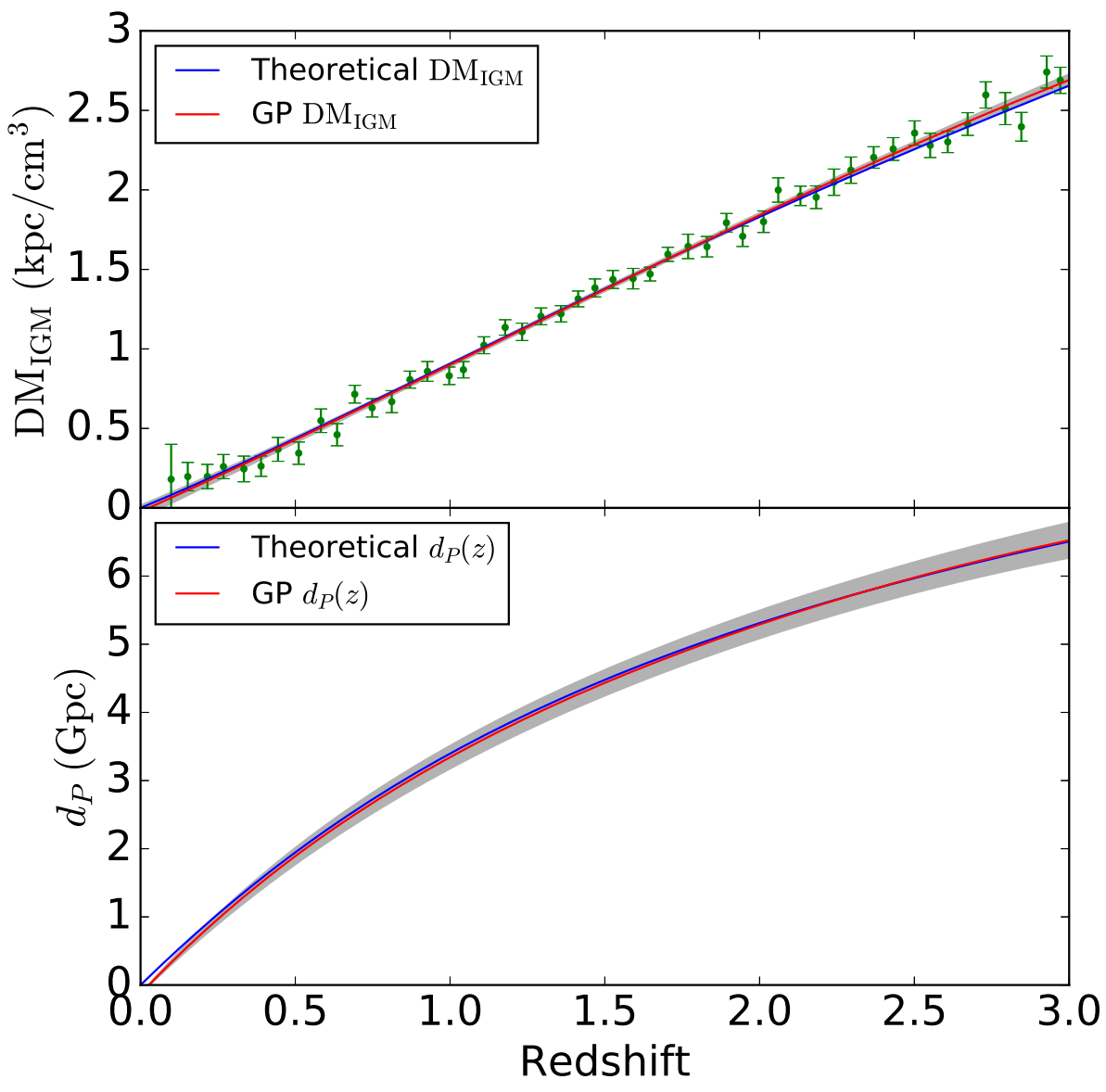

**Figure 9** Top panel shows the binned mock $\mathrm{DM_{IGM}}$ data with $1\sigma$ error bars. Bottom panel shows the $d_{\rm p}(z)$ function with its $1\sigma$ confidence region derived from Gaussian Process method and its theoretical function which is shown by red line. $\Omega_{\rm k} = 0$ is assumed. (See [252] for more details.)

### 7.4 Hubble Parameter $H(z)$

FRBs with redshift measurements can be used to measure Hubble parameter $H(z)$, which is a very important parameter in cosmology. In flat ΛCDM cosmology, $H(z)$ can be expresses as:

$$H(z) = H_0\sqrt{\Omega_\Lambda + \Omega_{\rm M}(1+z)^3}, \quad (90)$$

where $\Omega_\Lambda$ is the vacuum energy density fraction.

Wu et al. (2020) proposed a new model-independent method to calculate the Hubble parameter $H(z)$ [253]. The averaged redshifts $\langle z\rangle$ and dispersion measures $\langle \mathrm{DM_{IGM}}\rangle$ of FRBs are known. Differentiating equation (9), the Hubble parameter can be expressed as

$$H(z) = \frac{3c}{8\pi G m_p}\Omega_{\rm b}H_0^2 F(z)\frac{\Delta z}{\Delta \mathrm{DM_{IGM}}}, \quad (91)$$

where $\Delta z$ and $\Delta\mathrm{DM_{IGM}}$ are the differences of $z$ and $\mathrm{DM_{IGM}}$ between two adjacent bins, respectively. Figure 10 shows the simulation result. The error of derived $H(z)$ is about 6%. The DMs and redshifts of 500 simulated FRBs are used to measure $H(z)$, and the derived results are consistent with the theoretical values of $H(z)$. As more localized FRBs are observed, the calculation of Hubble parameters that rely on redshift will become possible and provide useful probes for cosmology.

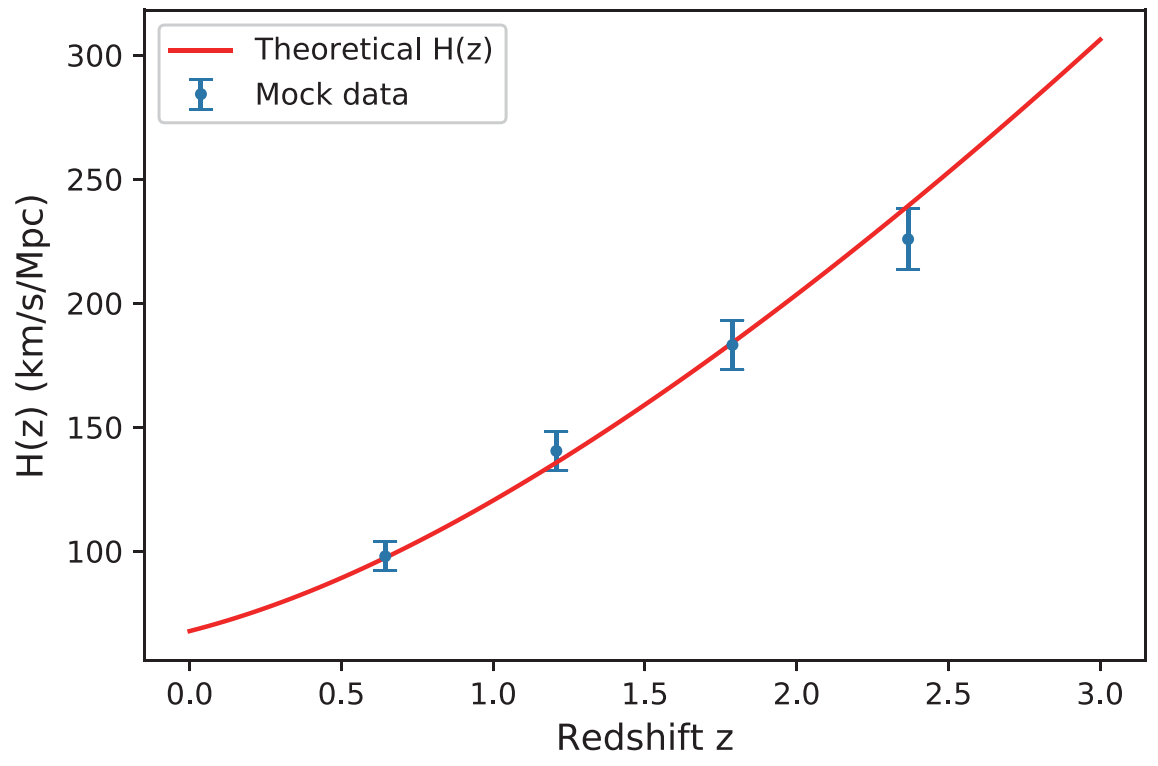

**Figure 10** $H(z)$ and $1\sigma$ errors derived from 500 mock FRBs are shown as blue dots. The red line represents the theoretical $H(z)$ function. The deviation between the simulated value of $H(z)$ and the theoretical one is about 6% at $z = 2.4$. (Adapted from [253].)

## 7.5 Dark Matter

Dark matter is abundant in the universe. Because it does not interact with electromagnetic fields, it is difficult to detect it directly. Massive compact halo objects (MACHOs) are the possible candidate for dark matter, and the fraction of dark matter $f_{\rm DM}$ in MACHO has received a lot of attention. Strong lensing of FRB by MACHO can be used to constrain $f_{\rm DM}$. When there is MACHO in the LOS, due to gravitational lensing effect, two images and the corresponding time interval will appear. For a MACHO with mass $M_{\rm L}$ as a point lens, the Einstein radius is

$$\theta_{\rm E} = 2\sqrt{\frac{GM_L}{c^2}\frac{D_{LS}}{D_S D_L}}, \tag{92}$$

where $D_S$, $D_L$ and $D_{LS}$ are the distances to the source, to the lens and between the source and the lens, respectively. The time delay between the two separated images is [254, 255]

$$\Delta t = \frac{4GM_L}{c^3}(1+z_L)\left[\frac{y}{2}\sqrt{y^2+4} + \log\left(\frac{\sqrt{y^2+4}+y}{\sqrt{y^2+4}-y}\right)\right], \tag{93}$$

where $y \equiv \beta/\theta_E$ and $\beta$ is the impact parameter.

Muñoz et al. (2016) proposed that the lensed FRBs with double peaks can be used to constrain the fraction of dark matter in MACHOs [254]. They constrained $f_{\rm DM}$ in MACHOs to $f_{\rm DM} \lesssim 0.08$ for $M_{\rm L} \gtrsim 20M_\odot$ if there are no FRBs microlensed. Wang & Wang (2018) proposed to probe the compact dark matter with FRBs gravitational lensed by MACHO binaries [256]. They concluded that $f_{\rm DM}$ in MACHO is constrained to $< 0.001$ when there were no search of multiple-peaked FRBs for time intervals larger than 1 ms and flux ratio less than $10^3$. These methods of using FRBs to find dark matter in MACHO is widely studied later [257-261].

## 7.6 Hubble Constant

The large event rate, small temporal duration and cosmological origin of FRBs suggests that FRBs can be a clean cosmological probe. Li et al. (2018) proposed that strongly lensed repeating FRBs can constrain Hubble constant $H_0$ [262]. The accurate time intervals between different images of lensed FRBs can establish connections with Hubble constant, which is a precision probe of the universe. The Hubble constant $H_0$ can be constrained with $\sim 0.91\%$ uncertainty by simulated 10 lensed FRB systems within the flat $\Lambda$CDM model. Compared with other possible constrains on $H_0$, lensed FRB has obvious advantages in ascendancy. In addition, cosmological curvature can also be derived using lensed FRB, which is independent of model [262]. The gravitational lensed FRBs can be used in various cosmological research in the future [263].

## 7.7 Weak Equivalence Principle

Einstein's Equivalence Principle (EEP) is the equivalence of gravitational and inertial mass, which is the foundation of the General Relativity. The verification of the EEP comes from the parameterized post-Newtonian (PPN) parameters, such as the parameter $\gamma$. Calculating the difference of the $\gamma$ value between different particles $\Delta\gamma$ can constrain the accuracy of the EEP. Now, the extragalactic transients have been used to the constrains on EEP or Einstein's Weak Equivalence Principle (WEP) [264-268].

Wei et al. (2015) proposed that FRBs with cosmological origin can be a good candidate for constraining the EEP through the association with GRB [269]. They obtained that the upper limit of the difference of the PPN parameter $\gamma$ is $[\gamma(1.23{\rm GHz}) - \gamma(1.45{\rm GHz})] < 4.36\times10^{-9}$, which can be expressed as

$$\Delta\gamma = \gamma_1 - \gamma_2 < (\Delta t_{\rm obs} - \Delta t_{\rm DM})\left(\frac{GM_{\rm MW}}{c^3}\right)^{-1}\ln^{-1}\left(\frac{d}{b}\right), \tag{94}$$

where $\Delta t_{\rm obs}$ is the observed time delay between two different energy bands, $\Delta t_{\rm DM}$ is the time delay contribution from the dispersion, $M_{\rm MW}$ is the Milky Way mass, $d$ is the luminosity distance from the source to the earth and $b$ is the impact parameter of light rays relative to the Milky Way center.

This is a strong restriction on the EEP, which is more accurate than the restriction using SN and GRB. Using the above method, Tingay & Kalpan (2016) used the first localized FRB 150418 alone to limit the EEP [270]. They obtained a limit of $\Delta\gamma < 1 - 2 \times 10^{-9}$, which is an order of magnitude better than the previous constraint [269].

### 7.8 Contamination from $\mathrm{DM_{host}}$ and IGM inhomogeneity

#### *7.8.1* $\mathrm{DM_{host}}$

In order to know the exact value of $\mathrm{DM_{IGM}}$, the other contributions to the observed DM should be well understood. In Eq. (7), $\mathrm{DM_{MW}}$ represents the total contribution from the Milky Way, which consists of two components, $\mathrm{DM_{MW}} = \mathrm{DM_{disk}} + \mathrm{DM_{halo}}$. The disk component $\mathrm{DM_{disk}}$ can be subtracted using the NE2001 model [33] or YMW16 model [34]. $\mathrm{DM_{halo}}$ is the part contributed by the halo of Milky Way, of which the typical value is $30 - 80$ pc cm$^{-3}$ [85, 271]. The source contribution $\mathrm{DM_{source}}$ depends on the environment of the progenitor. If FRBs are produced by the remnants from collapses of massive stars, the values of $\mathrm{DM_{source}}$ can be large [272]. Differently, if FRBs are produced by BNS mergers [273], $\mathrm{DM_{source}}$ is usually small [98, 274]. Once the central engine is known, $\mathrm{DM_{source}}$ can be derived analytically [98, 272, 275].

The contribution by host galaxies $\mathrm{DM_{host}}$ is poorly known. Xu & Han (2015) estimated the DM contributions by different types of galaxies using the scaled model of NE2001 [276]. Methods of constraining $\mathrm{DM_{host}}$ statistically have been discussed [277, 278]. Moreover, the IllustrisTNG simulation is used to derive the distributions of $\mathrm{DM_{host}}$ [279]. Figure 11 gives $\mathrm{DM_{host}}$ at different redshifts. For repeating FRBs, the median $\mathrm{DM_{host}}$ are $35(1+z)^{1.08}$ and $96(1+z)^{0.83}$ pc cm$^{-3}$ for FRBs like FRB 121102 and FRB 180916, respectively. The median of $\mathrm{DM_{host}}$ is about $30 - 70$ pc cm$^{-3}$ for non-repeating FRBs in the redshift range $z = 0.1 - 1.5$. In this case, the evolution of the median $\mathrm{DM_{host}}$ can be fitted as $33(1+z)^{0.84}$ pc cm$^{-3}$. The distributions of $\mathrm{DM_{host}}$ of repeating and non-repeating FRBs can be well fitted with the log-normal function

$$P(x;\mu,\sigma) = \frac{1}{x\sigma\sqrt{2\pi}} \exp\left(-\frac{(\ln x - \mu)^2}{2\sigma^2}\right). \quad (95)$$

The mean and variance of this distribution are $e^{\mu}$ and $e^{(2\mu+\sigma^2)}[e^{\sigma^2} - 1]$, respectively.

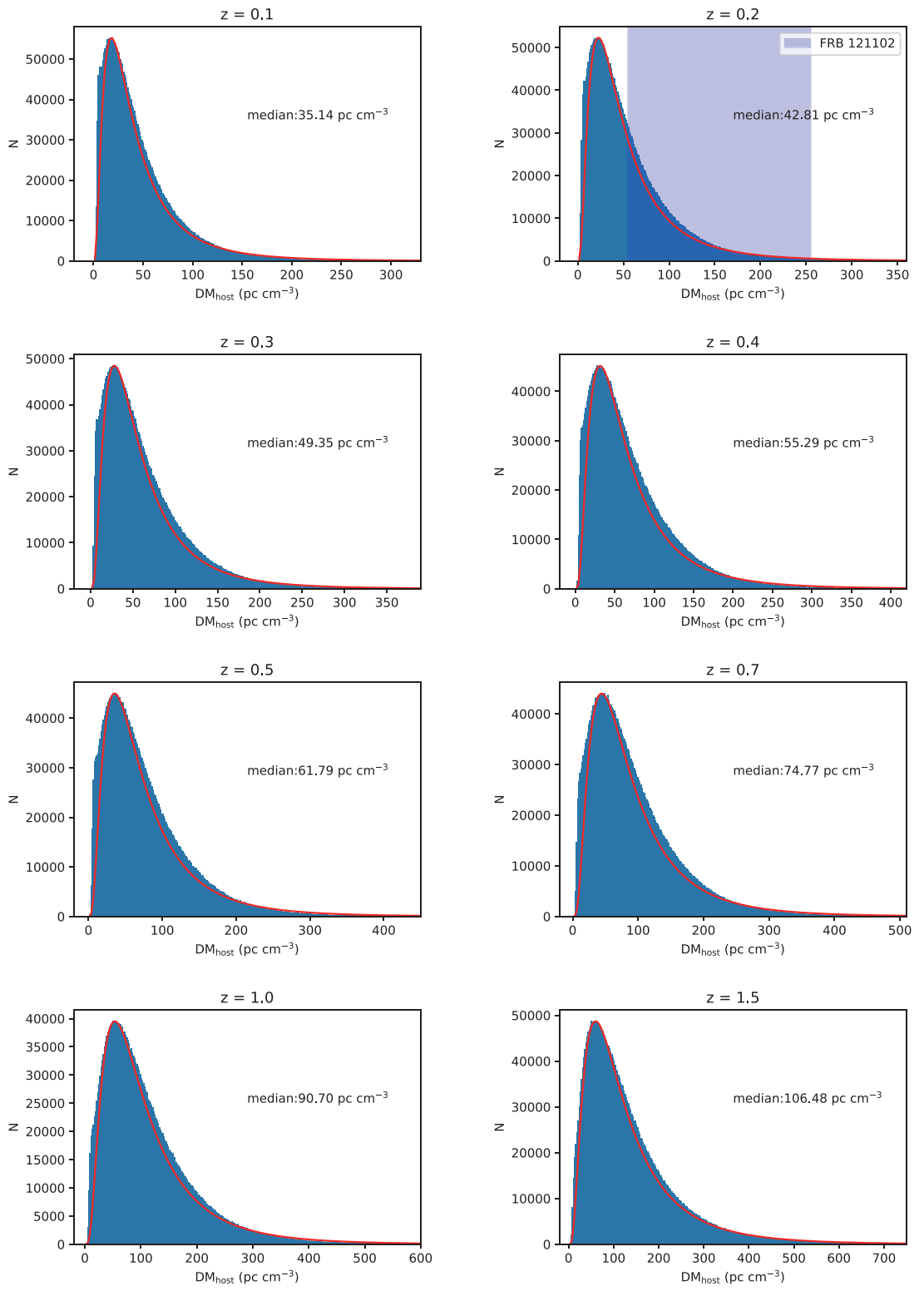

**Figure 11** The distributions of $\mathrm{DM_{host}}$ at different redshifts. The red lines are the best-fitting results using log-normal distribution. The blue shaded region in z = 0.2 panel is the $\mathrm{DM_{host}}$ for FRB 121102. (See [279] for more details.)

#### *7.8.2 IGM inhomogeneity*

McQuinn (2014) considered the effect of inhomogeneity with three models for halo gas profile of the ionized baryons [280]. Dolag et al. (2015) and Pol et al. (2019) studied $\mathrm{DM_{IGM}}$ with different cosmological simulations at low-redshift ($z < 2$) universe [271, 281, 282]. Efforts of tacking IGM inhomogeneity using the FRB data have been made [283, 284]. Jaroszynski (2019) used the Illustris simulation to estimate the $\mathrm{DM_{IGM}}$ and its scatters in the $z < 5$ universe [285]. Zhang et al. (2020) and Takahashi et al. (2020) used the IllustrisTNG simulation to realistically estimate the $\mathrm{DM_{IGM}}$ up to $z \sim 9$ [279, 286]. At low redshifts, the distributions of $\mathrm{DM_{IGM}}$ are shown in Figure 12. They are well fitted by [82]

$$p_{\mathrm{IGM}}(\Delta) = A\Delta^{-\beta} \exp[-\frac{(\Delta^{-\alpha} - C_0)^2}{2\alpha^2\sigma_{\mathrm{DM}}^2}], \quad \Delta > 0. \quad (96)$$

where $\Delta \equiv \mathrm{DM_{IGM}}/\langle\mathrm{DM_{IGM}}\rangle$, and $\beta$ is related to the inner density profile of gas ($\rho \propto r^{-\alpha}$) in halos. $\sigma_{\mathrm{DM}}$ is an effective standard deviation. $C_0$ can affect the horizontal position.

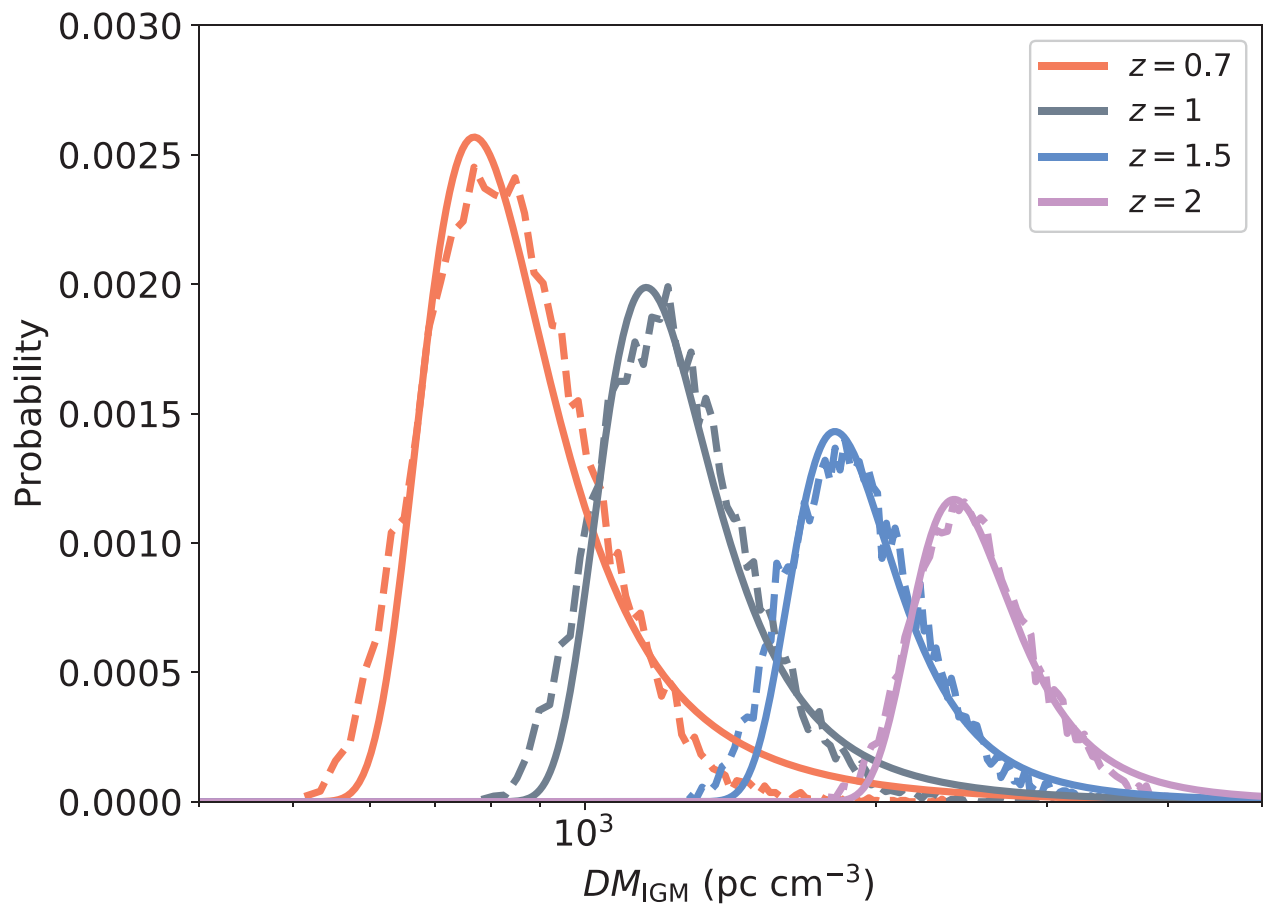

**Figure 12** The distributions of $\mathrm{DM_{IGM}}$ at different redshifts. The red lines are the best-fitting results using Eq. (96). (Adapted from [279].)

In order to obtain the reliable cosmological constraint from FRBs, the distributions of $\mathrm{DM_{host}}$ and $\mathrm{DM_{IGM}}$ must be properly handled [82]. The potential of FRBs as cosmological probes is huge, however, it is currently subject to limited sample of localized bursts. With more events coming soon, FRB could become a very powerful probe.

### 7.9 Constraints on photon mass

The rest mass of a photon is supposed to be strictly zero and the velocity of the photon is the speed of light. It is very important to test the correctness of this hypothesis experimentally, because it involves many fundamental physical problems. The arrival time of the lower frequency pulse of an FRB is later than the higher one due to plasma effect. The frequency dependence of the pulse arrival time of the FRB follows the $\nu^{-2}$ law, which can constrain the photon mass [287-289]. Considering two photons with different frequencies, the high frequency $\nu_{\mathrm{h}}$ and the low frequency $\nu_{\mathrm{l}}$. The photon mass $m_\gamma$ can be constrained as

$$m_\gamma = hc^{-2}\left[\frac{2H_0\Delta t_{m_\gamma}}{\left(\nu_{\mathrm{l}}^{-2}-\nu_{\mathrm{h}}^{-2}\right)H_1(z)}\right]^{1/2}, \tag{97}$$

where

$$H_1(z) = \int_0^z \frac{(1+z')^{-2}dz'}{\sqrt{\Omega_{\mathrm{m}}(1+z')^3+\Omega_\Lambda}}, \tag{98}$$

and $\Delta t_{m_\gamma}$ is the arrival time interval of the low frequency and the high frequency photons.

Based on the observations of FRB 150418 at $z = 0.492$, Wu et al. (2016) found an upper limit of $m_\gamma \leq 5.2 \times 10^{-47}$ g, which is a strict constraint on the photon mass [287]. The plasma effect and scattering due to multi-path propagation through the intervening material should be considered. Bonetti et al. (2016) derived the photon mass $m_\gamma \leq 3.2 \times 10^{-50}$ kg by considering the frequency dependence of dispersion [288]. Shao & Zhang (2017) developed a Bayesian framework to constrain the photon mass with a catalog of FRBs [289]. With the growth of FRB population, the constraint on the photon mass will be promising using this method.

### 7.10 Reionization history

The transition from neutral to ionized universe is called reionization. Usually, it is believed that the reionization of hydrogen occurs between $6 < z < 12$. For helium the range is $2 < z < 6$ [290]. When reionization happens is still unresolved in the modern cosmology. Extragalactic FRBs can be a good probe for tracing the history of the H and He II reionization using high DM contributed by IGM [32, 150, 242, 291]. According to Eq.(9), the ionization fractions of intergalactic hydrogen $X_{\mathrm{e,H}}$ and helium $X_{\mathrm{e,He}}$ can be derived when the redshifts and $\mathrm{DM_{IGM}}$ of FRBs are measured.

Zheng et al. (2014) proposed that FRBs can be the potential probe to study the history of He II reionization [242]. During the epoch of the He II reionization at redshift $z \sim 3$, the derivation of DM value decreased 8%. This indicates that DM will changes across the He II reionization epoch. It is clearly that a large FRB population is needed to study the He II reionization history. Caleb et al. (2019) simulated the FRB samples, and tracked the evolution history of reionization using the simulations of DM values of FRBs. They obtained the number of FRBs needed to distinguish reionization occurred at $z = 3$ or $z = 6$ through simulations. They concluded that more than 1100 FRBs were needed when the intergalactic medium IGM is homogeneous and more than 1600 FRBs needed when the IGM is inhomogeneous [150]. The joint observation of the next generation radio telescope will observe enough FRBs to probe the history of the He II reionization.

### 7.11 Intergalactic turbulence

The intergalactic turbulence is very common in the universe, which is related to the formation of large-scale structure of the universe. The observations and statistical analysis of the intergalactic turbulence are very useful in astronomical research [292]. The high DM of FRBs is mainly contributed by IGM, so the dispersion disturbance of $\mathrm{DM_{IGM}}$ can reflect the density fluctuation in the IGM. Thus, an FRB can be used to probe intergalactic turbulence. Recently, Xu & Zhang (2020) first proposed a statistical measurement of the intergalactic turbulence by using a population of FRBs [292]. The structure function of DMs can represent the multi-scale electron number density fluctuations from the intergalactic turbulence. Xu & Zhang (2020) calculated the structure formation value using the published FRB data, which has large amplitude and a Kolmogorov power-law scaling. They concluded that the intergalactic turbulence extended outward to 100 Mpc with a Kolmogorov power spectrum. The extensive application of using fast radio bursts to study the intergalactic turbulence depends on the accumulation of more FRB data in the future.

## 8 Summary and Future Prospects

The FRB field is very energetic and rapidly developing. Important findings keep coming quickly. The current status and major developments in understanding their inherent physics are summarized here.

(i) It is controversial whether all FRBs repeat. Several statistical properties of repeaters and apparent non-repeaters seem different. However, no clear sign of dichotomy has been found. It has been claimed that these differences could be owing to the diversity in repeating modes. The other viewpoint that some authors prefer is that nature is diverse, and two populations could have different physical origins. The existence of genuinely non-repeating FRBs could be finally confirmed if an FRB from a cataclysmic event (e.g., BNS merger) is observed. The large offset and host galaxy property of FRB 180924 and FRB 190523 seem to favor this hypothesis. As may be imagined, it will be a big breakthrough if FRB is associated with gravitational-wave signal and other EM counterparts.

(ii) The radiation mechanism of FRBs remains unknown. We have introduced five mechanisms, which do not explicitly define the whole mechanism. Currently, the coherent curvature emission and synchrotron maser emission from magnetized shocks are in the leading position. These two have been studied in great detail and applied in the case of FRB 121102 and FRB 200428. Under a few assumptions, both of them can explain the observational data well. Their predictions are different and verifiable in the future. Some pieces of emerging evidence, such as the FAST detection of diverse polarization angle swings from FRB 180301 [69] support the former emission mechanism, i.e., a pulsar's magnetospheric origin.

(iii) The magnetar origin of FRBs has been established through FRB 200428. However, the emission site has not yet been determined. Both close-in and far-away models can explain the observations well [216, 230]. Kirsten et al. (2020) reported a repetition of FRB 200428 with double peaks again after long-time monitoring [21], which makes the situation more interesting. However, conclusive evidence has not yet been found, which might be polarization property. Generally, far-away maser models predict that FRB is linearly polarized with a flat PA curve, while close-in models could have diverse PA swings according to the geometry of magnetic field and observing direction. This will come to a conclusion once the polarization information of FRB 200428 is obtained from the subsequent repeating bursts.

(iv) The size of the total FRB sample is still small. We definitely need more FRBs to precisely analyze the statistical properties. Especially, the number of localized FRBs is very limited. Hundreds of localized FRBs are needed to do fine cosmology. In this sense, the population study and cosmological application of FRBs are still at an early stage, and plenty of work needs to be done in the future. With the upgrade of current instruments and the upcoming new facilities, the total number should reach the order of a thousand in a short time.

Of course, certain open key questions in the FRB field remain to be addressed over the next few years. Do genuinely non-repeating FRBs exist? What is the real ratio of repeaters? Do all repeaters have a periodical activity? What is the form of the intrinsic luminosity function and redshift distribution? Is there any other physical-motivated criterion for classifying FRB? Except for magnetars, could any other progenitors produce FRBs? Are there multi-wavelength counterparts or multi-messenger signals accompanying FRBs? These questions are expected to be answered sooner or later, with future observations performed by high-sensitivity radio telescopes, such as FAST and SKA.

## References

*We would like to thank two anonymous referees for helpful comments. This work is supported by the National Key Research and Development Program of China (grant No. 2017YFA0402600) and the National Natural Science Foundation of China (grant No. 11833003, U1831207, 11903018 and 11851305). D.X. is also supported by the Natural Science Foundation for the Youth of Jiangsu Province (grant No. BK20180324).*

**Conflict of interest** The authors declare that they have no conflict of interest.

1 D. R. Lorimer, M. Bailes, M. A. McLaughlin, D. J. Narkevic, and F. Crawford, Science **318**(5851), 777 (Nov. 2007), 0709.4301.

2 E. F. Keane, M. Kramer, A. G. Lyne, B. W. Stappers, and M. A. McLaughlin, MNRAS, **415**(4), 3065 (Aug. 2011), 1104.2727.

3 S. Burke-Spolaor, M. Bailes, R. Ekers, J.-P. Macquart, and I. Crawford, Fronefield, ApJ, **727**(1), 18, 18 (Jan. 2011), 1009.5392.

4 D. Thornton, B. Stappers, M. Bailes, B. Barsdell, S. Bates, N. D. R. Bhat, M. Burgay, S. Burke-Spolaor, D. J. Champion, P. Coster, *et al.*, Science **341**(6141), 53 (Jul. 2013), 1307.1628.

5 L. G. Spitler, P. Scholz, J. W. T. Hessels, S. Bogdanov, A. Brazier, F. Camilo, S. Chatterjee, J. M. Cordes, F. Crawford, J. Deneva, *et al.*, Nature, **531**(7593), 202 (Mar. 2016), 1603.00581.

6 P. Scholz, L. G. Spitler, J. W. T. Hessels, S. Chatterjee, J. M. Cordes, V. M. Kaspi, R. S. Wharton, C. G. Bassa, S. Bogdanov, F. Camilo, *et al.*, ApJ, **833**(2), 177, 177 (Dec. 2016), 1603.08880.

7 S. Chatterjee, C. J. Law, R. S. Wharton, S. Burke-Spolaor, J. W. T. Hessels, G. C. Bower, J. M. Cordes, S. P. Tendulkar, C. G. Bassa, P. Demorest, *et al.*, Nature, **541**(7635), 58 (Jan. 2017), 1701.01098.

8 B. Marcote, Z. Paragi, J. W. T. Hessels, A. Keimpema, H. J. van Langevelde, Y. Huang, C. G. Bassa, S. Bogdanov, G. C. Bower, S. Burke-Spolaor, *et al.*, ApJL, **834**(2), L8, L8 (Jan. 2017), 1701.01099.

9 S. P. Tendulkar, C. G. Bassa, J. M. Cordes, G. C. Bower, C. J. Law, S. Chatterjee, E. A. K. Adams, S. Bogdanov, S. Burke-Spolaor, B. J. Butler, *et al.*, ApJL, **834**(2), L7, L7 (Jan. 2017), 1701.01100.

10 E. Petroff, E. D. Barr, A. Jameson, E. F. Keane, M. Bailes, M. Kramer, V. Morello, D. Tabbara, and W. van Straten, PASA, **33**, e045, e045 (Sep. 2016), 1601.03547.

11 CHIME/FRB Collaboration, M. Amiri, K. Bandura, M. Bhardwaj, P. Boubel, M. M. Boyce, P. J. Boyle, C. . Brar, M. Burhanpurkar, T. Cassanelli, *et al.*, Nature, **566**(7743), 235 (Jan. 2019), 1901.04525.

12 CHIME/FRB Collaboration, B. C. Andersen, K. Bandura, M. Bhardwaj, P. Boubel, M. M. Boyce, P. J. Boyle, C. Brar, T. Cassanelli, P. Chawla, *et al.*, ApJL, **885**(1), L24, L24 (Nov. 2019), 1908.03507.

13 E. Fonseca, B. C. Andersen, M. Bhardwaj, P. Chawla, D. C. Good, A. Josephy, V. M. Kaspi, K. W. Masui, R. Mckinven, D. Michilli, *et al.*, ApJL, **891**(1), L6, L6 (Mar. 2020), 2001.03595.

14 K. W. Bannister, A. T. Deller, C. Phillips, J. P. Macquart, J. X. Prochaska, N. Tejos, S. D. Ryder, E. M. Sadler, R. M. Shannon, S. Simha, *et al.*, Science **365**(6453), 565 (Aug. 2019), 1906.11476.

15 S. Bhandari, E. M. Sadler, J. X. Prochaska, S. Simha, S. D. Ryder, L. Marnoch, K. W. Bannister, J.-P. Macquart, C. Flynn, R. M. Shannon, *et al.*, ApJL, **895**(2), L37, L37 (Jun. 2020), 2005.13160.

16 CHIME/FRB Collaboration, M. Amiri, B. C. Andersen, K. M. Band ura, M. Bhardwaj, P. J. Boyle, C. Brar, P. Chawla, T. Chen, J. F. Cliche, *et al.*, Nature, **582**(7812), 351 (Jun. 2020), 2001.10275.

17 K. M. Rajwade, M. B. Mickaliger, B. W. Stappers, V. Morello, D. Agarwal, C. G. Bassa, R. P. Breton, M. Caleb, A. Karastergiou, E. F. Keane, *et al.*, MNRAS, **495**(4), 3551 (May 2020), 2003.03596.

18 M. Cruces, L. G. Spitler, P. Scholz, R. Lynch, A. Seymour, J. W. T. Hessels, C. Gouiffés, G. H. Hilmarsson, M. Kramer, and S. Munjal, MNRAS, **500**(1), 448 (Jan. 2021), 2008.03461.

19 C. D. Bochenek, V. Ravi, K. V. Belov, G. Hallinan, J. Kocz, S. R. Kulkarni, and D. L. McKenna, Nature, **587**(7832), 59 (Nov. 2020), 2005.10828.

20 CHIME/FRB Collaboration, B. Â. C. Andersen, K. Â. M. Band ura, M. Bhardwaj, A. Bij, M. Â. M. Boyce, P. Â. J. Boyle, C. Brar, T. Cassanelli, P. Chawla, *et al.*, Nature, **587**(7832), 54 (Nov. 2020).

21 F. Kirsten, M. P. Snelders, M. Jenkins, K. Nimmo, J. van den Eijnden, J. W. T. Hessels, M. P. Gawroński, and J. Yang, Nature Astronomy (Nov. 2020), 2007.05101.

22 C. K. Li, L. Lin, S. L. Xiong, M. Y. Ge, X. B. Li, T. P. Li, F. J. Lu, S. N. Zhang, Y. L. Tuo, Y. Nang, *et al.*, arXiv e-prints p. arXiv:2005.11071, arXiv:2005.11071 (May 2020), 2005.11071.

23 S. Mereghetti, V. Savchenko, C. Ferrigno, D. Götz, M. Rigoselli, A. Tiengo, A. Bazzano, E. Bozzo, A. Coleiro, T. J. L. Courvoisier, *et al.*, ApJL, **898**(2), L29, L29 (Aug. 2020), 2005.06335.

24 A. Ridnaia, D. Svinkin, D. Frederiks, A. Bykov, S. Popov, R. Aptekar, S. Golenetskii, A. Lysenko, A. Tsvetkova, M. Ulanov, *et al.*, arXiv e-prints p. arXiv:2005.11178, arXiv:2005.11178 (May 2020), 2005.11178.

25 M. Tavani, C. Casentini, A. Ursi, F. Verrecchia, A. Addis, L. A. Antonelli, A. Argan, G. Barbiellini, L. Baroncelli, G. Bernardi, *et al.*, arXiv e-prints p. arXiv:2005.12164, arXiv:2005.12164 (May 2020), 2005.12164.

26 J. I. Katz, Progress in Particle and Nuclear Physics **103**, 1 (Nov. 2018), 1804.09092.

27 S. B. Popov, K. A. Postnov, and M. S. Pshirkov, Physics Uspekhi **61**(10), 965 (Oct. 2018), 1806.03628.

28 E. Petroff, J. W. T. Hessels, and D. R. Lorimer, A&ARv, **27**(1), 4, 4 (May 2019), 1904.07947.

29 J. M. Cordes and S. Chatterjee, ARA&A, **57**, 417 (Aug. 2019), 1906.05878.

30 B. Zhang, Nature, **587**(7832), 45 (Nov. 2020), 2011.03500.

31 S. Chatterjee, arXiv e-prints p. arXiv:2012.10377, arXiv:2012.10377 (Dec. 2020), 2012.10377.

32 W. Deng and B. Zhang, ApJL, **783**(2), L35, L35 (Mar. 2014), 1401.0059.

33 J. M. Cordes and T. J. W. Lazio, arXiv e-prints pp. astro–ph/0207156, astro-ph/0207156 (Jul. 2002), astro-ph/0207156.

34 J. M. Yao, R. N. Manchester, and N. Wang, ApJ, **835**(1), 29, 29 (Jan. 2017), 1610.09448.

35 J. M. Cordes and M. A. McLaughlin, ApJ, **596**(2), 1142 (Oct. 2003), astro-ph/0304364.

36 O. Löhmer, M. Kramer, D. Mitra, D. R. Lorimer, and A. G. Lyne, ApJL, **562**(2), L157 (Dec. 2001), astro-ph/0111165.

37 S. Xu and B. Zhang, ApJ, **832**(2), 199, 199 (Dec. 2016), 1608.03930.

38 V. Ravi, MNRAS, **482**(2), 1966 (Jan. 2019), 1710.08026.

39 R. N. Manchester, G. B. Hobbs, A. Teoh, and M. Hobbs, AJ, **129**(4), 1993 (Apr. 2005), astro-ph/0412641.

40 J. M. Cordes, R. S. Wharton, L. G. Spitler, S. Chatterjee, and I. Wasserman, arXiv e-prints p. arXiv:1605.05890, arXiv:1605.05890 (May 2016), 1605.05890.

41 S. Xu and B. Zhang, ApJ, **835**(1), 2, 2 (Jan. 2017), 1610.03011.

42 H. Qiu, R. M. Shannon, W. Farah, J.-P. Macquart, A. T. Deller, K. W. Bannister, C. W. James, C. Flynn, C. K. Day, S. Bhandari, *et al.*, MNRAS, **497**(2), 1382 (Jul. 2020), 2006.16502.

43 B. J. Rickett, ARA&A, **28**, 561 (Jan. 1990).

44 R. Narayan, Philosophical Transactions of the Royal Society of London Series A **341**(1660), 151 (Oct. 1992).

45 J. M. Cordes and B. J. Rickett, ApJ, **507**(2), 846 (Nov. 1998).

46 V. Ravi, R. M. Shannon, M. Bailes, K. Bannister, S. Bhandari, N. D. R. Bhat, S. Burke-Spolaor, M. Caleb, C. Flynn, A. Jameson, *et al.*, Science **354**(6317), 1249 (Dec. 2016), 1611.05758.

47 V. Gajjar, A. P. V. Siemion, D. C. Price, C. J. Law, D. Michilli, J. W. T. Hessels, S. Chatterjee, A. M. Archibald, G. C. Bower, C. Brinkman, *et al.*, ApJ, **863**(1), 2, 2 (Aug. 2018), 1804.04101.

48 D. Simard and V. Ravi, ApJL, **899**(1), L21, L21 (Aug. 2020), 2006.13184.

49 P. Schneider, J. Ehlers, and E. E. Falco, *Gravitational Lenses* (1992).

50 A. W. Clegg, A. L. Fey, and T. J. W. Lazio, ApJ, **496**(1), 253 (Mar. 1998), astro-ph/9709249.

51 J. Wagner and X. Er, arXiv e-prints p. arXiv:2006.16263, arXiv:2006.16263 (Jun. 2020), 2006.16263.

52 J. M. Cordes, I. Wasserman, J. W. T. Hessels, T. J. W. Lazio, S. Chatterjee, and R. S. Wharton, ApJ, **842**(1), 35, 35 (Jun. 2017), 1703.06580.

53 X. Er and A. Rogers, MNRAS, **488**(4), 5651 (Oct. 2019), 1907.10787.

54 A. Rogers and X. Er, MNRAS, **485**(4), 5800 (Jun. 2019), 1903.06384.

55 X. Er, Y.-P. Yang, and A. Rogers, ApJ, **889**(2), 158, 158 (Feb. 2020), 2001.02100.

56 J. W. T. Hessels, L. G. Spitler, A. D. Seymour, J. M. Cordes, D. Michilli, R. S. Lynch, K. Gourdji, A. M. Archibald, C. G. Bassa, G. C. Bower, *et al.*, ApJL, **876**(2), L23, L23 (May 2019), 1811.10748.

57 C. K. Day, A. T. Deller, R. M. Shannon, H. Qiu, K. W. Bannister, S. Bhandari, R. Ekers, C. Flynn, C. W. James, J. P. Macquart, *et al.*, MNRAS, (Jul. 2020), 2005.13162.

58 W. Wang, B. Zhang, X. Chen, and R. Xu, ApJL, **876**(1), L15, L15 (May 2019), 1903.03982.

59 B. D. Metzger, B. Margalit, and L. Sironi, MNRAS, **485**(3), 4091 (May 2019), 1902.01866.

60 A. W. Hotan, M. Bailes, and S. M. Ord, MNRAS, **362**(4), 1267 (Oct. 2005).

61 J. P. Macquart, R. D. Ekers, I. Feain, and M. Johnston-Hollitt, ApJ, **750**(2), 139, 139 (May 2012), 1203.2706.

62 S. P. O'Sullivan, S. Brown, T. Robishaw, D. H. F. M. Schnitzeler, N. M. McClure-Griffiths, I. J. Feain, A. R. Taylor, B. M. Gaensler, T. L. Land ecker, L. Harvey-Smith, *et al.*, MNRAS, **421**(4), 3300 (Apr. 2012), 1201.3161.

63 M. Caleb, E. F. Keane, W. van Straten, M. Kramer, J. P. Macquart, M. Bailes, E. D. Barr, N. D. R. Bhat, S. Bhandari, M. Burgay, *et al.*, MNRAS, **478**(2), 2046 (Aug. 2018), 1804.09178.

64 D. Michilli, A. Seymour, J. W. T. Hessels, L. G. Spitler, V. Gajjar, A. M. Archibald, G. C. Bower, S. Chatterjee, J. M. Cordes, K. Gourdji, *et al.*, Nature, **553**(7687), 182 (Jan. 2018), 1801.03965.

65 K. Nimmo, J. W. T. Hessels, A. Keimpema, A. M. Archibald, J. M. Cordes, R. Karuppusamy, F. Kirsten, D. Z. Li, B. Marcote, and Z. Paragi, arXiv e-prints p. arXiv:2010.05800, arXiv:2010.05800 (Oct. 2020), 2010.05800.

66 S. Dai, J. Lu, C. Wang, W. Wang, R. Xu, Y. Yang, S. Zhang, G. Hobbs, D. Li, and R. Luo, arXiv e-prints p. arXiv:2011.03960, arXiv:2011.03960 (Nov. 2020), 2011.03960.

67 G. H. Hilmarsson, D. Michilli, L. G. Spitler, R. S. Wharton, P. Demorest, G. Desvignes, K. Gourdji, S. Hackstein, J. W. T. Hessels, K. Nimmo, *et al.*, arXiv e-prints p. arXiv:2009.12135, arXiv:2009.12135 (Sep. 2020), 2009.12135.

68 H. Cho, J.-P. Macquart, R. M. Shannon, A. T. Deller, I. S. Morrison, R. D. Ekers, K. W. Bannister, W. Farah, H. Qiu, M. W. Sammons, *et al.*, ApJL, **891**(2), L38, L38 (Mar. 2020), 2002.12539.

69 R. Luo, B. J. Wang, Y. P. Men, C. F. Zhang, J. C. Jiang, H. Xu, W. Y. Wang, K. J. Lee, J. L. Han, B. Zhang, *et al.*, Nature, **586**(7831), 693 (Oct. 2020), 2011.00171.

70 K. Murase, K. Kashiyama, and P. Mészáros, MNRAS, **461**(2), 1498 (Sep. 2016), 1603.08875.

71 B. Margalit and B. D. Metzger, ApJL, **868**(1), L4, L4 (Nov. 2018), 1808.09969.

72 Z. Y. Zhao, G. Q. Zhang, Y. Y. Wang, and F. Y. Wang, arXiv e-prints p. arXiv:2010.10702, arXiv:2010.10702 (Oct. 2020), 2010.10702.

73 Y.-P. Yang, B. Zhang, and Z.-G. Dai, ApJL, **819**(1), L12, L12 (Mar. 2016), 1602.05013.

74 Q.-C. Li, Y.-P. Yang, and Z.-G. Dai, ApJ, **896**(1), 71, 71 (Jun. 2020), 2004.12516.

75 P. Chawla, B. C. Andersen, M. Bhardwaj, E. Fonseca, A. Josephy, V. M. Kaspi, D. Michilli, Z. Pleunis, K. M. Bandura, C. G. Bassa, *et al.*, ApJL, **896**(2), L41, L41 (Jun. 2020), 2004.02862.

76 M. Pilia, M. Burgay, A. Possenti, A. Ridolfi, V. Gajjar, A. Corongiu, D. Perrodin, G. Bernardi, G. Naldi, G. Pupillo, *et al.*, ApJL, **896**(2), L40, L40 (Jun. 2020), 2003.12748.

77 E. Parent, P. Chawla, V. M. Kaspi, G. Y. Agazie, H. Blumer, M. DeCesar, W. Fiore, E. Fonseca, J. W. T. Hessels, D. L. Kaplan, *et al.*, ApJ, **904**(2), 92, 92 (Dec. 2020), 2008.04217.

78 I. Pastor-Marazuela, L. Connor, J. van Leeuwen, Y. Maan, S. ter Veen, A. Bilous, L. Oostrum, E. Petroff, S. Straal, D. Vohl, *et al.*, arXiv e-prints p. arXiv:2012.08348, arXiv:2012.08348 (Dec. 2020), 2012.08348.

79 Z. Pleunis, D. Michilli, C. G. Bassa, J. W. T. Hessels, A. Naidu, B. C. Andersen, P. Chawla, E. Fonseca, A. Gopinath, V. M. Kaspi, *et al.*, arXiv e-prints p. arXiv:2012.08372, arXiv:2012.08372 (Dec. 2020), 2012.08372.

80 L. J. M. Houben, L. G. Spitler, S. ter Veen, J. P. Rachen, H. Falcke, and M. Kramer, A&A, **623**, A42, A42 (Mar. 2019), 1902.01779.

81 B. Marcote, K. Nimmo, J. W. T. Hessels, S. P. Tendulkar, C. G. Bassa, Z. Paragi, A. Keimpema, M. Bhardwaj, R. Karuppusamy, V. M. Kaspi, *et al.*, Nature, **577**(7789), 190 (Jan. 2020), 2001.02222.

82 J. P. Macquart, J. X. Prochaska, M. McQuinn, K. W. Bannister, S. Bhandari, C. K. Day, A. T. Deller, R. D. Ekers, C. W. James, L. Marnoch, *et al.*, Nature, **581**(7809), 391 (May 2020), 2005.13161.

83 P. Kumar, R. M. Shannon, C. Flynn, S. Osłowski, S. Bhandari, C. K. Day, A. T. Deller, W. Farah, J. F. Kaczmarek, M. Kerr, *et al.*, MNRAS, **500**(2), 2525 (Jan. 2021), 2009.01214.

84 V. Ravi, M. Catha, L. D'Addario, S. G. Djorgovski, G. Hallinan, R. Hobbs, J. Kocz, S. R. Kulkarni, J. Shi, H. K. Vedantham, *et al.*, Nature, **572**(7769), 352 (Aug. 2019), 1907.01542.

85 J. X. Prochaska, J.-P. Macquart, M. McQuinn, S. Simha, R. M. Shannon, C. K. Day, L. Marnoch, S. Ryder, A. Deller, K. W. Bannister, *et al.*, Science **366**(6462), 231 (Oct. 2019), 1909.11681.

86 J. S. Chittidi, S. Simha, A. Mannings, J. X. Prochaska, M. Rafelski, M. Neeleman, J.-P. Macquart, N. Tejos, R. A. Jorgenson, S. D. Ryder, *et al.*, arXiv e-prints p. arXiv:2005.13158, arXiv:2005.13158 (May 2020), 2005.13158.

87 K. E. Heintz, J. X. Prochaska, S. Simha, E. Platts, W.-f. Fong, N. Tejos, S. D. Ryder, K. Aggerwal, S. Bhandari, C. K. Day, *et al.*, ApJ, **903**(2), 152, 152 (Nov. 2020), 2009.10747.

88 C. J. Law, B. J. Butler, J. X. Prochaska, B. Zackay, S. Burke-Spolaor, A. Mannings, N. Tejos, A. Josephy, B. Andersen, P. Chawla, *et al.*, ApJ, **899**(2), 161, 161 (Aug. 2020), 2007.02155.

89 S. B. Popov and K. A. Postnov, arXiv e-prints p. arXiv:1307.4924, arXiv:1307.4924 (Jul. 2013), 1307.4924.

90 Y. Lyubarsky, MNRAS, **442**, L9 (Jul. 2014), 1401.6674.

91 S. R. Kulkarni, E. O. Ofek, J. D. Neill, Z. Zheng, and M. Juric, ApJ, **797**(1), 70, 70 (Dec. 2014), 1402.4766.

92 J. I. Katz, ApJ, **826**(2), 226, 226 (Aug. 2016), 1512.04503.

93 B. D. Metzger, E. Berger, and B. Margalit, ApJ, **841**(1), 14, 14 (May 2017), 1701.02370.

94 A. M. Beloborodov, ApJL, **843**(2), L26, L26 (Jul. 2017), 1702.08644.

95 Y.-P. Yang and B. Zhang, ApJ, **868**(1), 31, 31 (Nov. 2018), 1712.02702.

96 G. Q. Zhang, S. X. Yi, and F. Y. Wang, ApJ, **893**(1), 44, 44 (Apr. 2020), 2003.01919.

97 B. Zhang, ApJL, **890**(2), L24, L24 (Feb. 2020), 2002.00335.

98 F. Y. Wang, Y. Y. Wang, Y.-P. Yang, Y. W. Yu, Z. Y. Zuo, and Z. G. Dai, ApJ, **891**(1), 72, 72 (Mar. 2020), 2002.03507.

99 S. Yamasaki, T. Totani, and K. Kiuchi, arXiv e-prints p. arXiv:2010.07796, arXiv:2010.07796 (Oct. 2020), 2010.07796.

100 Y. Li and B. Zhang, ApJL, **899**(1), L6, L6 (Aug. 2020), 2005.02371.

101 C. D. Bochenek, V. Ravi, and D. Dong, arXiv e-prints p. arXiv:2009.13030, arXiv:2009.13030 (Sep. 2020), 2009.13030.

102 M. Safarzadeh, J. X. Prochaska, K. E. Heintz, and W.-f. Fong, arXiv e-prints p. arXiv:2009.11735, arXiv:2009.11735 (Sep. 2020), 2009.11735.

103 F. Y. Wang and G. Q. Zhang, ApJ, **882**(2), 108, 108 (Sep. 2019), 1904.12408.

104 W. Lu and P. Kumar, MNRAS, **461**(1), L122 (Sep. 2016), 1605.04605.

105 L.-B. Li, Y.-F. Huang, Z.-B. Zhang, D. Li, and B. Li, Research in Astronomy and Astrophysics **17**(1), 6, 6 (Jan. 2017), 1602.06099.

106 F. Y. Wang and H. Yu, JCAP, **2017**(3), 023, 023 (Mar. 2017), 1604.08676.

107 J. P. Macquart and R. Ekers, MNRAS, **480**(3), 4211 (Nov. 2018),

1808.00908.
108 R. Luo, K. Lee, D. R. Lorimer, and B. Zhang, MNRAS, **481**(2), 2320 (Dec. 2018), 1808.09929.
109 X.-F. Cao, Y.-W. Yu, and X. Zhou, ApJ, **858**(2), 89, 89 (May 2018), 1803.06266.
110 W. Lu and A. L. Piro, ApJ, **883**(1), 40, 40 (Sep. 2019), 1903.00014.
111 Y. Cheng, G. Q. Zhang, and F. Y. Wang, MNRAS, **491**(1), 1498 (Jan. 2020), 1910.14201.
112 F. Y. Wang, G. Q. Zhang, and Z. G. Dai, arXiv e-prints p. arXiv:1903.11895, arXiv:1903.11895 (Mar. 2019), 1903.11895.
113 B. Zhang, ApJL, **867**(2), L21, L21 (Nov. 2018), 1808.05277.
114 C. J. Law, M. W. Abruzzo, C. G. Bassa, G. C. Bower, S. Burke-Spolaor, B. J. Butler, T. Cantwell, S. H. Carey, S. Chatterjee, J. M. Cordes, *et al.*, ApJ, **850**(1), 76, 76 (Nov. 2017), 1705.07553.
115 K. Gourdji, D. Michilli, L. G. Spitler, J. W. T. Hessels, A. Seymour, J. M. Cordes, and S. Chatterjee, ApJL, **877**(2), L19, L19 (Jun. 2019), 1903.02249.
116 Y. G. Zhang, V. Gajjar, G. Foster, A. Siemion, J. Cordes, C. Law, and Y. Wang, ApJ, **866**(2), 149, 149 (Oct. 2018), 1809.03043.
117 P. Scholz, S. Bogdanov, J. W. T. Hessels, R. S. Lynch, L. G. Spitler, C. G. Bassa, G. C. Bower, S. Burke-Spolaor, B. J. Butler, S. Chatterjee, *et al.*, ApJ, **846**(1), 80, 80 (Sep. 2017), 1705.07824.
118 G. Q. Zhang and F. Y. Wang, MNRAS, **487**(3), 3672 (Aug. 2019), 1906.01176.
119 E. GöğüŞ , P. M. Woods, C. Kouveliotou, J. van Paradijs, M. S. Briggs, R. C. Duncan, and C. Thompson, ApJL, **526**(2), L93 (Dec. 1999), astro-ph/9910062.
120 Z. Prieskorn and P. Kaaret, ApJ, **755**(1), 1, 1 (Aug. 2012), 1207.0045.
121 F. Y. Wang, G. Q. Zhang, and Z. G. Dai, arXiv e-prints p. arXiv:1903.11895, arXiv:1903.11895 (Mar. 2019), 1903.11895.
122 R. Luo, Y. Men, K. Lee, W. Wang, D. R. Lorimer, and B. Zhang, MNRAS, **494**(1), 665 (Mar. 2020), 2003.04848.
123 W. Lu, A. L. Piro, and E. Waxman, MNRAS, **498**(2), 1973 (Aug. 2020), 2003.12581.
124 W. Lu, P. Kumar, and B. Zhang, MNRAS, **498**(1), 1397 (Aug. 2020), 2005.06736.
125 R. C. Zhang, B. Zhang, Y. Li, and D. R. Lorimer, MNRAS, (Nov. 2020), 2011.06151.
126 F. Y. Wang and Z. G. Dai, Nature Physics **9**(8), 465 (Aug. 2013), 1308.1253.
127 N. Oppermann, H.-R. Yu, and U.-L. Pen, MNRAS, **475**(4), 5109 (Apr. 2018), 1705.04881.
128 A. Bera, S. Bhattacharyya, S. Bharadwaj, N. D. R. Bhat, and J. N. Chengalur, MNRAS, **457**(3), 2530 (Apr. 2016), 1601.05410.
129 M. Caleb, C. Flynn, M. Bailes, E. D. Barr, R. W. Hunstead, E. F. Keane, V. Ravi, and W. van Straten, MNRAS, **458**(1), 708 (May 2016), 1512.02738.
130 J. P. Macquart and R. D. Ekers, MNRAS, **474**(2), 1900 (Feb. 2018), 1710.11493.
131 Y. Niino, ApJ, **858**(1), 4, 4 (May 2018), 1801.06578.
132 A. Fialkov, A. Loeb, and D. R. Lorimer, ApJ, **863**(2), 132, 132 (Aug. 2018), 1711.04396.
133 N. Locatelli, M. Ronchi, G. Ghirlanda, and G. Ghisellini, A&A, **625**, A109, A109 (May 2019), 1811.10641.
134 S. Ando, JCAP, **2004**(6), 007, 007 (Jun. 2004), astro-ph/0405411.
135 D. Wanderman and T. Piran, MNRAS, **448**(4), 3026 (Apr. 2015), 1405.5878.
136 H. K. Vedantham, V. Ravi, G. Hallinan, and R. M. Shannon, ApJ, **830**(2), 75, 75 (Oct. 2016), 1606.06795.
137 N. Oppermann, L. D. Connor, and U.-L. Pen, MNRAS, **461**(1), 984 (Sep. 2016), 1604.03909.
138 E. Lawrence, S. Vander Wiel, C. Law, S. Burke Spolaor, and G. C. Bower, AJ, **154**(3), 117, 117 (Sep. 2017), 1611.00458.
139 C. Patel, D. Agarwal, M. Bhardwaj, M. M. Boyce, A. Brazier, S. Chatterjee, P. Chawla, V. M. Kaspi, D. R. Lorimer, M. A. McLaughlin, *et al.*, ApJ, **869**(2), 181, 181 (Dec. 2018), 1808.03710.
140 C. W. James, R. D. Ekers, J. P. Macquart, K. W. Bannister, and R. M. Shannon, MNRAS, **483**(1), 1342 (Feb. 2019), 1810.04357.
141 R. M. Shannon, J. P. Macquart, K. W. Bannister, R. D. Ekers, C. W. James, S. Osłowski, H. Qiu, M. Sammons, A. W. Hotan, M. A. Voronkov, *et al.*, Nature, **562**(7727), 386 (Oct. 2018).
142 C.-M. Deng, J.-J. Wei, and X.-F. Wu, Journal of High Energy Astrophysics **23**, 1 (Aug. 2019), 1811.09483.
143 M. Bhattacharya and P. Kumar, arXiv e-prints p. arXiv:1902.10225, arXiv:1902.10225 (Feb. 2019), 1902.10225.
144 J. P. Macquart, R. M. Shannon, K. W. Bannister, C. W. James, R. D. Ekers, and J. D. Bunton, ApJL, **872**(2), L19, L19 (Feb. 2019), 1810.04353.
145 D. Palaniswamy, Y. Li, and B. Zhang, ApJL, **854**(1), L12, L12 (Feb. 2018), 1703.09232.
146 L. Connor, MNRAS, **487**(4), 5753 (Aug. 2019), 1905.00755.
147 L. Connor and E. Petroff, ApJL, **861**(1), L1, L1 (Jul. 2018), 1804.00896.
148 P. Kumar, R. M. Shannon, S. Osłowski, H. Qiu, S. Bhandari, W. Farah, C. Flynn, M. Kerr, D. R. Lorimer, J. P. Macquart, *et al.*, ApJL, **887**(2), L30, L30 (Dec. 2019), 1908.10026.
149 M. Caleb, L. G. Spitler, and B. W. Stappers, Nature Astronomy **2**, 839 (Oct. 2018), 1811.00360.
150 M. Caleb, B. W. Stappers, K. Rajwade, and C. Flynn, MNRAS, **484**(4), 5500 (Apr. 2019), 1902.00272.
151 M. Caleb, B. W. Stappers, T. D. Abbott, E. D. Barr, M. C. Bezuidenhout, S. J. Buchner, M. Burgay, W. Chen, I. Cognard, L. N. Driessen, *et al.*, MNRAS, **496**(4), 4565 (Jun. 2020), 2006.08662.
152 E. Sobacchi, Y. Lyubarsky, A. M. Beloborodov, and L. Sironi, MNRAS, (Oct. 2020), 2010.08282.
153 D. W. Gardenier, L. Connor, J. van Leeuwen, L. C. Oostrum, and E. Petroff, arXiv e-prints p. arXiv:2012.02460, arXiv:2012.02460 (Dec. 2020), 2012.02460.
154 D. W. Gardenier and J. van Leeuwen, arXiv e-prints p. arXiv:2012.06396, arXiv:2012.06396 (Dec. 2020), 2012.06396.
155 L. Connor, M. C. Miller, and D. W. Gardenier, MNRAS, **497**(3), 3076 (Jul. 2020), 2003.11930.
156 S. Ai, H. Gao, and B. Zhang, arXiv e-prints p. arXiv:2007.02400, arXiv:2007.02400 (Jul. 2020), 2007.02400.
157 L. C. Oostrum, Y. Maan, J. van Leeuwen, L. Connor, E. Petroff, J. J. Attema, J. E. Bast, D. W. Gardenier, J. E. Hargreaves, E. Kooistra, *et al.*, A&A, **635**, A61, A61 (Mar. 2020), 1912.12217.
158 C. W. James, MNRAS, **486**(4), 5934 (Jul. 2019), 1902.04932.
159 C. W. James, S. Osłowski, C. Flynn, P. Kumar, K. Bannister, S. Bhandari, W. Farah, M. Kerr, D. R. Lorimer, J. P. Macquart, *et al.*, MNRAS, **495**(2), 2416 (May 2020), 1912.07847.
160 V. Ravi, Nature Astronomy **3**, 928 (Jul. 2019), 1907.06619.
161 K. Aggarwal, C. J. Law, S. Burke-Spolaor, G. Bower, B. J. Butler, P. Demorest, J. Linford, and T. J. W. Lazio, Research Notes of the American Astronomical Society **4**(6), 94, 94 (Jun. 2020), 2006.10513.
162 B. Zhang, Nature, **582**(7812), 344 (Jun. 2020).
163 Z. G. Dai and S. Q. Zhong, ApJL, **895**(1), L1, L1 (May 2020), 2003.04644.
164 W.-M. Gu, T. Yi, and T. Liu, MNRAS, **497**(2), 1543 (Jul. 2020), 2002.10478.
165 K. Ioka and B. Zhang, ApJL, **893**(1), L26, L26 (Apr. 2020), 2002.08297.
166 M. Lyutikov, M. V. Barkov, and D. Giannios, ApJL, **893**(2), L39, L39 (Apr. 2020), 2002.01920.
167 S. B. Popov, Research Notes of the American Astronomical Society **4**(6), 98, 98 (Jun. 2020), 2006.13037.
168 X. Zhang and H. Gao, MNRAS, **498**(1), L1 (Jun. 2020), 2006.10328.
169 S. Du, W. Wang, X. Wu, and R. Xu, MNRAS, **500**(4), 4678 (Jan. 2021), 2004.11223.

170 F. Mottez, P. Zarka, and G. Voisin, arXiv e-prints p. arXiv:2002.12834, arXiv:2002.12834 (Feb. 2020), 2002.12834.
171 Z. G. Dai, J. S. Wang, X. F. Wu, and Y. F. Huang, ApJ, **829**(1), 27, 27 (Sep. 2016), 1603.08207.
172 Y. Levin, A. M. Beloborodov, and A. Bransgrove, ApJL, **895**(2), L30, L30 (Jun. 2020), 2002.04595.
173 D. N. Sob'yanin, MNRAS, **497**(1), 1001 (Jul. 2020), 2007.01616.
174 H. Yang and Y.-C. Zou, ApJL, **893**(2), L31, L31 (Apr. 2020), 2002.02553.
175 J. J. Zanazzi and D. Lai, ApJL, **892**(1), L15, L15 (Mar. 2020), 2002.05752.
176 W.-C. Chen, PASJ, **72**(4), L8, L8 (Aug. 2020), 2006.01552.
177 H. Tong, W. Wang, and H.-G. Wang, Research in Astronomy and Astrophysics **20**(9), 142, 142 (Sep. 2020), 2002.10265.
178 J. I. Katz, MNRAS, **494**(1), L64 (May 2020), 1912.00526.
179 P. Beniamini, Z. Wadiasingh, and B. D. Metzger, MNRAS, **496**(3), 3390 (Jun. 2020), 2003.12509.
180 D. B. Melrose, Reviews of Modern Plasma Physics **1**(1), 5, 5 (Dec. 2017), 1707.02009.
181 P. Kumar, W. Lu, and M. Bhattacharya, MNRAS, **468**(3), 2726 (Jul. 2017), 1703.06139.
182 G. Ghisellini and N. Locatelli, A&A, **613**, A61, A61 (Jun. 2018), 1708.07507.
183 J. I. Katz, MNRAS, **481**(3), 2946 (Dec. 2018), 1803.01938.
184 D. B. Melrose, ARA&A, **29**, 31 (Jan. 1991).
185 D. B. Melrose and M. E. Gedalin, ApJ, **521**(1), 351 (Aug. 1999).
186 E. Waxman, ApJ, **842**(1), 34, 34 (Jun. 2017), 1703.06723.
187 G. Ghisellini, MNRAS, **465**(1), L30 (Feb. 2017), 1609.04815.
188 W. Lu and P. Kumar, MNRAS, **477**(2), 2470 (Jun. 2018), 1710.10270.
189 G. B. Rybicki and A. P. Lightman, *Radiative processes in astrophysics* (1979).
190 K. S. Cheng and J. L. Zhang, ApJ, **463**, 271 (May 1996).
191 D. Viganò, D. F. Torres, K. Hirotani, and M. E. Pessah, MNRAS, **447**(2), 1164 (Feb. 2015), 1411.5836.
192 S. R. Kelner, A. Y. Prosekin, and F. A. Aharonian, AJ, **149**(1), 33, 33 (Jan. 2015), 1501.04994.
193 V. L. Ginzburg and V. V. Zhelezniakov, ARA&A, **13**, 511 (Jan. 1975).
194 R. Buschauer and G. Benford, MNRAS, **177**, 109 (Oct. 1976).
195 G. Benford and R. Buschauer, MNRAS, **179**, 189 (Apr. 1977).
196 A. Saggion, A&A, **44**, 285 (Nov. 1975).
197 A. F. Cheng and M. A. Ruderman, ApJ, **212**, 800 (Mar. 1977).
198 A. Kaganovich and Y. Lyubarsky, ApJ, **721**(2), 1164 (Oct. 2010), 1008.4922.
199 V. L. Ginzburg and V. V. Zhelezniakov, Astronomicheskii Zhurnal, **35**, 694 (Jan. 1958).
200 V. S. Beskin, A. V. Gurevich, and I. N. Istomin, Ap&SS, **146**(2), 205 (Jul. 1988).
201 J. P. Wild, S. F. Smerd, and A. A. Weiss, ARA&A, **1**, 291 (Jan. 1963).
202 V. L. Ginzburg, *Applications of electrodynamics in theoretical physics and astrophysics.* (1989).
203 V. V. Zheleznyakov, *Radiation in Astrophysical Plasmas*, vol. 204 (1996).
204 V. V. Zheleznyakov, G. Thejappa, S. A. Koryagin, and R. G. Stone, Washington DC American Geophysical Union Geophysical Monograph Series **119**, 57 (Jan. 2000).
205 A. Sagiv and E. Waxman, ApJ, **574**(2), 861 (Aug. 2002), astro-ph/0202337.
206 G. Ghisellini and R. Svensson, MNRAS, **252**, 313 (Oct. 1991).
207 J. D. Jackson, *Classical Electrodynamics, 3rd Edition* (1998).
208 D. Alsop and J. Arons, Physics of Fluids **31**(4), 839 (Apr. 1988).
209 M. Hoshino and J. Arons, Physics of Fluids B **3**(3), 818 (Mar. 1991).
210 Y. A. Gallant, M. Hoshino, A. B. Langdon, J. Arons, and C. E. Max, ApJ, **391**, 73 (May 1992).
211 I. Plotnikov and L. Sironi, MNRAS, **485**(3), 3816 (May 2019), 1901.01029.
212 E. Amato and J. Arons, ApJ, **653**(1), 325 (Dec. 2006), astro-ph/0609034.
213 M. Hoshino, J. Arons, Y. A. Gallant, and A. B. Langdon, ApJ, **390**, 454 (May 1992).
214 Y.-P. Yang, J.-P. Zhu, B. Zhang, and X.-F. Wu, ApJL, **901**(1), L13, L13 (Sep. 2020), 2006.03270.
215 B. Margalit, P. Beniamini, N. Sridhar, and B. D. Metzger, ApJL, **899**(2), L27, L27 (Aug. 2020), 2005.05283.
216 D. Xiao and Z.-G. Dai, ApJL, **904**(1), L5, L5 (Nov. 2020), 2010.14787.
217 E. Platts, A. Weltman, A. Walters, S. P. Tendulkar, J. E. B. Gordin, and S. Kandhai, PhR, **821**, 1 (Aug. 2019), 1810.05836.
218 S. Burke-Spolaor, Nature Astronomy **2**, 845 (Oct. 2018), 1811.00194.
219 A. M. Beloborodov, ApJ, **896**(2), 142, 142 (Jun. 2020), 1908.07743.
220 P. Kumar and Ž. Bošnjak, MNRAS, **494**(2), 2385 (Mar. 2020), 2004.00644.
221 R. Sari and T. Piran, ApJL, **455**, L143 (Dec. 1995), astro-ph/9508081.
222 R. D. Blandford and C. F. McKee, Physics of Fluids **19**, 1130 (Aug. 1976).
223 Y.-W. Yu, Y.-C. Zou, Z.-G. Dai, and W.-F. Yu, MNRAS, (Oct. 2020), 2006.00484.
224 Q. Wu, G. Q. Zhang, F. Y. Wang, and Z. G. Dai, ApJL, **900**(2), L26, L26 (Sep. 2020), 2008.05635.
225 J.-S. Wang, ApJ, **900**(2), 172, 172 (Sep. 2020), 2006.14503.
226 W.-Y. Wang, R. Xu, and X. Chen, ApJ, **899**(2), 109, 109 (Aug. 2020), 2005.02100.
227 K. Ioka, ApJL, **904**(2), L15, L15 (Dec. 2020), 2008.01114.
228 S. A. Colgate and A. G. Petschek, ApJ, **248**, 771 (Sep. 1981).
229 J. J. Geng and Y. F. Huang, ApJ, **809**(1), 24, 24 (Aug. 2015), 1502.05171.
230 Z. G. Dai, ApJL, **897**(2), L40, L40 (Jul. 2020), 2005.12048.
231 Z.-N. Liu, W.-Y. Wang, Y.-P. Yang, and Z.-G. Dai, arXiv e-prints p. arXiv:2010.14379, arXiv:2010.14379 (Oct. 2020), 2010.14379.
232 Z. G. Dai and S. Q. Zhong, ApJL, **895**(1), L1, L1 (May 2020), 2003.04644.
233 J.-J. Geng, B. Li, L.-B. Li, S.-L. Xiong, R. Kuiper, and Y.-F. Huang, ApJL, **898**(2), L55, L55 (Aug. 2020), 2006.04601.
234 M. Bhattacharya, P. Kumar, and E. V. Linder, arXiv e-prints p. arXiv:2010.14530, arXiv:2010.14530 (Oct. 2020), 2010.14530.
235 M. Fukugita, C. J. Hogan, and P. J. E. Peebles, ApJ, **503**(2), 518 (Aug. 1998), astro-ph/9712020.
236 J. M. Shull, B. D. Smith, and C. W. Danforth, ApJ, **759**(1), 23, 23 (Nov. 2012), 1112.2706.
237 R. J. Cooke, M. Pettini, and C. C. Steidel, ApJ, **855**(2), 102, 102 (Mar. 2018), 1710.11129.
238 Planck Collaboration, P. A. R. Ade, N. Aghanim, M. Arnaud, M. Ashdown, J. Aumont, C. Baccigalupi, A. J. Banday, R. B. Barreiro, J. G. Bartlett, *et al.*, A&A, **594**, A13, A13 (Sep. 2016), 1502.01589.
239 Z. Li, H. Gao, J.-J. Wei, Y.-P. Yang, B. Zhang, and Z.-H. Zhu, ApJ, **876**(2), 146, 146 (May 2019), 1904.08927.
240 J.-J. Wei, Z. Li, H. Gao, and X.-F. Wu, JCAP, **2019**(9), 039, 039 (Sep. 2019), 1907.09772.
241 D.-C. Qiang and H. Wei, JCAP, **2020**(4), 023, 023 (Apr. 2020), 2002.10189.
242 Z. Zheng, E. O. Ofek, S. R. Kulkarni, J. D. Neill, and M. Juric, ApJ, **797**(1), 71, 71 (Dec. 2014), 1409.3244.
243 J. B. Muñoz and A. Loeb, PhRvD, **98**(10), 103518, 103518 (Nov. 2018), 1809.04074.
244 A. Walters, Y.-Z. Ma, J. Sievers, and A. Weltman, PhRvD, **100**(10), 103519, 103519 (Nov. 2019), 1909.02821.
245 B. Zhou, X. Li, T. Wang, Y.-Z. Fan, and D.-M. Wei, PhRvD, **89**(10), 107303, 107303 (May 2014), 1401.2927.
246 B. Liu, Z. Li, H. Gao, and Z.-H. Zhu, PhRvD, **99**(12), 123517,

123517 (Jun. 2019), 1907.10488.
247 H. Gao, Z. Li, and B. Zhang, ApJ, **788**(2), 189, 189 (Jun. 2014), 1402.2498.
248 J.-J. Wei, X.-F. Wu, and H. Gao, ApJL, **860**(1), L7, L7 (Jun. 2018), 1805.12265.
249 P. Kumar and E. V. Linder, PhRvD, **100**(8), 083533, 083533 (Oct. 2019), 1903.08175.
250 Z.-W. Zhao, Z.-X. Li, J.-Z. Qi, H. Gao, J.-F. Zhang, and X. Zhang, ApJ, **903**(2), 83, 83 (Nov. 2020), 2006.01450.
251 S. Weinberg, *Gravitation and Cosmology: Principles and Applications of the General Theory of Relativity* (1972).
252 H. Yu and F. Y. Wang, A&A, **606**, A3, A3 (Sep. 2017), 1708.06905.
253 Q. Wu, H. Yu, and F. Y. Wang, ApJ, **895**(1), 33, 33 (May 2020), 2004.12649.
254 J. B. Muñoz, E. D. Kovetz, L. Dai, and M. Kamionkowski, PhRvL, **117**(9), 091301, 091301 (Aug. 2016), 1605.00008.
255 L. Dai and W. Lu, ApJ, **847**(1), 19, 19 (Sep. 2017), 1706.06103.
256 Y. K. Wang and F. Y. Wang, A&A, **614**, A50, A50 (Jun. 2018), 1801.07360.
257 R. G. Landim, European Physical Journal C **80**(10), 913, 913 (Oct. 2020), 2005.08621.
258 R. Laha, PhRvD, **102**(2), 023016, 023016 (Jul. 2020).
259 A. Katz, J. Kopp, S. Sibiryakov, and W. Xue, MNRAS, **496**(1), 564 (Jun. 2020), 1912.07620.
260 K. Liao, S. B. Zhang, Z. Li, and H. Gao, ApJL, **896**(1), L11, L11 (Jun. 2020), 2003.13349.
261 M. W. Sammons, J.-P. Macquart, R. D. Ekers, R. M. Shannon, H. Cho, J. X. Prochaska, A. T. Deller, and C. K. Day, ApJ, **900**(2), 122, 122 (Sep. 2020), 2002.12533.
262 Z.-X. Li, H. Gao, X.-H. Ding, G.-J. Wang, and B. Zhang, Nature Communications **9**, 3833, 3833 (Sep. 2018), 1708.06357.
263 A. Walters, A. Weltman, B. M. Gaensler, Y.-Z. Ma, and A. Witzemann, ApJ, **856**(1), 65, 65 (Mar. 2018), 1711.11277.
264 H. Yu and F. Y. Wang, European Physical Journal C **78**(9), 692, 692 (Sep. 2018), 1801.01257.
265 J.-J. Wei, J.-S. Wang, H. Gao, and X.-F. Wu, ApJL, **818**(1), L2, L2 (Feb. 2016), 1601.04145.
266 N. Xing, H. Gao, J.-J. Wei, Z. Li, W. Wang, B. Zhang, X.-F. Wu, and P. Mészáros, ApJL, **882**(1), L13, L13 (Sep. 2019), 1907.00583.
267 X.-F. Wu, J.-J. Wei, M.-X. Lan, H. Gao, Z.-G. Dai, and P. Mészáros, PhRvD, **95**(10), 103004, 103004 (May 2017), 1703.09935.
268 A. Nusser, ApJL, **821**(1), L2, L2 (Apr. 2016), 1601.03636.
269 J.-J. Wei, H. Gao, X.-F. Wu, and P. Mészáros, PhRvL, **115**(26), 261101, 261101 (Dec. 2015), 1512.07670.
270 S. J. Tingay and D. L. Kaplan, ApJL, **820**(2), L31, L31 (Apr. 2016), 1602.07643.
271 K. Dolag, B. M. Gaensler, A. M. Beck, and M. C. Beck, MNRAS, **451**(4), 4277 (Aug. 2015), 1412.4829.
272 A. L. Piro, ApJL, **824**(2), L32, L32 (Jun. 2016), 1604.04909.
273 J.-S. Wang, Y.-P. Yang, X.-F. Wu, Z.-G. Dai, and F.-Y. Wang, ApJL, **822**(1), L7, L7 (May 2016), 1603.02014.
274 B. Margalit, E. Berger, and B. D. Metzger, ApJ, **886**(2), 110, 110 (Dec. 2019), 1907.00016.
275 Y.-P. Yang and B. Zhang, ApJ, **847**(1), 22, 22 (Sep. 2017), 1707.02923.
276 J. Xu and J. L. Han, Research in Astronomy and Astrophysics **15**(10), 1629, 1629 (Oct. 2015), 1504.00200.
277 Y.-P. Yang and B. Zhang, ApJL, **830**(2), L31, L31 (Oct. 2016), 1608.08154.
278 Y.-P. Yang, R. Luo, Z. Li, and B. Zhang, ApJL, **839**(2), L25, L25 (Apr. 2017), 1701.06465.
279 G. Q. Zhang, H. Yu, J. H. He, and F. Y. Wang, ApJ, **900**(2), 170, 170 (Sep. 2020), 2007.13935.
280 M. McQuinn, ApJL, **780**(2), L33, L33 (Jan. 2014), 1309.4451.
281 N. Pol, M. T. Lam, M. A. McLaughlin, T. J. W. Lazio, and J. M. Cordes, ApJ, **886**(2), 135, 135 (Dec. 2019), 1903.07630.
282 W. Zhu and L.-L. Feng, arXiv e-prints p. arXiv:2011.08519, arXiv:2011.08519 (Nov. 2020), 2011.08519.
283 Y. Li, B. Zhang, K. Nagamine, and J. Shi, ApJL, **884**(1), L26, L26 (Oct. 2019), 1906.08749.
284 S. Simha, J. N. Burchett, J. X. Prochaska, J. S. Chittidi, O. Elek, N. Tejos, R. Jorgenson, K. W. Bannister, S. Bhandari, C. K. Day, *et al.*, ApJ, **901**(2), 134, 134 (Oct. 2020), 2005.13157.
285 M. Jaroszynski, MNRAS, **484**(2), 1637 (Apr. 2019), 1812.11936.
286 R. Takahashi, K. Ioka, A. Mori, and K. Funahashi, arXiv e-prints p. arXiv:2010.01560, arXiv:2010.01560 (Oct. 2020), 2010.01560.
287 X.-F. Wu, S.-B. Zhang, H. Gao, J.-J. Wei, Y.-C. Zou, W.-H. Lei, B. Zhang, Z.-G. Dai, and P. Mészáros, ApJL, **822**(1), L15, L15 (May 2016), 1602.07835.
288 L. Bonetti, J. Ellis, N. E. Mavromatos, A. S. Sakharov, E. K. Sarkisyan-Grinbaum, and A. D. A. M. Spallicci, Physics Letters B **757**, 548 (Jun. 2016), 1602.09135.
289 L. Shao and B. Zhang, PhRvD, **95**(12), 123010, 123010 (Jun. 2017), 1705.01278.
290 Planck Collaboration, N. Aghanim, Y. Akrami, M. Ashdown, J. Aumont, C. Baccigalupi, M. Ballardini, A. J. Banday, R. B. Barreiro, N. Bartolo, *et al.*, A&A, **641**, A6, A6 (Sep. 2020), 1807.06209.
291 E. V. Linder, PhRvD, **101**(10), 103019, 103019 (May 2020), 2001.11517.
292 S. Xu and B. Zhang, ApJL, **898**(2), L48, L48 (Aug. 2020), 2007.04089.